\documentclass[]{jfm}

\usepackage{graphicx}
\usepackage{newtxtext}
\usepackage{newtxmath}
\usepackage{natbib}
\usepackage{graphicx}
\usepackage{xcolor}
\usepackage{hyperref}
\usepackage{subcaption}
\usepackage{longtable,tabularx}
\usepackage{float}
\usepackage{tcolorbox}
\usepackage{placeins}
\hypersetup{
    colorlinks = true,
    urlcolor   = blue,
    citecolor  = black,
}

\newcommand{\RomanNumeralCaps}[1]
\nolinenumbers

\title{Effects of Spanwise Curvature on Hypersonic Boundary
Layer Transition in G{\"o}rtler-driven Flows}

\author{Andrew Welter\aff{1}
  \corresp{\email{awelter@mit.edu}},
  Ngoc Cuong Nguyen\aff{1},
  Nicholas Webber\aff{2},
  Mark Gragston\aff{2},
  Jaime Peraire\aff{1},
  \and Wesley Harris\aff{1}}
  
\affiliation{\aff{1}Massachusetts Institute of Technology, 77 Massachusetts Ave., Cambridge, MA, USA 
\aff{2}University of Tennessee Space Institute, 411 B. H. Goethert Parkway, Tullahoma, TN, USA}

\begin{document}
\maketitle

\begin{abstract}
The influence of spanwise convex curvature on hypersonic boundary-layer transition over a streamwise-concave surface is investigated using direct numerical simulation and Mach 7 experiments. A family of smooth doubly curved geometries with fixed streamwise concavity and increasing spanwise convex curvature is considered to isolate the effects of three-dimensional wall curvature on instability dynamics of a centrifugally unstable boundary layer. Increasing spanwise curvature delays transition while preserving a Görtler-dominated route to breakdown. As spanwise curvature increases, the boundary layer becomes thinner, reducing the effective Görtler number and weakening the centrifugal amplification of streamwise vortices. Consequently, the growth of the primary Görtler instability is reduced, the vortical structures remain closer to the wall, and the onset of secondary instability and turbulent breakdown is displaced downstream. Spectral proper orthogonal decomposition confirms that low-frequency vortical dynamics remain dominant in controlling transition, with sinuous secondary instability of the Görtler vortices providing the principal route to breakdown. Meanwhile, boundary-layer thinning modifies the acoustic instability dynamics: the second Mack mode shifts to higher frequencies and higher-frequency supersonic disturbances become increasingly prominent near breakdown as spanwise curvature increases. These acoustic disturbances remain subdominant and do not replace the Görtler–sinuous mechanism as the primary transition pathway. The numerical and experimental results exhibit consistent trends in transition location, thermal-streak organisation and disturbance evolution, supporting a common physical interpretation. These findings demonstrate that spanwise convex curvature has a stabilising influence on hypersonic Görtler-dominated boundary layers, primarily by suppressing centrifugal amplification and delaying secondary instability, while leaving the transition pathway qualitatively unchanged.

\end{abstract}

\begin{keywords}
Authors should not enter keywords on the manuscript, as these must be chosen by the author during the online submission process and will then be added during the typesetting process (see \href{https://www.cambridge.org/core/journals/journal-of-fluid-mechanics/information/list-of-keywords}{Keyword PDF} for the full list).  Other classifications will be added at the same time.
\end{keywords}

{\bf MSC Codes }  {\it(Optional)} Please enter your MSC Codes here

\section{Introduction}
\label{sec:headings}

Accurate prediction of hypersonic boundary-layer transition remains a central challenge in vehicle design because the onset of turbulence produces substantial increases in aerodynamic heating and wall shear, with direct consequences for thermal-protection-system design and thermal margins \citep{Sandhametal2014,Schneider2001,ZhongWang2012}. Errors in transition prediction can therefore lead either to unnecessarily conservative thermal-protection-system designs or to underestimation of peak thermal loads \citep{LI2026111144}. This challenge becomes increasingly important as hypersonic vehicles move beyond canonical and simple geometries towards more aerodynamically efficient configurations, for which boundary layers develop over surfaces with curvature in multiple directions.

Hypersonic boundary-layer transition can involve a broad range of instability mechanisms. In addition to acoustic instabilities such as Mack modes, curved hypersonic boundary layers may support centrifugal and crossflow instabilities, exhibit modified receptivity to freestream disturbances and sustain nonlinear interactions between vortical and acoustic modes \citep{Chen_Xi_Ren_Fu_2022,Craig_Humble_Hofferth_Saric_2019,Viaro_Ricco_20192,Viaro_Ricco_2019,10.1063/5.0028698,XuRiccoDuan2024}. Surface curvature is particularly important because it can modify both the mean boundary-layer development and the amplification and interaction of these instability mechanisms. Streamwise concave curvature promotes G{\"o}rtler instability through centrifugal effects, leading to the amplification of streamwise-oriented vortical structures, whereas spanwise curvature alters the transverse geometry of the boundary layer and can modify its mean-flow characteristics and instability dynamics. These effects are of practical relevance because inward-turning inlets and highly curved external surfaces can improve aerodynamic performance while also modifying transition and surface heating \citep{FlahertyAustin2010,berry2001hypersonic}.

Most previous studies of curvature-induced transition in hypersonic boundary layers have focused on streamwise concavity in canonical configurations such as concave ramps and flared cones, where the amplification of G{\"o}rtler vortices and their subsequent secondary instability have been investigated extensively \citep{chen2019hypersonic,li2022secondary}. Much less is known about the independent influence of spanwise curvature, particularly spanwise convex curvature, despite its presence together with streamwise curvature on many three-dimensional hypersonic surfaces \citep{Malik1991AxisymmetricStability,XuRiccoDuan2024}. The present study addresses this gap by isolating the effect of spanwise convex curvature on a G{\"o}rtler-dominated hypersonic boundary layer while holding the streamwise concavity fixed. Specifically, we study how spanwise curvature modifies the amplification of the primary G{\"o}rtler instability, the development of secondary instability, the relative prominence of acoustic disturbances and, ultimately, the location of boundary-layer transition.

\subsection{The G{\"o}rtler Instability}

G{\"o}rtler instability arises when a boundary layer develops over a streamwise-concave surface and the associated streamline curvature produces a destabilising centrifugal effect \citep{gortler1954three}. Spanwise variations in the disturbance field can then amplify into streamwise-oriented, counter-rotating vortex pairs. These vortices belong to the broader class of centrifugal instabilities that includes Taylor and Dean modes \citep{drazin2004hydrodynamic}. Although the classical theory was developed for incompressible flow, the same centrifugal mechanism persists in compressible and hypersonic boundary layers, where G{\"o}rtler vortices have been observed and analysed in canonical configurations such as concave ramps and flared cones \citep{chen2019hypersonic,li2022secondary}.

As the primary G{\"o}rtler vortices amplify, they redistribute momentum within the boundary layer and generate streamwise streaks of alternating high- and low-speed fluid. At sufficiently large amplitudes, the resulting distorted flow can become susceptible to secondary instabilities that provide a route to turbulent breakdown. In hypersonic boundary layers, these secondary instabilities coexist with compressibility-induced acoustic instabilities, including Mack modes, but arise through distinct physical mechanisms. The former develop on the finite-amplitude vortices and streaks generated by centrifugal amplification, whereas the latter are acoustic disturbances supported by the compressible boundary layer. Previous studies of hypersonic G{\"o}rtler-dominated transition have identified sinuous secondary instability as an important mechanism leading to breakdown, although other secondary disturbances may also arise depending on the flow and disturbance environment \citep{Souza2017}.

The development of G{\"o}rtler vortices also depends strongly on receptivity, through which external or wall-based disturbances excite the centrifugal instability. Relevant excitation mechanisms include freestream vortical disturbances, wall blowing and suction, distributed roughness and synthetic freestream turbulence \citep{xu2024excitation,chen2019hypersonic,bottaro1997gortler,nguyen2023implicit}. The nature and amplitude of this forcing influence the initial vortex structure and its subsequent amplification, thereby affecting the onset of secondary instability and transition. Conversely, thermal forcing and surface modifications such as grooves have been investigated as means of suppressing G{\"o}rtler-vortex amplification and delaying breakdown \citep{elhady1981growth,kahawita1977influence,park2025delaying,Huang2023,HUANG2025204335}.

While these studies have established the principal mechanisms governing G{\"o}rtler-dominated transition over streamwise-concave surfaces, much less is known about how they are modified when the surface is also curved in the spanwise direction. In particular, spanwise convex curvature can alter the boundary-layer development and thereby modify the centrifugal amplification of the primary G{\"o}rtler vortices, their secondary instability and their interaction with acoustic disturbances. Understanding how these effects combine to impact the eventual transition and breakdown to turbulence is a central objective of the present study.

\subsection{Spanwise Curvature and Hypersonics}



In hypersonic flow, G{\"o}rtler instability is particularly important because the streamwise vortices generated by centrifugal amplification can produce substantial local heat-transfer augmentation before the boundary layer becomes fully turbulent. This behaviour has been documented in flared-cone experiments, where G{\"o}rtler vortices and their subsequent breakdown are associated with pronounced surface-heating signatures \citep{doi:10.2514/1.A34285}. Concave ramps and flared cones have therefore become important canonical configurations for studying curvature-induced transition in hypersonic boundary layers \citep{HUANG2025204335,DuanWangZhong2013,Wang2021}. In these flows, centrifugal disturbances may coexist with first- and second-mode instabilities, such that the transition pathway depends on the relative amplification of the different instability mechanisms, as well as on the disturbance environment and their nonlinear evolution \citep{XuRiccoDuan2024,pruett1998dns,laible2011numerical,balakumar2015receptivity,hader2016laminar}.

Acoustic instabilities remain an important component of hypersonic boundary-layer dynamics even when centrifugal disturbances dominate the vortical response. In particular, the second Mack mode is a trapped acoustic instability associated with the sonic-line structure of the boundary layer and is generally more strongly amplified than Tollmien–Schlichting waves at hypersonic speeds \citep{10.1063/1.5077007}. G{\"o}rtler vortices can modify the environment in which these acoustic disturbances evolve by redistributing momentum and producing spanwise variations in the local boundary-layer thickness and mean-flow structure \citep{Wang2021}. Centrifugal and acoustic instabilities may therefore coexist and interact through the distorted boundary layer, with their relative importance depending on the geometry, disturbance environment and downstream evolution of the flow.

Most previous studies of curvature-induced hypersonic transition have focused on streamwise concavity, which provides the centrifugal mechanism responsible for G{\"o}rtler instability. Three-dimensional surfaces, however, may possess substantial curvature in both the streamwise and spanwise directions. While streamwise concavity directly drives centrifugal amplification, spanwise curvature modifies the three-dimensional development of the boundary layer and can thereby alter its susceptibility to both vortical and acoustic instabilities. The independent influence of spanwise curvature, particularly spanwise convex curvature, remains much less well understood than that of streamwise concavity. By varying the spanwise convex curvature while holding the streamwise concavity fixed, the present study isolates this geometric effect and examines how it modifies the instability dynamics and transition of a G{\"o}rtler-dominated hypersonic boundary layer.

\subsection{Mangler--Levy--Lees transformation}

To interpret how spanwise convex curvature modifies the mean hypersonic boundary layer, we employ the Mangler--Levy--Lees (MLL) transformation in the locally axisymmetric form used by \citep{Malik1991AxisymmetricStability}, following the compressible transformation of \citep{Lees1956}. In the present problem, the streamwise concavity responsible for the primary G{\"o}rtler instability is held fixed, while the transverse radius is varied in a controlled comparison to isolate the influence of transverse, or spanwise, curvature on the mean boundary-layer development.

Let \(x\) and \(y\) denote the streamwise and wall-normal coordinates, respectively, with \(y>0\) directed away from the wall, and let \((\bar U,\bar V,\bar T,\bar\rho,\bar p)\) denote the steady mean flow. Under the compressible boundary-layer approximation, the governing equations for a locally axisymmetric geometry are
\begin{align}
\bar\rho \bar U \frac{\partial \bar U}{\partial x}
+
\bar\rho \bar V \frac{\partial \bar U}{\partial y}
&=
-\frac{d\bar p_e}{dx}
+
\frac{1}{r}\frac{\partial}{\partial y}
\left(
r\bar\mu\frac{\partial\bar U}{\partial y}
\right),
\label{eq:BL_momentum}
\\
\frac{\partial}{\partial x}(r\bar\rho\bar U)
+
\frac{\partial}{\partial y}(r\bar\rho\bar V)
&=0,
\label{eq:BL_continuity}
\\
\bar\rho c_p\bar U\frac{\partial\bar T}{\partial x}
+
\bar\rho c_p\bar V\frac{\partial\bar T}{\partial y}
&=
\frac{1}{r}\frac{\partial}{\partial y}
\left(
r\bar k\frac{\partial\bar T}{\partial y}
\right)
+
\bar\mu
\left(
\frac{\partial\bar U}{\partial y}
\right)^2
+
\bar U\frac{d\bar p_e}{dx},
\label{eq:BL_energy}
\end{align}
together with
\begin{equation}
\bar p=\bar p_e=\bar\rho R\bar T,
\qquad
\frac{\partial\bar p}{\partial y}=0,
\label{eq:BL_EOS}
\end{equation}
and
\begin{equation}
\bar k=\frac{\bar\mu c_p}{Pr},
\label{eq:k_Pr}
\end{equation}
where \(\bar\mu=\bar\mu(\bar T)\). These equations assume a thin boundary layer, negligible streamwise viscous diffusion, a calorically perfect gas, and constant \(Pr\).

The geometric radius is written as
\begin{equation}
r(x,y)=r_0(x)+y\cos\theta,
\label{eq:r_geom}
\end{equation}
where \(r_0(x)\) is the local transverse radius of curvature at the wall. For the local transverse-curvature model considered here, \(\theta=0\), so that \(r=r_0+y\). With the sign convention adopted here, the spanwise convex curvature is
\begin{equation}
\kappa_z=\frac{1}{r_0},
\end{equation}
and increasing spanwise convex curvature therefore corresponds to decreasing \(r_0\). This transverse curvature is distinct from the streamwise concavity that drives the G{\"o}rtler instability.

Following \cite{Lees1956}, the transformed coordinates are defined by
\begin{equation}
\xi(x)=\int_0^x
\bar\rho_e\bar\mu_e\bar U_e r_0^{\,2}\,dx',
\qquad
\eta(x,y)=
\frac{\bar U_e}{\sqrt{2\xi}}
\int_0^y
\bar\rho(x,y')r(x,y')\,dy',
\label{eq:MLL_coords}
\end{equation}
where subscript \(e\) denotes the boundary-layer-edge value. Introducing
\begin{equation}
F=\frac{\bar U}{\bar U_e},
\qquad
G=\frac{\bar T}{\bar T_e},
\label{eq:FG_def}
\end{equation}
and using the inviscid edge relations
\begin{equation}
\frac{d\bar p_e}{dx}
=
-\bar\rho_e\bar U_e\frac{d\bar U_e}{dx},
\qquad
c_p\frac{d\bar T_e}{dx}
=
-\bar U_e\frac{d\bar U_e}{dx},
\label{eq:edge_relations}
\end{equation}
one obtains
\begin{equation}
c_p\bar T_e+\frac{\bar U_e^2}{2}
=
\mathrm{constant},
\qquad
\frac{\bar\rho_e}{\bar\rho}
=
\frac{\bar T}{\bar T_e}
=G.
\label{eq:edge_energy_density}
\end{equation}
The transformed wall-normal velocity is defined by
\begin{equation}
V=
\frac{2\xi}
{\bar\rho_e\bar\mu_e\bar U_e r_0^{\,2}}
\left[
F\left.\frac{\partial\eta}{\partial x}\right|_y
+
\frac{\bar\rho\bar V r}{\sqrt{2\xi}}
\right].
\label{eq:V_def}
\end{equation}
The locally axisymmetric compressible boundary-layer equations then take the MLL form \citep{Lees1956,Malik1991AxisymmetricStability}
\begin{align}
\frac{\partial}{\partial\eta}
\left(
a_1\frac{\partial F}{\partial\eta}
\right)
-
V\frac{\partial F}{\partial\eta}
+
a_2(G-F^2)
&=
2\xi F\frac{\partial F}{\partial\xi},
\label{eq:MLL_momentum}
\\
\frac{\partial V}{\partial\eta}+F
&=
-2\xi\frac{\partial F}{\partial\xi},
\label{eq:MLL_continuity}
\\
\frac{\partial}{\partial\eta}
\left(
a_3\frac{\partial G}{\partial\eta}
\right)
-
V\frac{\partial G}{\partial\eta}
+
a_4
\left(
\frac{\partial F}{\partial\eta}
\right)^2
&=
2\xi F\frac{\partial G}{\partial\xi},
\label{eq:MLL_energy}
\end{align}
where
\begin{equation}
a_1=(1+\chi)^2
\frac{\bar\rho\bar\mu}
{\bar\rho_e\bar\mu_e},
\qquad
a_2=\frac{2\xi}{\bar U_e}
\frac{d\bar U_e}{d\xi},
\qquad
a_3=\frac{a_1}{Pr},
\qquad
a_4=(\gamma-1)M_e^2a_1,
\label{eq:a1234}
\end{equation}
with
\begin{equation}
\chi=\frac{y\cos\theta}{r_0},
\qquad
M_e=\frac{\bar U_e}{\sqrt{\gamma R\bar T_e}}.
\label{eq:chi_def}
\end{equation}
For an impermeable isothermal wall, the boundary conditions are
\begin{equation}
F(0)=0,
\qquad
V(0)=0,
\qquad
G(0)=G_w(\xi)=\frac{T_w}{T_e(\xi)},
\label{eq:BC_wall_iso}
\end{equation}
together with
\begin{equation}
F(\infty)=1,
\qquad
G(\infty)=1.
\label{eq:BC_outer}
\end{equation}
Thus, even for constant dimensional wall temperature \(T_w\), the transformed wall-temperature ratio \(G_w\) varies downstream if \(T_e\) varies with \(x\).

The MLL transformation also provides a useful interpretation of the influence of transverse curvature on the physical boundary-layer thickness. For \(\theta=0\), the wall-normal mapping becomes
\begin{equation}
\eta=
\frac{\bar U_e}{\sqrt{2\xi}}
\int_0^y
\bar\rho(x,y')\left(r_0+y'\right)\,dy'.
\label{eq:eta_exact_curved}
\end{equation}
Approximating the density locally by a representative boundary-layer value \(\bar\rho_*\) gives
\begin{equation}
\eta
\approx
\frac{\bar\rho_*\bar U_e}{\sqrt{2\xi}}
\left(
r_0y+\frac{y^2}{2}
\right).
\label{eq:eta_thinning}
\end{equation}
For a local comparison between nearby geometries at the same physical streamwise station, assume that the edge-flow histories are the same and that \(r_0\) is approximately constant over the relevant upstream interval. Defining
\begin{equation}
I(x)=\int_0^x
\bar\rho_e\bar\mu_e\bar U_e\,dx',
\label{eq:I_def}
\end{equation}
the transformed streamwise coordinate becomes
\begin{equation}
\xi=r_0^2 I(x).
\label{eq:xi_scaling}
\end{equation}
Equation \eqref{eq:eta_thinning} then takes the approximate form
\begin{equation}
\eta
\approx
C(x)
\left(
y+\frac{y^2}{2r_0}
\right),
\qquad
C(x)=
\frac{\bar\rho_*\bar U_e}{\sqrt{2I(x)}}.
\label{eq:eta_thinning2}
\end{equation}
Under the assumptions of this controlled comparison, \(C(x)\) is approximately independent of \(r_0\). At fixed \(x\) and \(\eta\), decreasing \(r_0\) increases the curvature contribution \(y^2/(2r_0)\); the corresponding physical wall-normal coordinate \(y\) must therefore decrease. Thus, increasing spanwise convex curvature produces a coordinate-mapping tendency toward a thinner physical boundary layer for otherwise comparable transformed profiles.

This scaling analysis is approximate because transverse curvature also enters the transformed transport coefficient \(a_1\) and can therefore modify the transformed velocity, temperature, and density profiles. Variations in \(r_0\) may also modify the edge state and pressure-gradient history unless these quantities are explicitly held fixed. The result should therefore not be interpreted as a general consequence of the MLL transformation for arbitrary geometries, or as proof that every integral measure of boundary-layer thickness must decrease. Rather, it provides a physical scaling for the closely related geometries considered here. The resulting change in boundary-layer thickness can, in turn, modify the characteristic scales governing both centrifugal and acoustic instabilities, thereby linking spanwise curvature to the instability dynamics and transition behavior.

\section{Numerical methodology}

The simulations are performed with Exasim \citep{Exasim}, using a high-order hybridizable discontinuous Galerkin (HDG) discretization \citep{Peraire2010a,Nguyen2015c,Vila-Perez2021} of the three-dimensional compressible Navier--Stokes equations together with diagonally implicit Runge--Kutta (DIRK) time integration \citep{Nguyen2012,Ciuca2020} and a Newton--Krylov solution strategy \citep{Fernandez2017a,HOSKIN2024100999,WELTER2026118951}. The HDG formulation is attractive for the present DNS because it combines the high-order accuracy and geometric flexibility of discontinuous Galerkin methods with hybridization and static condensation, so that the global implicit solve is posed on trace unknowns defined on the mesh skeleton rather than on the full set of element-interior degrees of freedom \citep{Nguyen2012,Nguyen2015c,Vila-Perez2021}. This reduction in global coupling is particularly useful in hypersonic transitional boundary layers, where instability amplification, localized compressive structures and strong near-wall gradients must all be resolved on highly stretched meshes while retaining a robust fully implicit solve. The use of an implicit DIRK integrator, together with Newton linearization and an Additive Schwarz-preconditioned Krylov method for the condensed trace system, alleviates the severe time-step restriction that would otherwise be imposed by the smallest near-wall elements while preserving the high-order spatial resolution needed for direct numerical simulation \citep{Fernandez2017a,HOSKIN2024100999,WELTER2026118951}.

\subsection{Compressible Navier--Stokes equations}

The governing equations are the dimensional compressible Navier--Stokes equations written in conservative form as
\begin{equation}
\frac{\partial \boldsymbol{U}}{\partial t}
+
\nabla\cdot \boldsymbol{F}^{\mathrm{inv}}(\boldsymbol{U})
-
\nabla\cdot \boldsymbol{F}^{\mathrm{vis}}(\boldsymbol{U},\nabla\boldsymbol{U})
=
\boldsymbol{S}_g,
\label{eq:NS_conservative}
\end{equation}
where $\boldsymbol{U}
= (\rho  \, \rho u  \, \rho v  \, \rho w \, \rho E)^T$.
Here, \(\rho\) is the density, \((u,v,w)\) are the velocity components, and
\begin{equation}
E
=
e+\frac{1}{2}\left(u^2+v^2+w^2\right)
\end{equation}
is the total energy per unit mass. For a calorically perfect gas,
\begin{equation}
p
=
(\gamma-1)\rho e
=
(\gamma-1)
\left[
\rho E
-
\frac{1}{2}\rho\left(u^2+v^2+w^2\right)
\right],
\label{eq:ideal_gas}
\end{equation}
where \(\gamma\) is the ratio of specific heats. The inviscid flux and the viscous flux are given by
\begin{equation}
\boldsymbol{F}^{\mathrm{inv}}(\boldsymbol{U})
=
\begin{bmatrix}
\rho \boldsymbol{v} \\
\rho u\boldsymbol{v}+p\boldsymbol{e}_x \\
\rho v\boldsymbol{v}+p\boldsymbol{e}_y \\
\rho w\boldsymbol{v}+p\boldsymbol{e}_z \\
(\rho E+p)\boldsymbol{v}
\end{bmatrix} \qquad \boldsymbol{F}^{\mathrm{vis}}(\boldsymbol{U},\nabla\boldsymbol{U})
=
\begin{bmatrix}
\boldsymbol{0} \\
\boldsymbol{\tau}_{x} \\
\boldsymbol{\tau}_{y} \\
\boldsymbol{\tau}_{z} \\
\boldsymbol{\tau}\boldsymbol{v}-\boldsymbol{q}
\end{bmatrix}, 
\label{eq:inviscid_flux}
\end{equation}
where $\boldsymbol{v}=(u,v,w)^T$ and \((\boldsymbol{e}_x,\boldsymbol{e}_y,\boldsymbol{e}_z)\) are the Cartesian basis vectors, the Newtonian stress tensor is
\begin{equation}
\boldsymbol{\tau}
=
\mu
\left[
\nabla\boldsymbol{v}
+
\nabla\boldsymbol{v}^{T}
-
\frac{2}{3}
(\nabla\cdot\boldsymbol{v})\boldsymbol{I}
\right],
\label{eq:stress_tensor}
\end{equation}
and the Fourier heat flux is
\begin{equation}
\boldsymbol{q}
=
-k\nabla T,
\qquad
k
=
\frac{\mu c_p}{Pr}.
\label{eq:heat_flux}
\end{equation}
The dynamic viscosity \(\mu\) is prescribed by Sutherland's law, and the Prandtl number \(Pr\) is taken to be constant.

For the flat configuration, the equations are solved in Cartesian form and \(\boldsymbol{S}_g=\boldsymbol{0}\). For the spanwise-curved configurations, the equivalent cylindrical-coordinate form is employed, with \(\boldsymbol{S}_g\) representing the geometric source terms associated with the cylindrical metric \citep{Sandberg2007GovEqNS}. The streamwise concavity responsible for centrifugal destabilization is represented through the body geometry, while spanwise convex curvature enters through the cylindrical geometry and the resulting mean-flow development.

At the inflow boundary, the mean hypersonic inflow state is prescribed together with the synthetic disturbance field described in the following subsection. A no-slip isothermal condition is imposed at the wall. Upstream of the leading edge, a symmetry condition is applied over the portion of the boundary preceding the body surface. The homogeneous lateral direction is treated periodically: the spanwise direction is periodic for the flat configuration, while the corresponding azimuthal direction is periodic for the spanwise-curved configurations. A supersonic outflow condition is imposed at the downstream boundary, consistent with related hypersonic transition studies on flared-cone-type configurations \citep{balakumar2015receptivity,hader2016laminar}. Boundary conditions are imposed weakly through the numerical traces and fluxes, consistent with the standard hybridized discontinuous Galerkin method for compressible flow \citep{Nguyen2012,Vila-Perez2021}. 


\subsection{Numerical discretization}

The governing equations are discretized in space using a high-order hybridizable discontinuous Galerkin (HDG) method \citep{Peraire2010a,Nguyen2012,Nguyen2015c,Vila-Perez2021}. The second-order viscous operator is written as a first-order system by introducing the gradient of the conservative state as an auxiliary variable. Within each curved element, the conservative state and its gradient are approximated by discontinuous polynomial fields of the same degree, while a single-valued polynomial approximation of the conservative state, referred to as the trace variable, is defined on the element faces. In the present simulations, quadratic polynomial approximations are employed for both the element-interior and trace variables. The physical geometry is represented using curved element mappings of the same polynomial order, allowing the discretization to conform accurately to the blunt leading edge and the spanwise-curved surfaces.

The elemental conservation and auxiliary-gradient equations are coupled to the trace variable through numerical fluxes containing the inviscid and viscous contributions together with a stabilization term that penalizes differences between the element-interior and trace states. The trace solution is determined by enforcing continuity of the normal numerical flux across interior faces. Consequently, conservation is satisfied locally within each element and the normal numerical flux is conservative across the mesh skeleton. At physical boundaries, the corresponding boundary conditions are imposed weakly through the numerical fluxes. For the spanwise-curved configurations, the cylindrical metric terms and associated geometric source terms are incorporated consistently into the elemental residuals. The resulting formulation therefore retains the local conservation, high-order accuracy, and geometric flexibility characteristic of discontinuous Galerkin methods while reducing the number of globally coupled degrees of freedom \citep{Peraire2010a,Nguyen2015c,Vila-Perez2021}.

A central feature of HDG is that global coupling is confined to the trace unknowns on the mesh skeleton. For a prescribed trace state, the conservative and auxiliary-gradient variables can be determined independently within each element. During an implicit solve, the nonlinear HDG residual is linearized by Newton’s method. At each Newton iteration, the linearized element-interior variables are eliminated locally by static condensation, yielding a global Schur-complement system involving only the trace correction. After the trace system is solved, the corrections to the conservative and gradient variables are recovered independently within each element. Thus, although the full discretization contains a very large number of degrees of freedom, only the face-based trace variables participate directly in the globally coupled linear solve. The element-local operations required for condensation and recovery can be performed concurrently, which is particularly advantageous for the highly parallel GPU computations considered here \citep{Nguyen2012,HOSKIN2024100999,WELTER2026118951}.

Temporal discretization is performed using a three-stage, third-order diagonally implicit Runge–Kutta scheme, denoted DIRK(3,3) \citep{Nguyen2012,Ciuca2020}. The Runge–Kutta stages are advanced sequentially, with the converged solutions from preceding stages providing the known contributions to the current stage. At every stage, the coupled conservation, auxiliary-gradient, and trace-transmission residuals are converged using Newton iterations. Each Newton linearization is statically condensed as described above, and the resulting global trace system is solved using restarted GMRES with an Additive Schwarz preconditioner \citep{Fernandez2017a,HOSKIN2024100999,WELTER2026118951}. The Krylov operator and subsequent recovery of the element-interior corrections reuse the element-local factorizations constructed during the current Newton iteration. This organization concentrates interelement communication in the trace solve while retaining the predominantly element-local computational structure of HDG, enabling efficient execution on the large distributed GPU systems used for the present DNS.

The combination of high-order spatial approximation and implicit time integration is particularly important for the present hypersonic boundary-layer calculations. Strong wall-normal mesh stretching is required to resolve the viscous and thermal boundary layers, while the streamwise and lateral resolution must be sufficient to capture the growth of instability waves and the emergence of three-dimensional structures during transition. The implicit DIRK formulation avoids a time-step restriction based solely on the smallest near-wall element, whereas the low-dissipation high-order HDG discretization provides the spatial resolution required to follow instability amplification over long propagation distances. Numerical dissipation associated with shock regularization is introduced separately and only in localized compressive regions, as described in the following subsection. The polynomial degree, time step, mesh resolution, and other case-specific discretization parameters used in the simulations are reported in the simulation setup.

\subsection{Shock capturing}

The simulations contain both smooth instability dynamics and localized compressive structures. To regularize the latter while preserving the resolution of instability waves and vortical motions in smooth regions, the discretization is augmented with a localized scalar artificial-viscosity field, following the broader use of artificial-viscosity shock treatment in compressible DG and HDG discretizations \citep{Moro2011a,Vila-Perez2021,HOSKIN2024100999,Nguyen2023c,nguyen2023optimal}. The sensor is built from the local compressive strength of the velocity field \citep{Moro2016} and is modulated by a Ducros-type vortical-suppression factor \citep{Fernandez2018}:
\begin{equation}
\mathcal{D}
=
\frac{
(\nabla\cdot\boldsymbol{v})^2
}{
(\nabla\cdot\boldsymbol{v})^2
+
|\nabla\times\boldsymbol{v}|^2
+
\varepsilon
},
\label{eq:ducros_switch}
\end{equation}
where \(\varepsilon\) is a small positive constant introduced to avoid division by zero. The switch \(\mathcal{D}\) approaches unity in strongly compressive regions with weak rotational content and decreases toward zero in vortical flow, thereby suppressing spurious activation of artificial viscosity in the instability-dominated portions of the boundary layer. In practice, the artificial viscosity is activated only when \(\nabla\cdot\boldsymbol{v}<0\), so expansive motions do not contribute to the shock sensor.

The resulting scalar viscosity enters the discretization through the diffusive operator as an additional dissipative contribution localized near shocklets, compression waves and other steep compressive structures. Its role is purely numerical: it broadens under-resolved compressive features sufficiently to prevent non-physical oscillations while remaining small in smooth regions of the flow. In this sense, the stabilization acts as a shock-regularization mechanism for the compressible Navier--Stokes discretization. The magnitude of the added dissipation is therefore controlled by local compression and by the Ducros factor in \eqref{eq:ducros_switch}, so that the method preferentially damps unresolved compressive structures without intentionally altering the resolved laminar--turbulent transition dynamics away from them.

\subsection{Synthetic freestream disturbance}\label{sec:synthetic}

Transition is initiated by superimposing a synthetic velocity-disturbance field on the mean inlet state. The perturbations are generated with a random Fourier construction following \cite{Kraichnan1970,Bechara1994,BaillyJuve1999} and the implementation adopted in related hypersonic-transition simulations \citep{HOSKIN2024100999}. The fluctuating velocity field is written as
\begin{equation}
\boldsymbol{v}'(\boldsymbol{x},t)
=
2
\sum_{n=1}^{N}
\hat{u}_n\boldsymbol{\sigma}_n
\cos
\left[
\boldsymbol{k}_n\cdot
\left(
\boldsymbol{x}-\boldsymbol{v}_c t
\right)
+
\psi_n
+
\omega_n t
\right],
\label{eq:synthetic_velocity}
\end{equation}
where \(\boldsymbol{k}_n\) is the wave vector of the \(n\)th mode, \(\boldsymbol{\sigma}_n\) its polarization vector, \(\psi_n\) a random phase, and \(\omega_n\) its angular frequency. The convective velocity is taken to be \(\boldsymbol{v}_c=\boldsymbol{u}_{\infty}\), so the synthetic field is advected at the mean freestream speed. The vectors \(\boldsymbol{k}_n\) are sampled over wave-number shells and the corresponding polarization vectors are chosen to satisfy
\begin{equation}
\boldsymbol{\sigma}_n\cdot\boldsymbol{k}_n=0,
\end{equation}
so that the synthesized velocity field is solenoidal mode by mode. The random phases and wave-vector orientations therefore produce a broadband perturbation field with no preferred lateral direction at the level of the underlying synthetic spectrum.

The modal amplitudes are prescribed from a target isotropic energy spectrum according to
\begin{equation}
\hat{u}_n
=
\sqrt{E(k_n)\Delta k_n},
\label{eq:mode_amplitude}
\end{equation}
where \(k_n=|\boldsymbol{k}_n|\) and \(\Delta k_n\) is the corresponding wave-number interval. In practice, the wave numbers are distributed over a finite range spanning the energetic and dissipative parts of the prescribed spectrum. A von K\'arm\'an--Pao form is used,
\begin{equation}
E(k)
=
\alpha
\frac{I^2u_\infty^2}{k_e}
\frac{(k/k_e)^4}
{\left[1+(k/k_e)^2\right]^{17/6}}
\exp
\left[
-2(k/k_\eta)^2
\right],
\label{eq:vk_pao}
\end{equation}
where \(I\) is the prescribed disturbance-intensity parameter, \(k_e\) is the energetic wave number associated with the largest energy-containing eddies, and \(k_\eta\) is the Kolmogorov wave number associated with the smallest represented scales \citep{BaillyJuve1999,HOSKIN2024100999}. This construction provides a controlled broadband disturbance environment at the inlet while avoiding the introduction of a single imposed instability mode.


Only the velocity components are perturbed at the inlet. The density and temperature retain their mean inflow values, and the total energy is updated consistently with the perturbed velocity through \eqref{eq:ideal_gas}. The inlet condition therefore introduces a prescribed disturbance environment for receptivity and transition studies without modifying the mean thermodynamic state. 

The disturbance intensity was deliberately set to a low value, $I = 0.025\%$. Computationally, the intensity level is observed to shift the transition location but not to alter the instability mechanisms or the curvature trends that are the focus of this study. The chosen level lies in a regime where further reduction of the intensity has no significant effect on the solution, so that the computed transition dynamics are not tied to a particular forcing amplitude. No attempt is made to reproduce the disturbance environment of the wind tunnel, which is substantially different, and nominally noisier (\S\ref{subsec:ES-Facility}); the role of the experiments is to confirm the main findings of the simulations, at least qualitatively, rather than to provide a direct quantitative match.
\subsection{Simulation Parameters}
\begin{figure}
     \centering
     \begin{subfigure}[b]{0.48\textwidth}
         \centering
         \includegraphics[width=\textwidth]{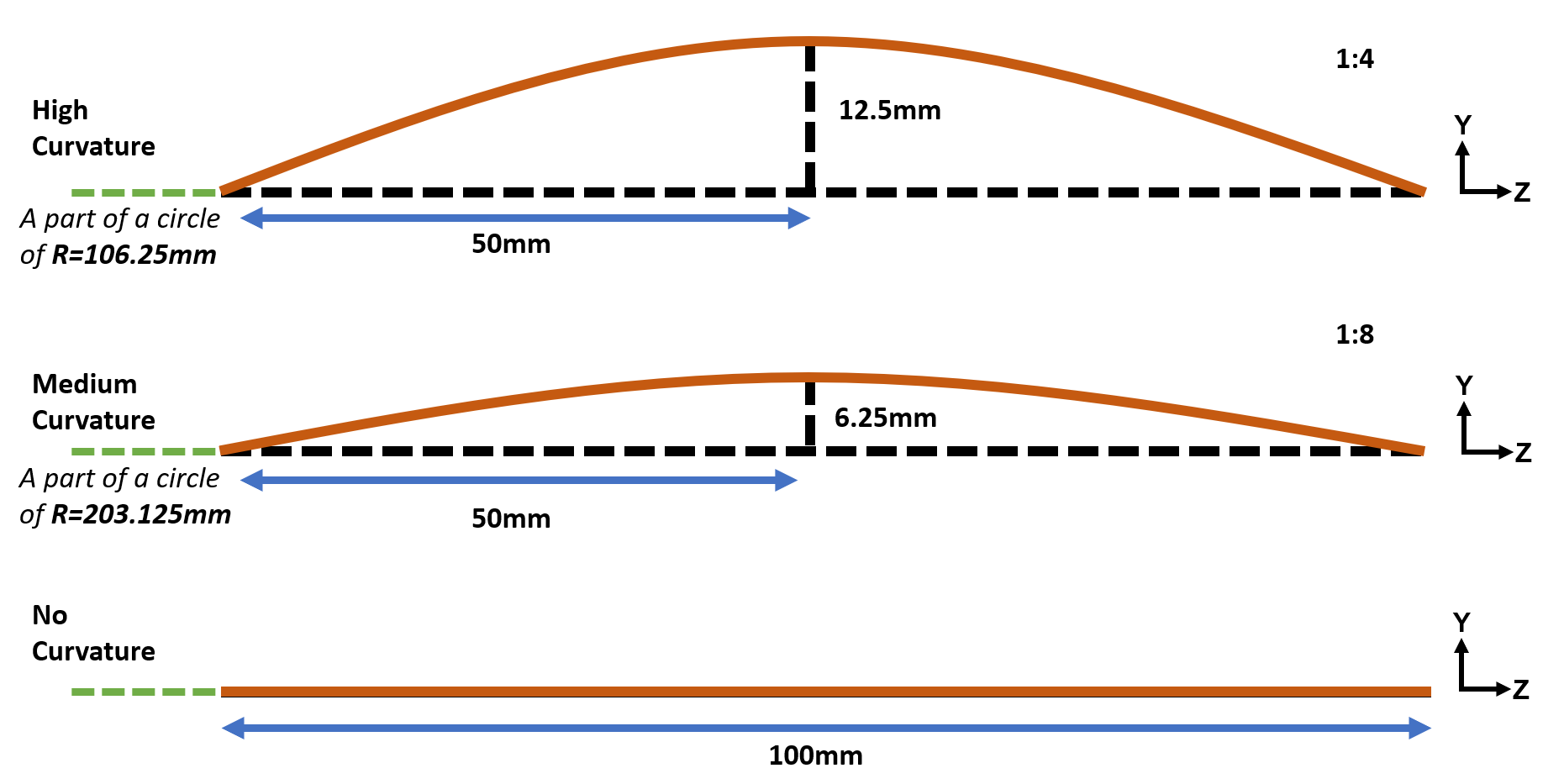}
         \label{fig:spanwise_geom}
     \end{subfigure}
     \hfill
     \begin{subfigure}[b]{0.48\textwidth}
         \centering
         \includegraphics[width=\textwidth]{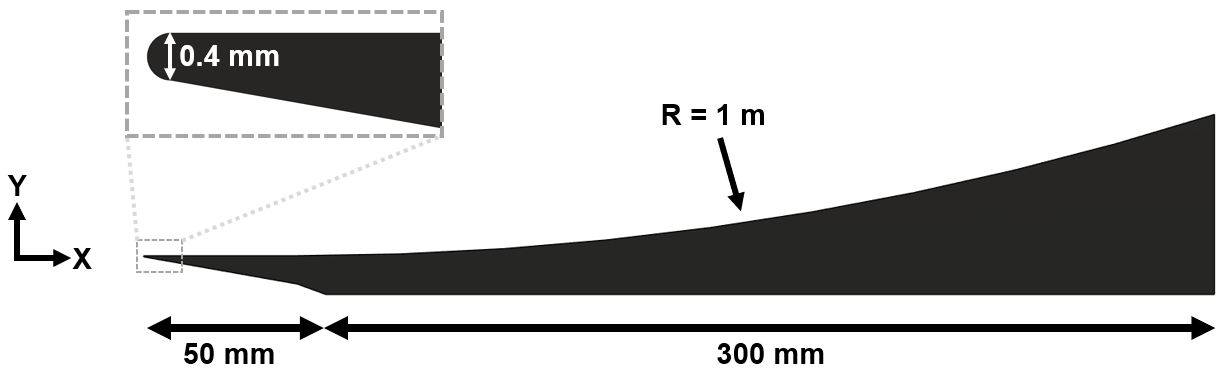}
         \label{fig:streamwise_geom}
     \end{subfigure}
     \caption{Spanwise geometry definitions (left), streamwise geometry definition (right).}
     \label{fig:geometry_definitions}
\end{figure}

Three configurations were simulated using the geometry shown in Figure~\ref{fig:geometry_definitions} with transverse curvature parameters specified in Table~\ref{tab:transverse_curvature}. All cases shared the same streamwise profile, including the blunt leading edge, while only the spanwise curvature was varied. The flow field was computed in Exasim \citep{Exasim} using the hybridizable discontinuous Galerkin (HDG) method applied to the unsteady compressible Navier--Stokes equations. The flat-span case was solved in Cartesian coordinates, whereas the curved cases were solved in cylindrical coordinates to conform to the geometry. Freestream conditions were selected to match the UTSI experiments, with $M_{\infty}=7$, $Re_{\infty}=8.088\times10^6\,\mathrm{m^{-1}}$, and $T_{\infty}=51.86\,\mathrm{K}$. The wall was treated as no-slip and isothermal at $296\,\mathrm{K}$. Air was modeled as a calorically perfect gas with $\gamma=1.4$, $Pr=0.71$, and Sutherland-law viscosity under frozen-chemistry conditions.

\begin{table}
  \centering
  \caption{Radius of curvature and local transverse curvature parameter for all profiles.}
  \label{tab:transverse_curvature}
  \begin{tabular}{lccc}
    \hline
    Profile & $r_0$ (mm) & $1/r_0$ (m$^{-1}$) & Local $\chi$ \\
    \hline
    High curvature   & 106.25   & 9.411765 & $\chi_H = \dfrac{y\cos\theta}{106.25}$ \\[4pt]
    Medium curvature & 203.125  & 4.923077 & $\chi_M = \dfrac{y\cos\theta}{203.125}$ \\[4pt]
    No curvature     & $\infty$ & 0        & $\chi_F = 0$ \\
    \hline
  \end{tabular}
\end{table}

Direct numerical simulation was used to resolve receptivity, instability growth, and transition onset without relying on a turbulence model. The computational domain included the blunt leading edge, supersonic inflow and outflow boundaries, and periodic lateral boundaries. For the curved cases, the spanwise direction was extruded to a constant arc length of $30\,\mathrm{mm}$. The span was chosen to be sufficiently large to prevent contamination from periodic images.

The mesh strategy emphasized the leading edge and near-wall region, where receptivity and early instability growth are most sensitive to resolution. All cases used quadratic hexahedral elements with strong local refinement near the wall and in the leading-edge region. The final three-dimensional grids contained approximately $1.18\times10^7$ elements, corresponding to roughly $1.6\times10^9$ degrees of freedom. Simulations were initialized from converged laminar precursor solutions and then advanced with implicit time integration. A localized artificial-viscosity sensor was used only in strongly compressive regions to stabilize shock-containing portions of the flow while minimizing dissipation in the smooth boundary layer. 

The problem is temporally discretized with a DIRK(3,3) time-marching scheme, using a nondimensional timestep of $\Delta t = 2e^{-4}$ performing 9200 timesteps. Each run required roughly 6,600 node-hours across 1,536 MI250X GPUs on Frontier and 1,024 MI300A GPUs on Tuolumne.

\begin{table}
\centering
\caption{Key simulation parameters.}
\label{tab:key_setup}
\small

\begin{tabular}{ll}
\hline
Parameter & Value \\ \hline
Solver/discretization & Exasim, HDG \\
Governing equations & 3D unsteady compressible Navier--Stokes \\
Coordinate system & Cartesian (flat case); cylindrical (curved cases) \\
Freestream Mach number, $M_{\infty}$ & 7.0 \\
Unit Reynolds number, $Re_{\infty}$ & $8.088\times10^{6}\,\mathrm{m^{-1}}$ \\
Freestream temperature, $T_{\infty}$ & $51.86\,\mathrm{K}$ \\
Wall temperature, $T_w$ & $296\,\mathrm{K}$ \\
Gas model & Calorically perfect gas, $\gamma=1.4$, $Pr=0.71$ \\
Disturbance forcing & Synthetic turbulence, $I=0.025\%$ \\
Mesh & $\sim 1.18\times10^7$ quadratic hexahedral elements \\
Boundary conditions & Supersonic inflow/outflow, isothermal no-slip wall, periodic sides \\ \hline
\end{tabular}
\end{table}

\subsection{SPOD Methodology}

Spectral proper orthogonal decomposition (SPOD) was used to identify dominant coherent structures in the unsteady flow field, following the Welch-based method of \cite{Towne_Schmidt_Colonius_2018}. The procedure was applied independently to the spanwise velocity fluctuation $w'$ and pressure fluctuation $p'$, with $w=(\rho w)/\rho$ recovered from the conserved variables and fluctuations defined relative to the time average, $a'(\mathbf{x},t_k)=a(\mathbf{x},t_k)-a_{\mathrm{TA}}(\mathbf{x})$ for $a\in\{w,p\}$.

Snapshots restricted to a prescribed spatial mask, $\mathbf{a}'(t_k)\in\mathbb{R}^N$, were assembled into $\mathbf{A}=[\mathbf{a}'(t_1),\dots,\mathbf{a}'(t_K)]$ and partitioned into $N_b$ overlapping, Hamming-windowed blocks of length $n_{\mathrm{fft}}$ with overlap $n_{\mathrm{ovlp}}$. Fourier-transformed block realizations at frequency $\omega_j$ were collected into $\mathbf{Q}(\omega_j)\in\mathbb{C}^{N\times N_b}$, giving the cross-spectral density matrix
\[
\mathbf{S}(\omega_j)=\frac{1}{N_b}\mathbf{Q}^*(\omega_j)\mathbf{Q}(\omega_j).
\]
The eigenvalue problem $\mathbf{S}(\omega_j)\mathbf{v}_m(\omega_j)=\lambda_m(\omega_j)\mathbf{v}_m(\omega_j)$, with $\lambda_1\ge\lambda_2\ge\cdots$, yields SPOD modes $\boldsymbol{\phi}_m(\omega_j)=\mathbf{Q}(\omega_j)\mathbf{v}_m(\omega_j)$. The leading seven modes were retained at each frequency for both $w'$ and $p'$; $n_{\mathrm{fft}}=512$, $n_{\mathrm{ovlp}}=384$ (75\% overlap, Hamming window).

\begin{figure}
\centering\includegraphics[alt={sample image},width=.65\textwidth]{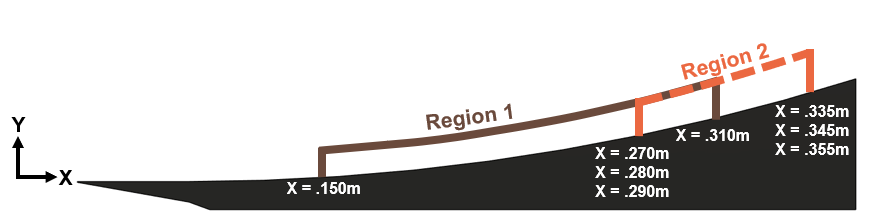} 
\caption{Locations of Region 1 and Region 2 sampled with SPOD}
\label{spodlocations}
\end{figure}

For each geometry, 8 streamwise slices were sampled over 5399 snapshots ($t^*=1.08$) from a converged transient simulation, at two spatial windows (Figure~\ref{spodlocations}): Region 1, fixed at $x=[0.150,0.310]$ upstream of breakdown, and Region 2, immediately upstream of turbulence, shifted to account for geometric differences — $x=[0.270,0.335]$ (no curvature), $[0.280,0.345]$ (medium curvature), $[0.290,0.355]$ (high curvature).

For relative sonic lines, the disturbance phase speed was extracted from the leading SPOD pressure mode $\hat{\psi}_p(\mathbf{x},f)$ by sampling along a wall-parallel arc through the energetic wavepacket, unwrapping the interpolated phase, and fitting $\phi(s)\approx\alpha s+\phi_0$; the wavenumber $\alpha$ gives $c_\phi=\omega/|\alpha|$. Arc location and fitting window were selected via automated sweep favoring high modal amplitude, low residual, and robustness of $c_\phi$. The relative Mach number,
\[
\bar{M}=\frac{u(y)-c_\phi}{a(y)},
\]
identifies regions supersonic relative to the disturbance.

\section{Experimental Setup}
\label{sec:exp_setup}

\subsection{Facility}
\label{subsec:ES-Facility}
Complementary measurements were made in the University of Tennessee Space Institute (UTSI) Mach 7 Ludwieg tube \citep{gragston2023design}. The facility has a nozzle exit dimension of 45.72 cm by 45.72 cm (18 inches by 18 inches) with three steady wave passages, each of approximately 80 ms in duration. A fast-acting valve is used to actuate the tunnel and precisely control the test conditions. The nominal conditions for these experiments are provided in Table \ref{table:conditions}. As reported in  \cite{webber2026experimental}, recent pitot survey results indicate that the second steady wave passage (used in this work) has pitot pressure fluctuations of about 2.1\% (reported as standard deviation divided by the mean) based on 50 kHz measurements.

\begin{table}
\centering
\caption{A summary of nominal tunnel conditions for the experiments.}
\begin{tabular}{cc}
\hline
\multicolumn{2}{c}{Tunnel Conditions} \\ \hline
Total Temperature [K] & 560 \\
Total Pressure [MPa] & 1.65 \\
Mach Number & 7 \\
Freestream $Re/x$ [$\times$ 10\textsuperscript{6} m\textsuperscript{-1}] & 8.11 \\
Steady Wave Passage Used (of 3) & 2 \\ \hline
\end{tabular}
\label{table:conditions}
\end{table}

\subsection{Model geometries}
\label{subsec:ES-Models}

\begin{figure}
    \centering
    
    \begin{subfigure}[t]{\textwidth}
        \centering
        \includegraphics[width=0.48\textwidth]{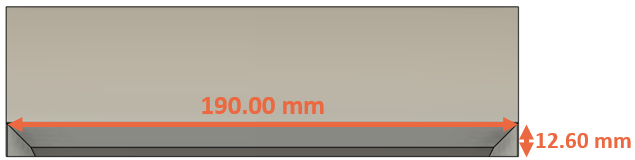}
        \hfill
        \includegraphics[width=0.48\textwidth]{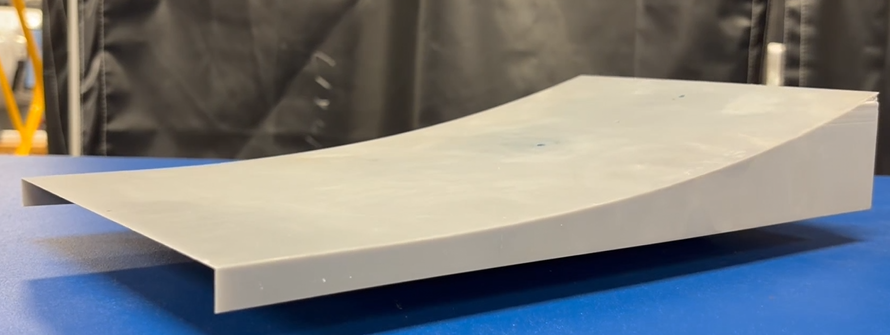}
        \caption{No spanwise curvature}
        \label{fig:sub1}
    \end{subfigure}
    
    \vspace{1em} 
    
    \begin{subfigure}[t]{\textwidth}
        \centering
        \includegraphics[width=0.48\textwidth]{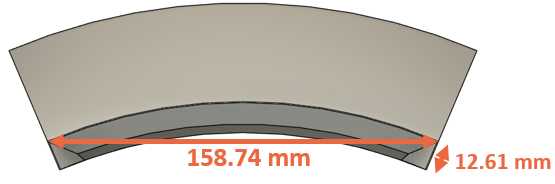}
        \hfill
        \includegraphics[width=0.48\textwidth]{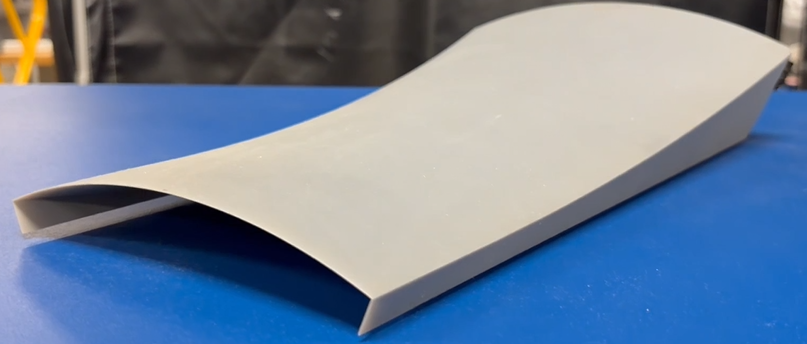}
        \caption{Medium spanwise curvature}
        \label{fig:sub2}
    \end{subfigure}
    
    \vspace{1em} 
    
    \begin{subfigure}[t]{\textwidth}
        \centering
        \includegraphics[width=0.48\textwidth]{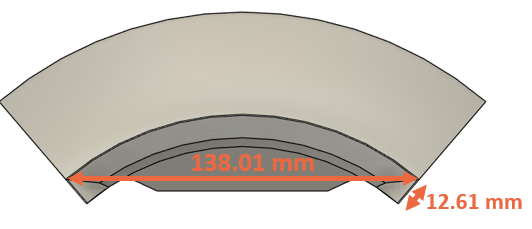}
        \hfill
        \includegraphics[width=0.48\textwidth]{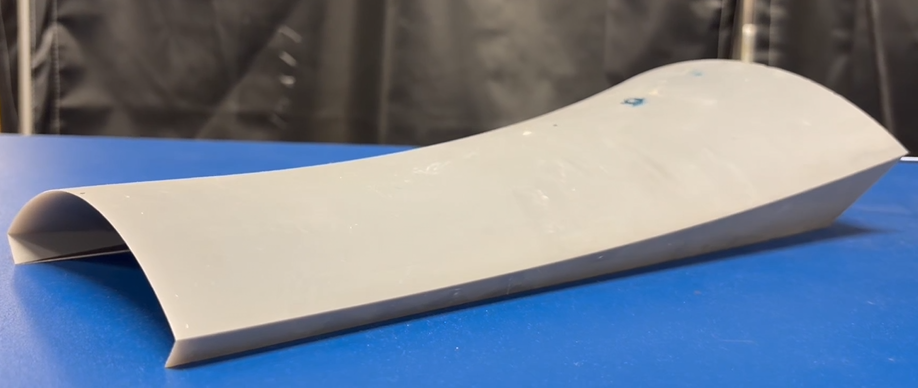}
        \caption{High spanwise curvature}
        \label{fig:sub3}
    \end{subfigure}
    
    \caption{Front view of all test geometries}
    \label{fig:models_front_view}
\end{figure}

All test articles were 3D-printed from Formlabs Grey V5 resin, whose thermal properties are denoted in Table \ref{table:thermalprop}. The zero-curvature baseline model was produced with a length of $340\,\mathrm{mm}$, while the medium and high spanwise curvature geometries measure $330\,\mathrm{mm}$ in length due to printer constraints. Detailed frontal dimensions and isometric shots are provided in Figure~\ref{fig:models_front_view}.

\subsection{Schlieren imaging}
\label{subsec:ES-Schlieren}

High-speed schlieren imaging of the flow around the model was also done to observe features for comparison with computational modeling. The schlieren imaging system was configured in a Z-type arrangement utilizing two f/8 mirrors with a focal length of 2.54 m. A CAVILUX HF 280 W laser was implemented as the light source operating at 20 kHz with a pulse duration of 100 ns. A Photron SA-Z camera was used to image this and was equipped with an 80-200 mm Sigma zoom lens.

\subsection{Infrared thermography}
\label{subsec:ES-IR}
A FLIR X8581 3-5 µm camera operating at a frame rate of 180 Hz and exposure time of 1.5 ms was used for IR thermography. The camera was positioned to view the model through an uncoated calcium fluoride window (Edmund Optics \#17-605) and was equipped with a 50 mm lens.

Prior to running, a two-point non-uniformity correction (NUC) was performed inside the tunnel section using cold and heated silicon carbide plates (McMaster \#6718N116) wrapped in black film (3M Wrap Film Series 1080). The temperature calibration was also done within the tunnel using the wrapped silicon carbide plate, which was heated using a hot plate and monitored with a precision thermocouple.

Heat flux calculations were performed using the generalized Cook-Felderman formulation developed by \cite{liu2010analytical}, which was implemented using the approach discussed by \cite{edwards2025adhesive}. Note that this approach reduces to the typical Cook-Felderman method for situations involving a single homogeneous material, which was the case for these experiments.

\begin{table}
\centering
\caption{A table showing the thermal properties of the model material.}
\begin{tabular}{cc}
\hline
Material  & Formlabs Gray Resin \\
Density $\rho$ [kg/m\textsuperscript{3}] & 1110 \\
Thermal Conductivity $k$ [W/m/K]  & 0.247 \\
Thermal Diffusivity $\alpha_T$ [mm\textsuperscript{2}/s] & {0.146} \\ \hline
\end{tabular}
\label{table:thermalprop}
\end{table}


\subsection{PCB Pressure Sensors}
\label{subsec:ES-PCBs}
Pressure fluctuation measurements were made using sensors (PCB Piezotronics \#132B38) on the zero and high-spanwise curvature models. The signals from the sensors were conditioned using the PCB Piezotronics signal conditioner and amplifier. The sensors were sampled at 1.25 MHz using a National Instruments data acquisition system integrated with the tunnel.

\section{Computational results}
\label{sec:comp_results}

This section presents the direct numerical simulation and experimental results for the flat, low-curvature and high-curvature configurations, with emphasis on how spanwise convex curvature modifies the transition process when the streamwise concavity is held fixed. The discussion proceeds from global indicators of transition, including wall heat transfer and surface signatures, to the underlying instability dynamics inferred from the instantaneous flow fields, boundary-layer statistics and modal analysis. Particular attention is given to the relative evolution of G\"ortler-dominated disturbances, secondary breakdown and compressible instability content, so that the mechanisms responsible for the observed shift in transition can be identified and related to the curvature-induced changes in the mean boundary-layer structure.

\subsection{Streamwise and spanwise structure}
\label{subsec:streamwise_spanwise}

\begin{figure}
    \centering
    
    \begin{subfigure}{0.48\textwidth}
        \centering
        \includegraphics[width=\textwidth]{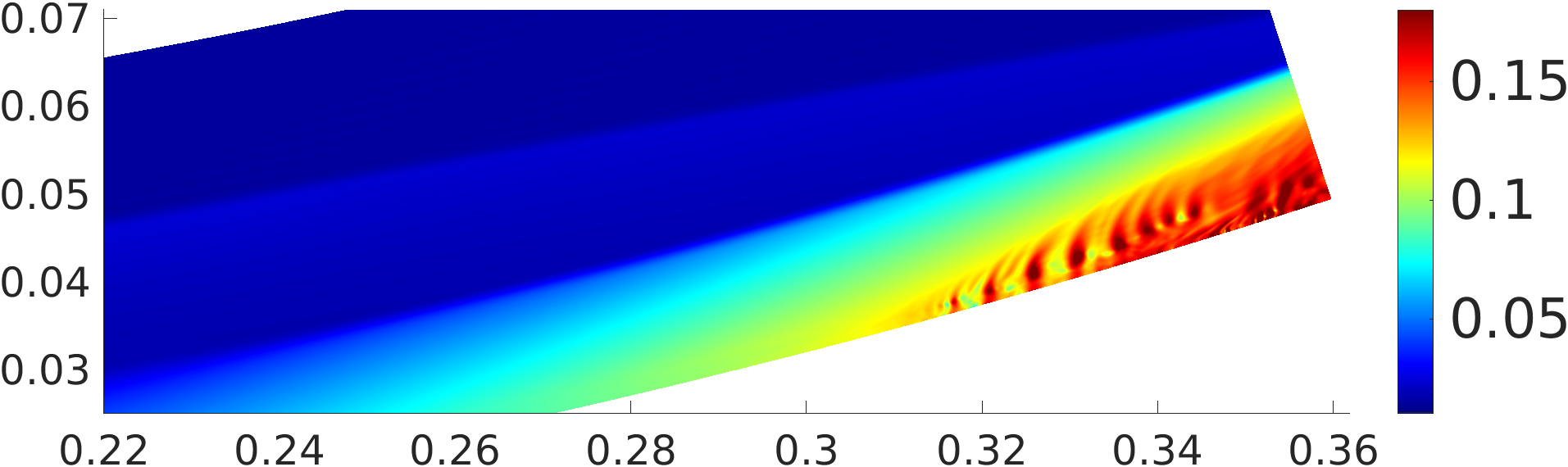}
        \caption{No curvature: pressure}
    \end{subfigure}
    \hfill
    \begin{subfigure}{0.48\textwidth}
        \centering
        \includegraphics[width=\textwidth]{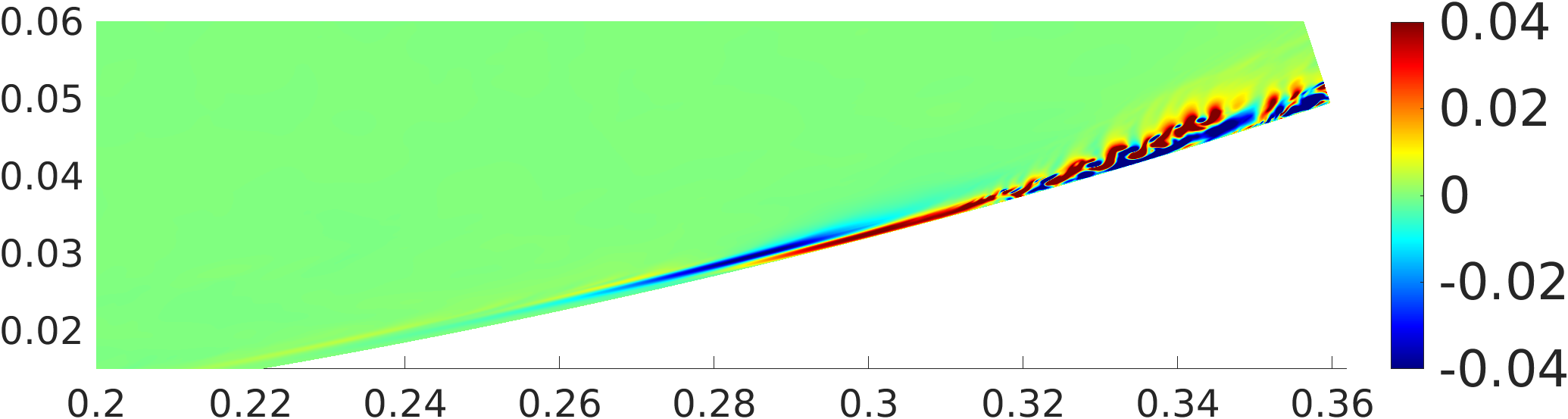}
        \caption{No curvature: $w$-velocity}
    \end{subfigure}

    \vspace{1em}

    \begin{subfigure}{0.48\textwidth}
        \centering
        \includegraphics[width=\textwidth]{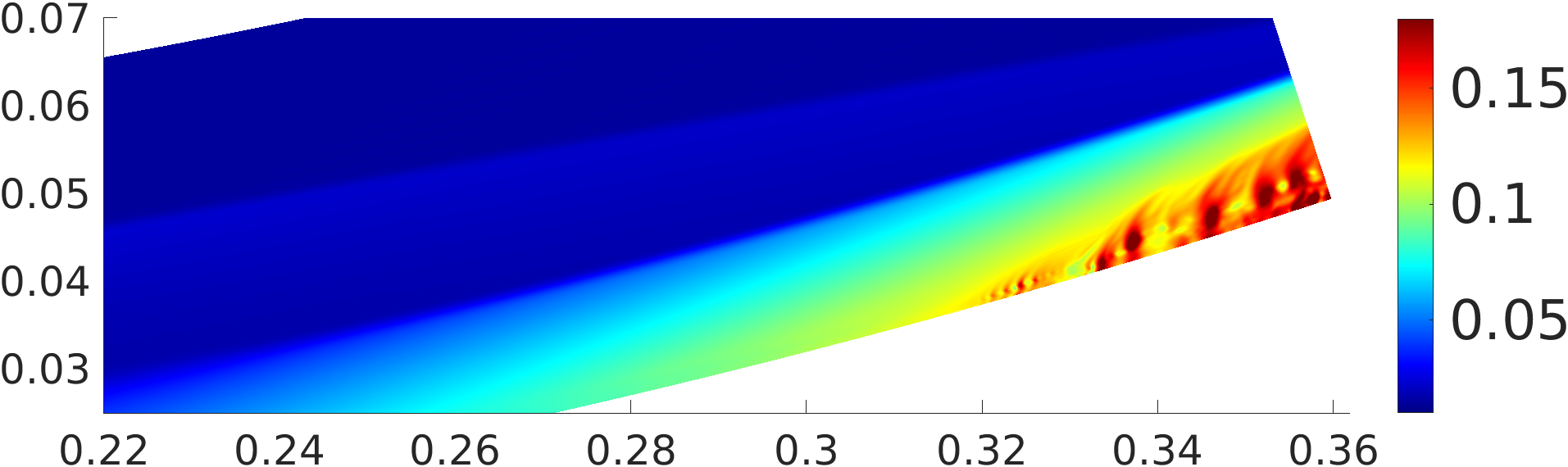}
        \caption{Medium curvature: pressure}
    \end{subfigure}
    \hfill
    \begin{subfigure}{0.48\textwidth}
        \centering
        \includegraphics[width=\textwidth]{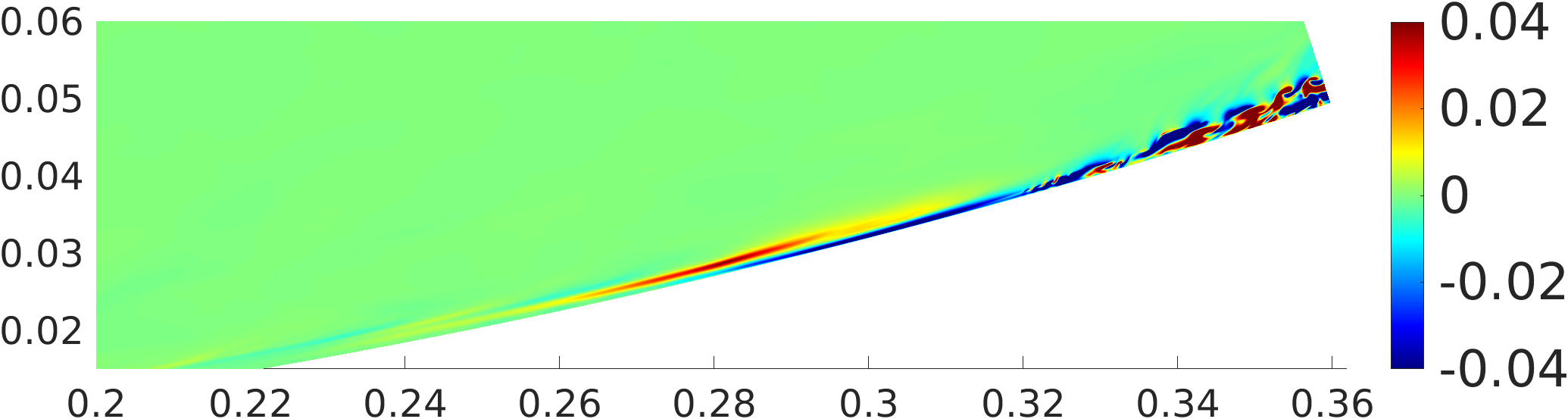}
        \caption{Medium curvature: $w$-velocity}
    \end{subfigure}

    \vspace{1em}

    \begin{subfigure}{0.48\textwidth}
        \centering
        \includegraphics[width=\textwidth]{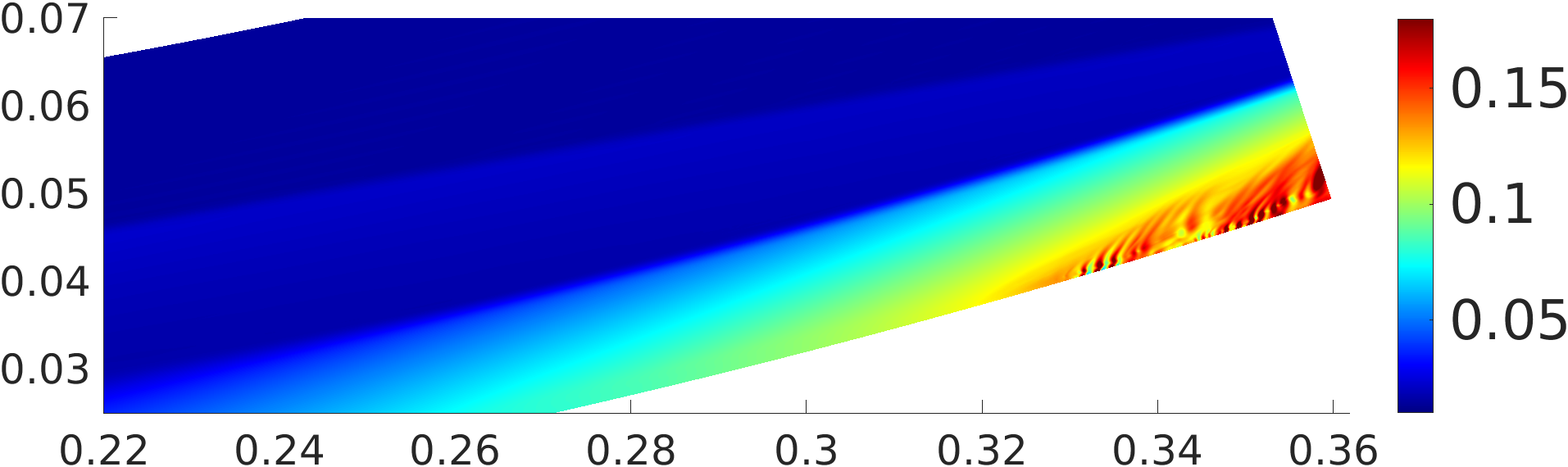}
        \caption{High curvature: pressure}
    \end{subfigure}
    \hfill
    \begin{subfigure}{0.48\textwidth}
        \centering
        \includegraphics[width=\textwidth]{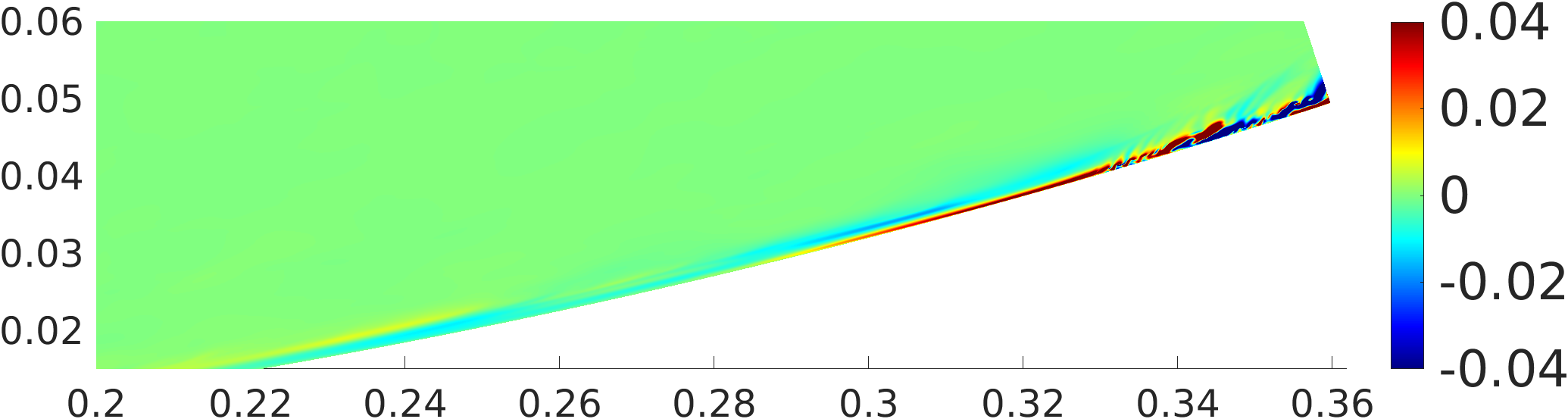}
        \caption{High curvature: $w$-velocity}
    \end{subfigure}

    \caption{Instantaneous streamwise ($x$--$y$) cuts of non-dimensional pressure (left column) and $w$-velocity (right column) for the flat, medium-curvature, and high-curvature configurations at timestep 9200.}
    \label{fig:pressure_velocity_grid}
\end{figure}

The effect of spanwise convex curvature is first evident in the instantaneous streamwise organization of the disturbance field. Figure~\ref{fig:pressure_velocity_grid} shows $x$--$y$ cuts of pressure and spanwise velocity at the same late stage of the simulations. In all three configurations, the dominant large-scale structures are elongated streamwise streaks associated with G\"ortler vortices, together with localized compressive activity that intensifies as the vortices deform and break down. The important change with curvature is not the appearance of a different transition pathway, but a systematic downstream displacement and weakening of the same one.

The flat case exhibits the earliest loss of coherence. Its pressure field contains the broadest region of strong unsteadiness, together with pronounced radiating wave activity downstream of the onset of breakdown. The corresponding $w$-velocity field shows large-amplitude rope-like structures that undergo rapid deformation and fragmentation, indicating that the primary vortices have already entered a strongly nonlinear stage. In the medium-curvature case these signatures persist, but they are displaced downstream and remain organized over a longer streamwise extent. The high-curvature case is more coherent still: the streaks remain narrower, their waviness develops later, and the downstream region occupied by intense pressure fluctuations is reduced. Taken together, these streamwise cuts indicate that increasing spanwise curvature delays the amplification and subsequent breakdown of the centrifugal structures that dominate transition in the flat configuration.

\begin{figure}
    \centering

    \begin{subfigure}{0.48\textwidth}
        \centering
        \includegraphics[width=\linewidth]{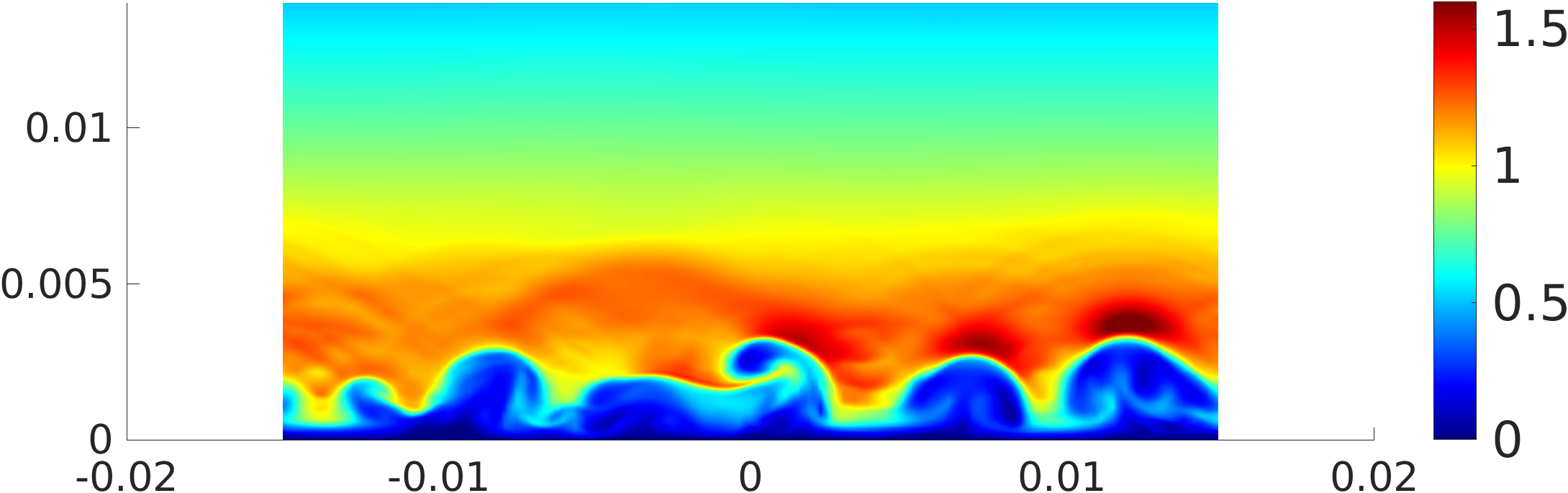}
        \caption{No curvature, $x=331\,\mathrm{mm}$}
    \end{subfigure}
    \hfill
    \begin{subfigure}{0.48\textwidth}
        \centering
        \includegraphics[width=\linewidth]{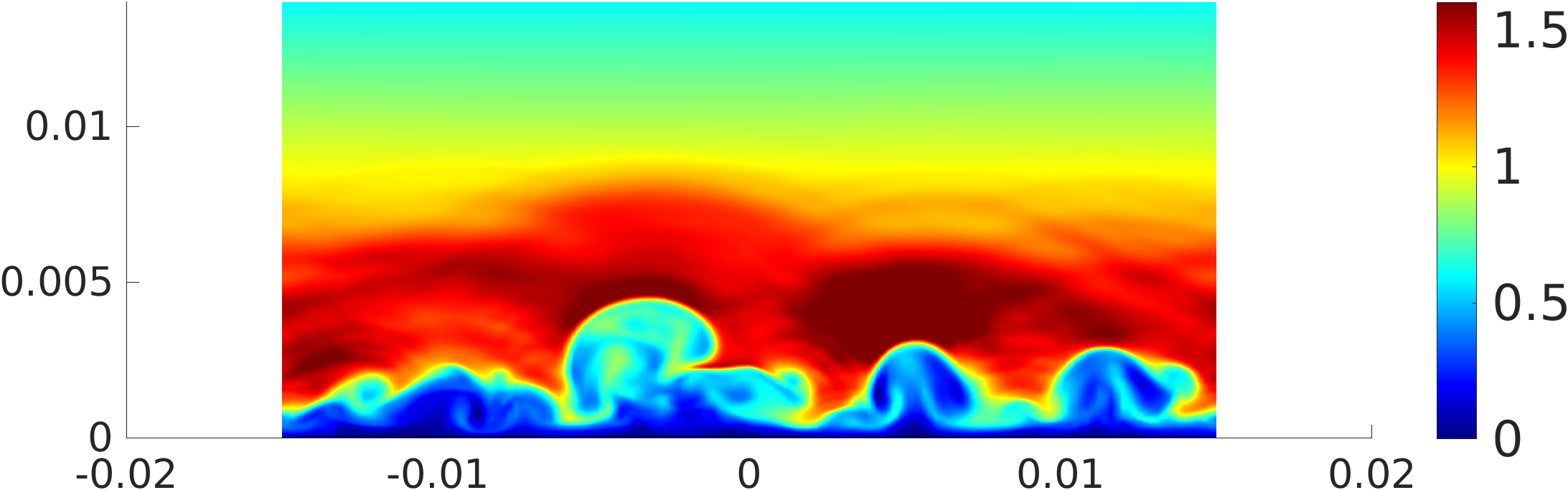}
        \caption{No curvature, $x=348\,\mathrm{mm}$}
    \end{subfigure}

    \vspace{1pt}

    \begin{subfigure}{0.48\textwidth}
        \centering
        \includegraphics[width=\linewidth]{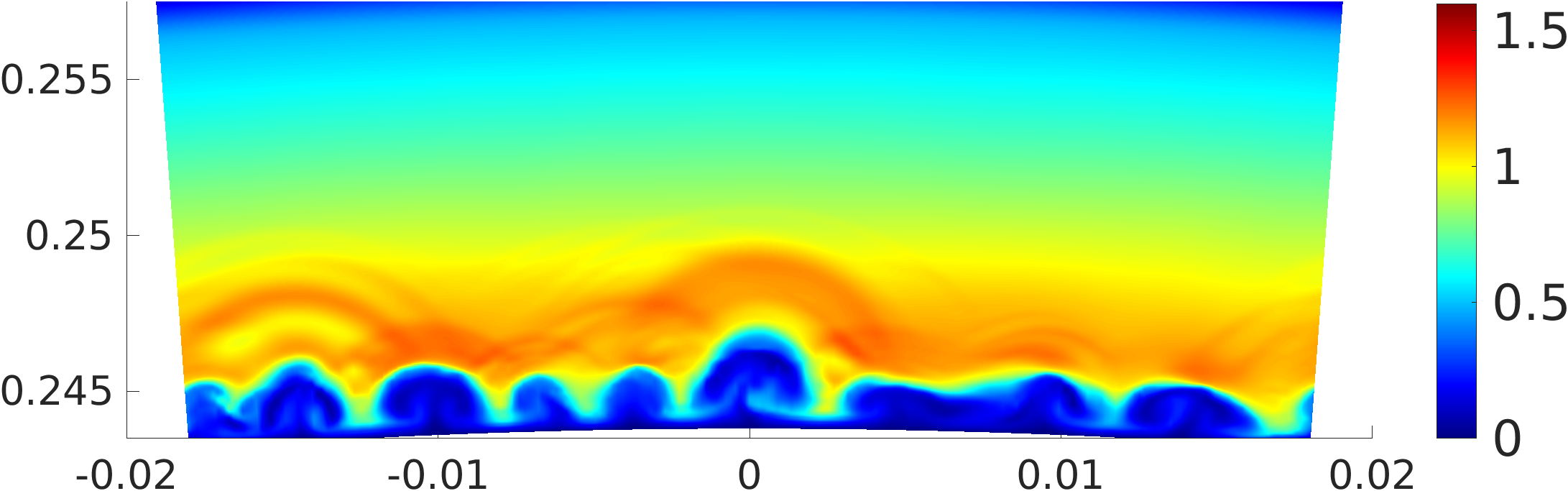}
        \caption{Medium curvature, $x=331\,\mathrm{mm}$}
    \end{subfigure}
    \hfill
    \begin{subfigure}{0.48\textwidth}
        \centering
        \includegraphics[width=\linewidth]{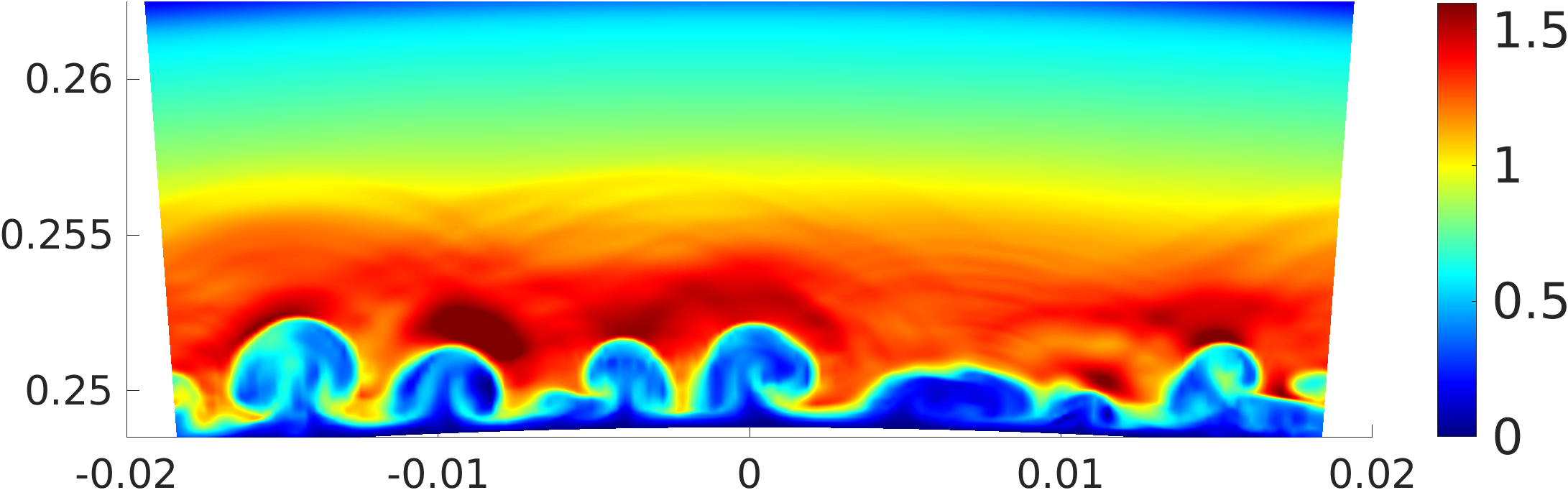}
        \caption{Medium curvature, $x=348\,\mathrm{mm}$}
    \end{subfigure}

    \vspace{1pt}

    \begin{subfigure}{0.48\textwidth}
        \centering
        \includegraphics[width=\linewidth]{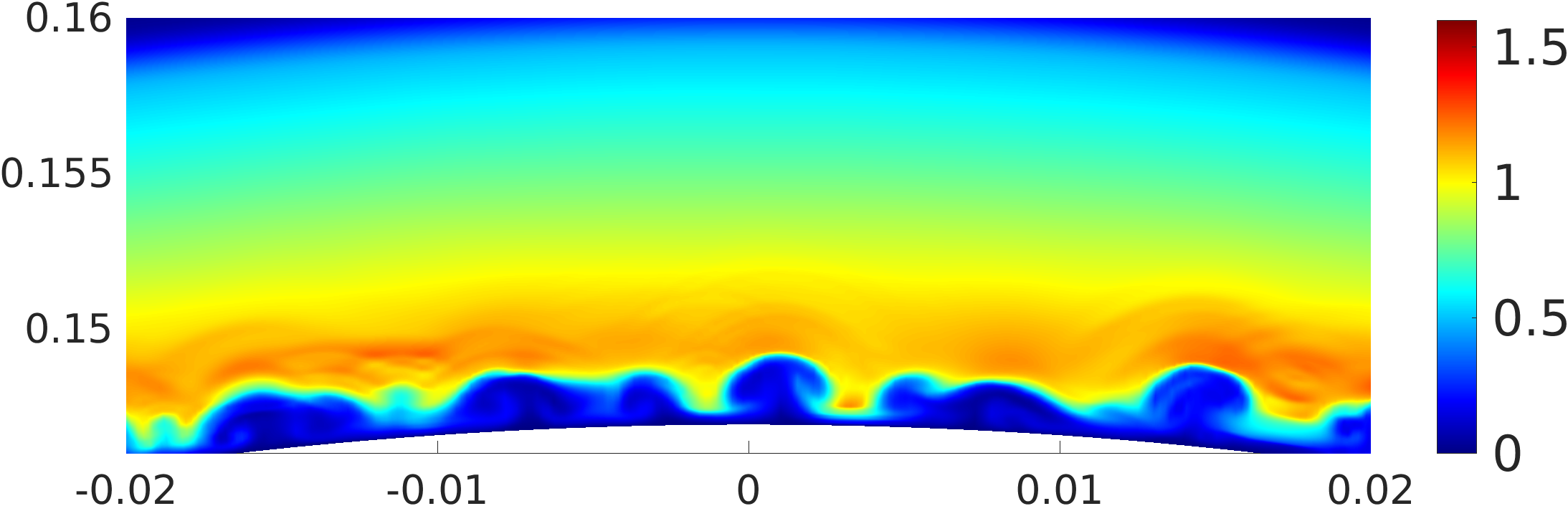}
        \caption{High curvature, $x=331\,\mathrm{mm}$}
    \end{subfigure}
    \hfill
    \begin{subfigure}{0.48\textwidth}
        \centering
        \includegraphics[width=\linewidth]{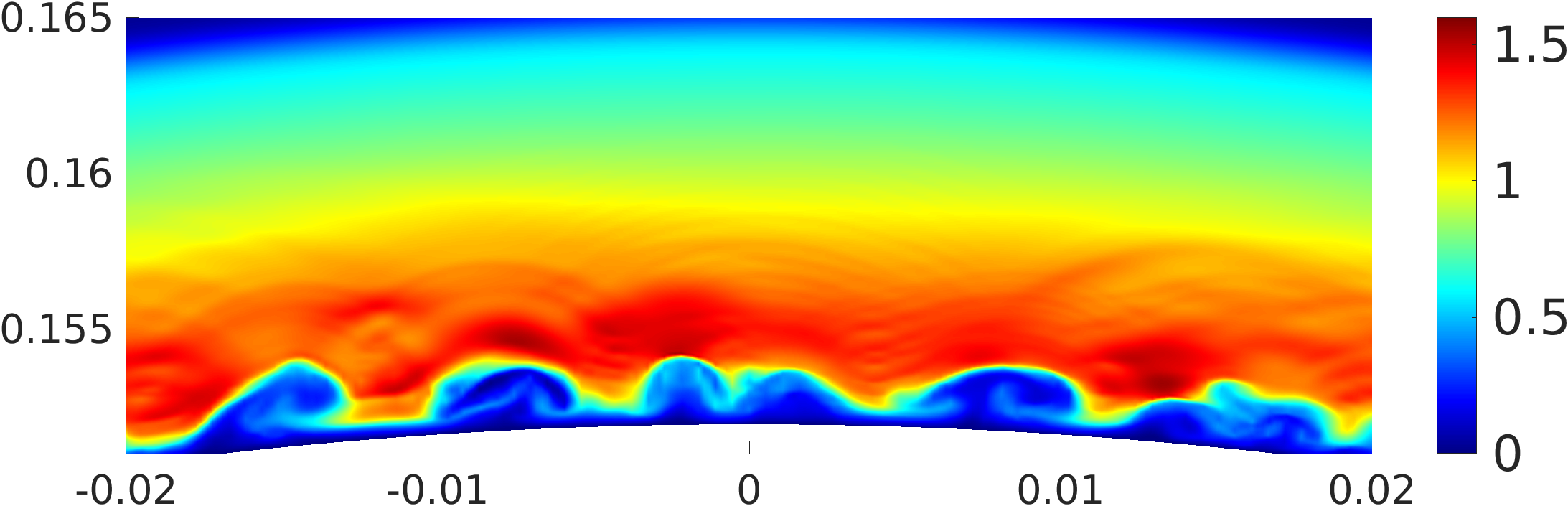}
        \caption{High curvature, $x=348\,\mathrm{mm}$}
    \end{subfigure}

    \caption{Spanwise ($z$--$y$) cuts of instantaneous vertical momentum, $\rho v$, across the three configurations. The mushroom-like heads identify G\"ortler vortices and the waviness of the stems is consistent with sinuous secondary instability.}
    \label{fig:spanwise_comparison_momentum}
\end{figure}

The spanwise cuts in Figure~\ref{fig:spanwise_comparison_momentum} clarify the underlying vortex topology. The plotted quantity is the instantaneous wall-normal momentum, $\rho v$, on $y$--$z$ planes at two representative streamwise stations. In every configuration the cross-sections display the characteristic mushroom-shaped heads of counter-rotating G\"ortler vortices, confirming that the streamwise streaks in figure~\ref{fig:pressure_velocity_grid} are the surface signature of a centrifugal instability. The more downstream cuts also show lateral waviness of the vortex stems, which is consistent with the onset of sinuous secondary instability.

These spanwise views again show a monotonic curvature effect. In the flat case the mushroom heads are broader, the vortices extend farther from the wall, and the stem deformation is strongest, all of which are consistent with a more amplified and more weakly constrained vortical field. The medium-curvature case retains the same qualitative structure but with reduced wall-normal extent and less pronounced distortion. In the high-curvature case the vortices remain noticeably more compact and closer to the wall at both stations, and the downstream section shows a less advanced state of secondary deformation. The evidence from the streamwise and spanwise planes is therefore consistent: spanwise convex curvature does not eliminate the G\"ortler route to transition, but weakens the vortices, reduces their wall-normal growth, and postpones the secondary breakdown that ultimately produces the turbulent state. This interpretation is reinforced below through wall-based diagnostics, boundary-layer-thickness measurements, and modal analysis.

\subsection{Wall signatures of transition}
\label{subsec:wall_signatures}

\begin{figure}
     \centering
     
     \begin{subfigure}[b]{0.48\textwidth}
         \centering
         \includegraphics[width=\textwidth]{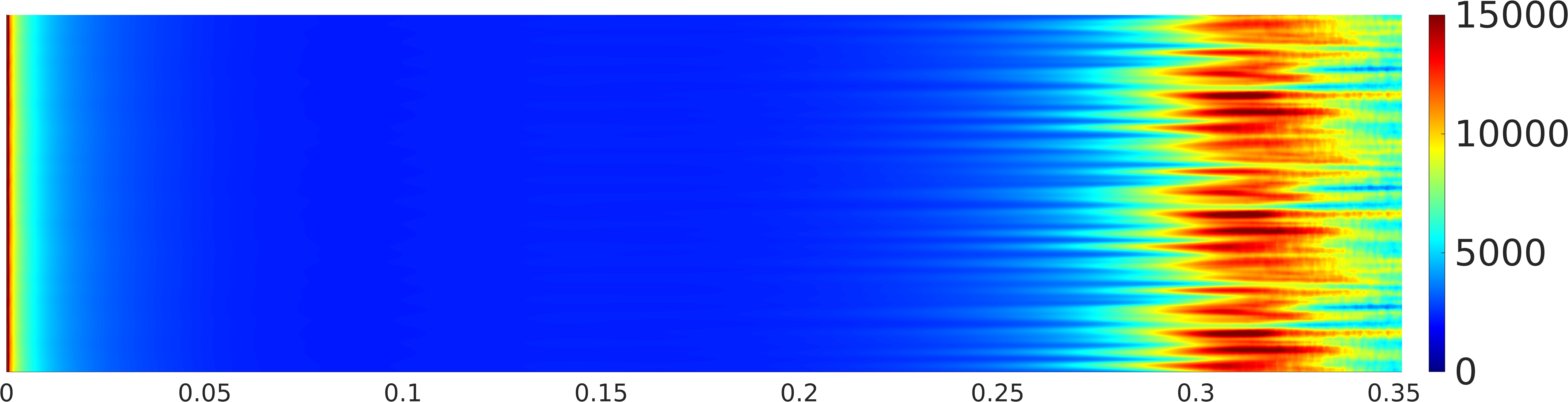}
         \caption{TA: no spanwise curvature}
     \end{subfigure}
     \hfill
     \begin{subfigure}[b]{0.48\textwidth}
         \centering
         \includegraphics[width=\textwidth]{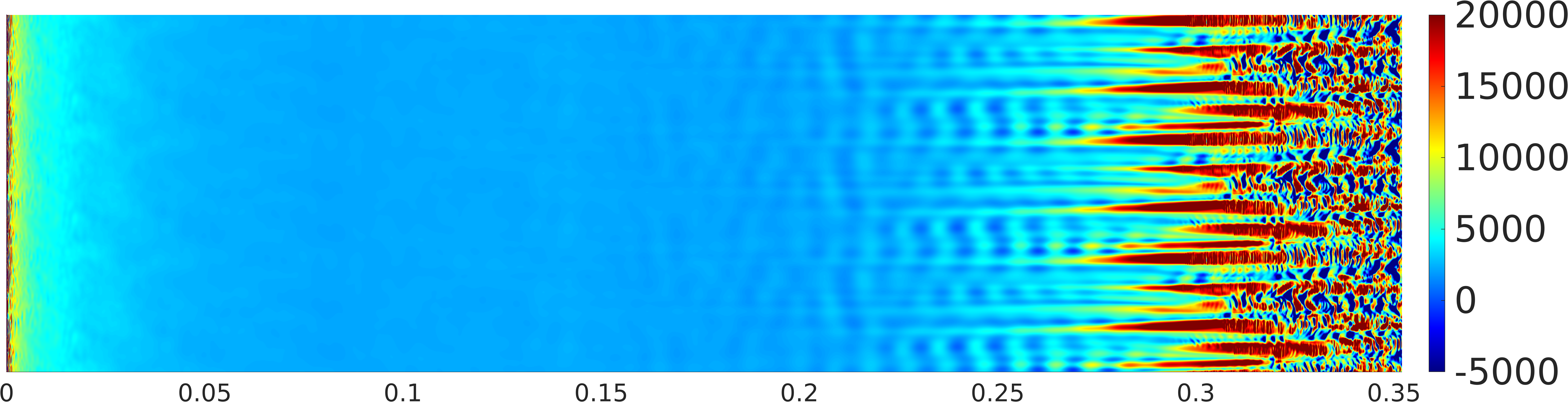}
         \caption{Inst.: no spanwise curvature}
     \end{subfigure}

     \vspace{10pt}

     \begin{subfigure}[b]{0.48\textwidth}
         \centering
         \includegraphics[width=\textwidth]{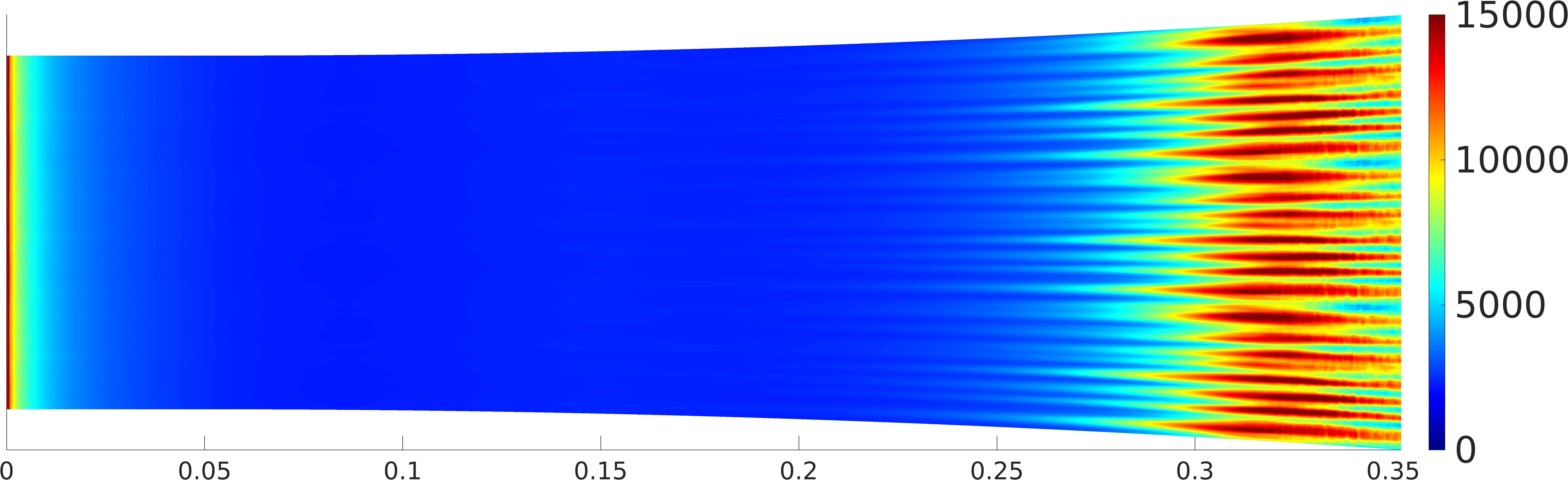}
         \caption{TA: medium curvature}
     \end{subfigure}
     \hfill
     \begin{subfigure}[b]{0.48\textwidth}
         \centering
         \includegraphics[width=\textwidth]{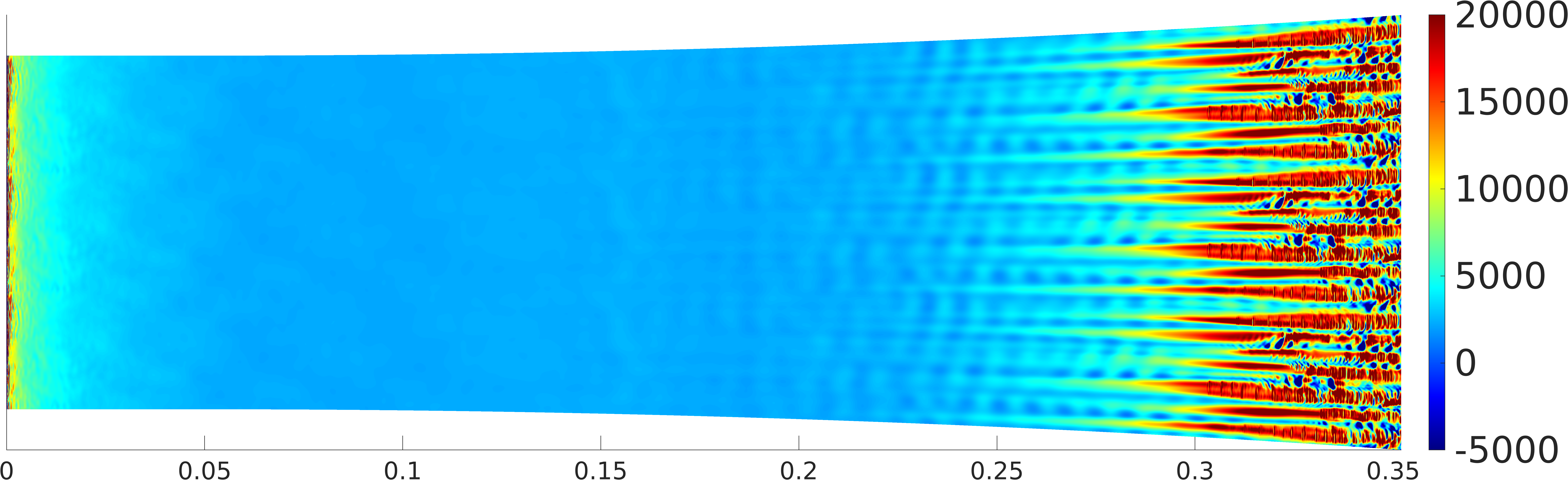}
         \caption{Inst.: medium curvature}
     \end{subfigure}

     \vspace{10pt}

     \begin{subfigure}[b]{0.48\textwidth}
         \centering
         \includegraphics[width=\textwidth]{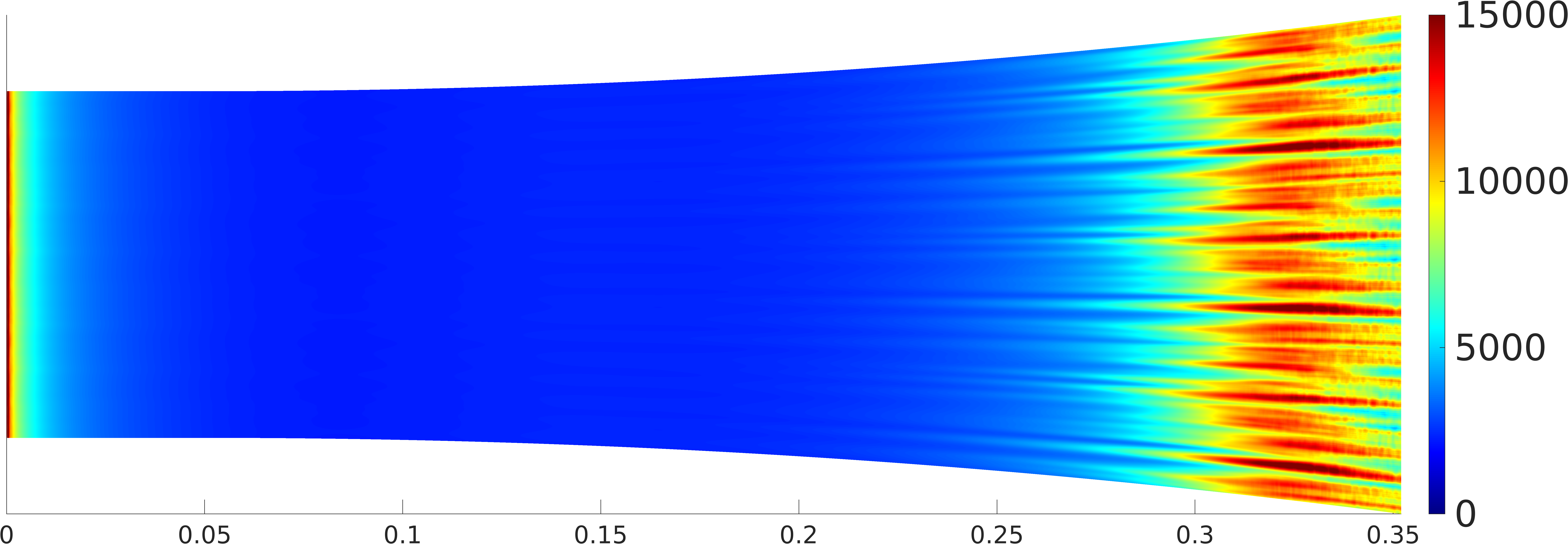}
         \caption{TA: high curvature}
     \end{subfigure}
     \hfill
     \begin{subfigure}[b]{0.48\textwidth}
         \centering
         \includegraphics[width=\textwidth]{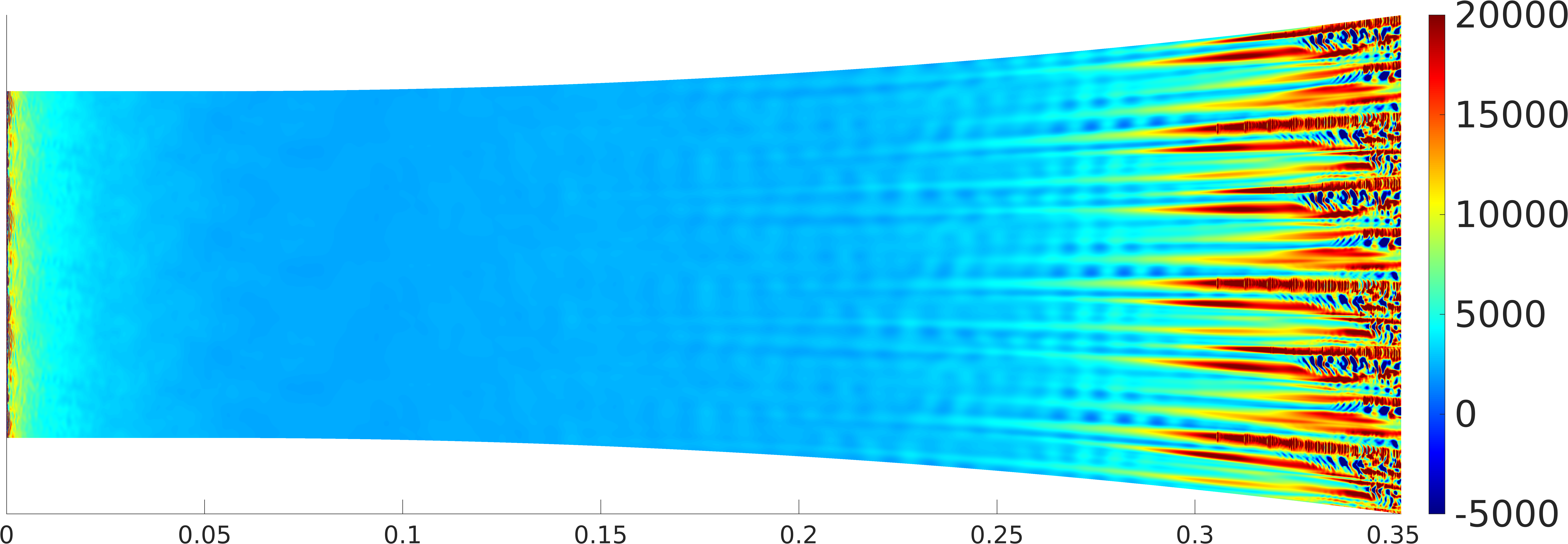}
         \caption{Inst.: high curvature}
     \end{subfigure}
     
     \caption{Comparison of time-averaged (left column) and instantaneous (right column) wall heat flux $(W/m^2)$ for all curvature cases. Instantaneous fields correspond to timestep 9200.}
     \label{fig:heatflux_main}
\end{figure}

The wall heat-flux fields in Figure~\ref{fig:heatflux_main} provide the clearest surface-level evidence of how spanwise convex curvature modifies the transition process. In all three configurations, the dominant footprint is a system of elongated thermal streaks aligned with the streamwise direction. These streaks are the wall manifestation of the G\"ortler vortices identified in Figures~\ref{fig:pressure_velocity_grid} and \ref{fig:spanwise_comparison_momentum}: low- and high-heat-flux bands form beneath the alternating upwash and downwash motions, and their downstream evolution records the growth, distortion and eventual breakdown of the vortical field. The heat-flux maps therefore compress the three-dimensional instability dynamics into a surface diagnostic that is especially useful for comparing the onset and streamwise extent of transition among the three geometries.

The time-averaged fields isolate the persistent organization of this response. In the flat case, the thermal streaks intensify rapidly and merge into a broad high-heat-flux region beginning near $x\approx 0.31$ m, indicating that strong vortex amplification and breakdown already occur relatively far upstream. The medium-curvature case exhibits the same sequence of events, but the streak amplification is displaced downstream and the high-heat-flux zone is narrower and less intense. The high-curvature case delays this evolution further still: coherent streaks persist farther downstream before giving way to a more limited region of elevated heat flux. Thus, increasing spanwise curvature does not alter the qualitative wall signature of a G\"ortler-dominated transition, but it moves the entire sequence of streak amplification, merger and turbulent heating downstream.

The instantaneous fields sharpen this trend by revealing the degree of spatial intermittency within the breakdown region. In the flat configuration, the downstream portion of the streak pattern has already lost much of its regularity and is replaced by a dense, irregular field of localized high- and low-heat-flux patches, consistent with fully developed secondary breakdown. The medium-curvature case remains more structured, with turbulent spots and fragmented streaks confined to a smaller downstream interval. The high-curvature case is more ordered still, retaining long, sharply defined streaks over most of the plate and showing the most limited area of fully disordered heat-flux activity near the downstream end of the domain. Based on these wall signatures, the effective transition location shifts from approximately $x=310\,\mathrm{mm}$ in the flat case to about $x=320\,\mathrm{mm}$ and $x=325\,\mathrm{mm}$ in the medium- and high-curvature cases, respectively. The wall data therefore corroborate the picture emerging from the interior flow fields: spanwise convex curvature weakens the same centrifugal-instability pathway and postpones the onset of intense surface heating associated with turbulent breakdown.

\subsection{Mechanism of delayed transition}
\label{subsec:mechanism}

\begin{figure}
\centering
    \includegraphics[width=.8\textwidth]{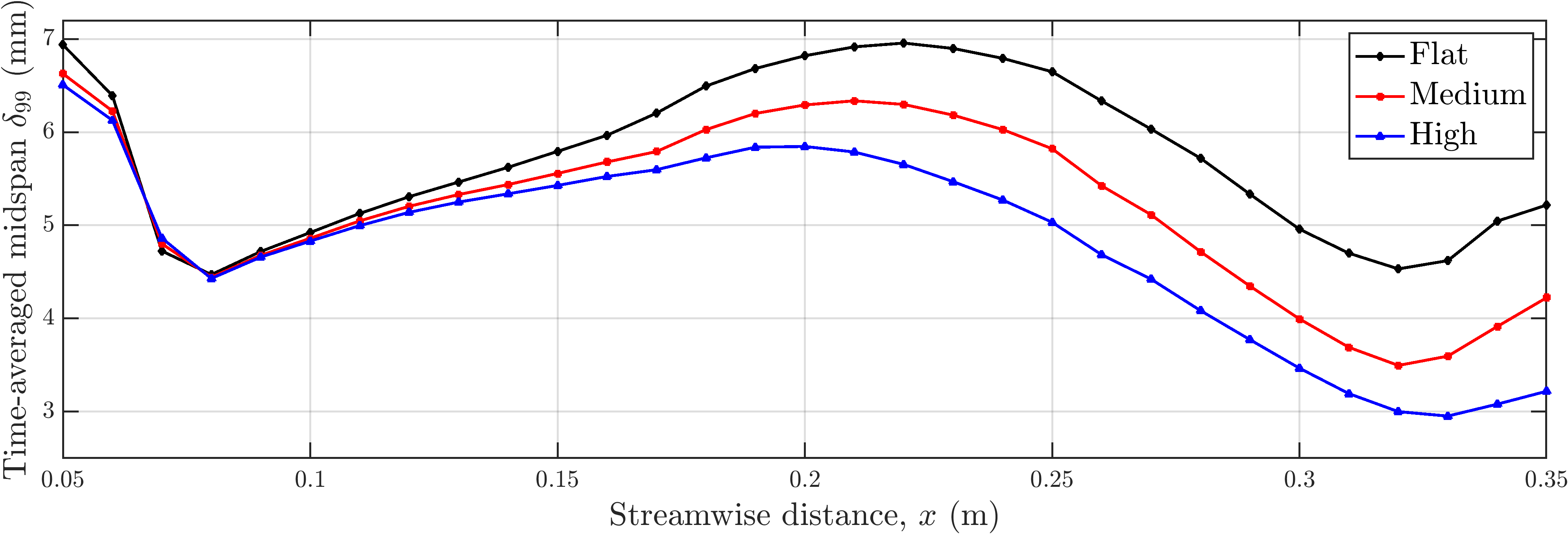}
    \caption{Time-averaged boundary-layer thickness for all cases, sampled at the spanwise midpoint.}
    \label{fig:BLthicknesscomparison}
\end{figure}

The results above indicate that increasing spanwise convex curvature delays transition without changing its qualitative route. The question is therefore not why one case is G\"ortler-dominated and another is not, but why the same centrifugal-instability pathway is systematically weakened as the transverse curvature increases. The principal mechanism is the curvature-induced reduction in effective boundary-layer thickness. As discussed earlier through the Mangler--Levy--Lees scaling, stronger spanwise convex curvature maps a comparable transformed profile onto a thinner physical boundary layer. In the present DNS, this trend appears directly in Figure~\ref{fig:BLthicknesscomparison}: after the common upstream minimum near the nose region, the three thickness curves separate monotonically, with the flat case remaining thickest, the low-curvature case intermediate, and the high-curvature case thinnest over nearly the entire downstream development. The separation grows progressively with $x$ and is largest in the same downstream region where the wall signatures and instantaneous fields show the onset of breakdown.

This thickness reduction provides a direct explanation for the weakening of the vortical branch. With the streamwise concavity held fixed, the effective centrifugal amplification scales with a G\"ortler number of the form $G\propto {\delta}^{1.5}$, so reducing the boundary-layer thickness lowers the instability parameter seen by the primary streamwise vortices. The consequence is not merely a smaller vortex amplitude at a single station, but a cumulative downstream effect: weaker primary growth leaves the vortices closer to the wall, reduces their wall-normal penetration, and delays the stage at which they become susceptible to strong sinuous deformation and secondary breakdown. This interpretation is consistent with the interior flow fields in \S\ref{subsec:streamwise_spanwise}, where increasing spanwise curvature produces more compact vortices and postpones their fragmentation, and with the wall heat-flux maps in \S\ref{subsec:wall_signatures}, where the onset of irregular high-heat-flux patches shifts downstream as the curvature increases.

\begin{figure}
\centering
    \includegraphics[width=.8\textwidth]{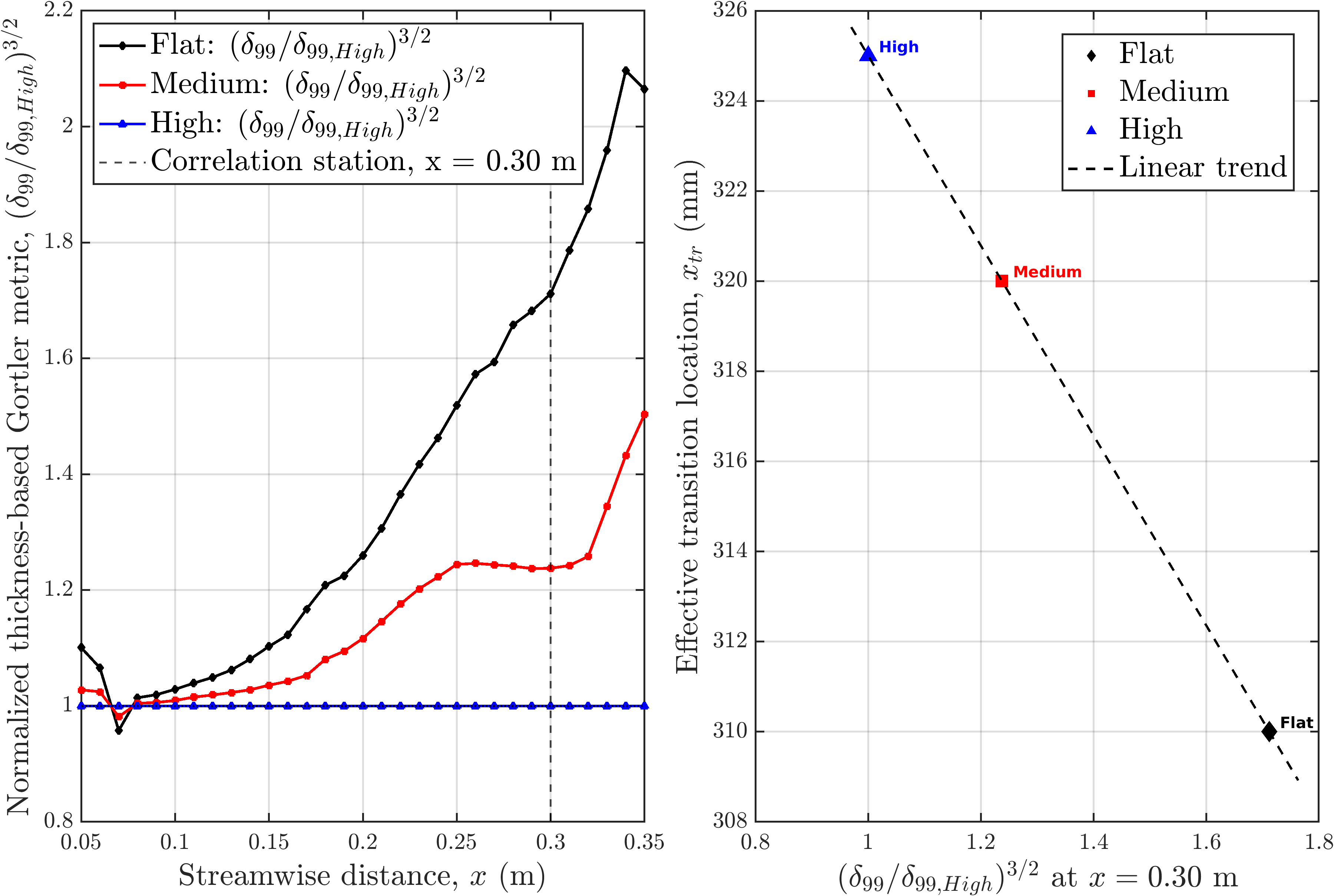}
    \caption{Thickness-based G\"ortler scaling and its relationship to the effective transition location. The metric is normalized by the high-curvature configuration and evaluated as $(\delta_{99,i}/\delta_{99,\mathrm{High}})^{3/2}$. At $x=300~\mathrm{mm}$, larger values of the normalized metric correspond systematically to earlier effective transition locations.}
    \label{fig:BLthicknessperaire}
\end{figure}

This connection can be quantified using the measured boundary-layer
thicknesses. Since the streamwise radius of curvature is the same for all
three configurations, $R=1~\mathrm{m}$ over $0.05\leq x\leq0.35~\mathrm{m}$,
and the extracted edge velocities differ negligibly, the ratio of the
thickness-based G\"ortler metrics may be approximated as
\begin{equation}
    \frac{G_{\delta,i}}{G_{\delta,\mathrm{High}}}
    =
    \frac{U_{e,i}}{U_{e,\mathrm{High}}}
    \left(
        \frac{\delta_{99,i}}
             {\delta_{99,\mathrm{High}}}
    \right)^{3/2}
    \simeq
    \left(
        \frac{\delta_{99,i}}
             {\delta_{99,\mathrm{High}}}
    \right)^{3/2}.
    \label{eq:normalized_gortler_scaling}
\end{equation}
Here, the high-curvature configuration is used only to normalize the
relative metric. As shown in Figure~\ref{fig:BLthicknessperaire}, evaluation
at $x=300~\mathrm{mm}$, immediately upstream of the earliest effective
transition location, gives normalized values of approximately $1.70$,
$1.23$, and $1.00$ for the flat-, medium-, and high-curvature
configurations, respectively. Thus, relative to the high-curvature case,
the thickness contribution to the G\"ortler metric is approximately
$70\%$ larger in the flat case and $23\%$ larger in the
medium-curvature case. These values follow the opposite ordering of the
effective transition locations, which shift from approximately
$x_{\mathrm{tr}}=310~\mathrm{mm}$ in the flat case to
$320~\mathrm{mm}$ and $325~\mathrm{mm}$ in the medium- and
high-curvature cases, respectively. The configuration with the largest
thickness-based centrifugal-instability metric therefore transitions
first, whereas the configuration with the smallest metric transitions
last. Although the comparison involves only three configurations and
should consequently be interpreted as a quantitative trend rather than
a statistical correlation, it demonstrates that applying the expected
$\delta_{99}^{3/2}$ scaling to the measured boundary-layer modification
is consistent with both the magnitude and ordering of the observed
transition delay.

Table~\ref{gortlertabel} supports the same interpretation quantitatively. The minimum values of $G_{fs,\theta\mathrm{Full}}$ differ only weakly among the three cases, indicating that the initial instability environment is broadly comparable near the upstream portion of the domain. The maximum value, however, decreases monotonically from $317.63$ in the flat configuration to $277.60$ in the low-curvature case and $193.07$ in the high-curvature case. Thus the principal influence of spanwise curvature is not to alter the sign or basic nature of the centrifugal forcing, but to suppress its downstream amplification by modifying the mean boundary-layer development. This is precisely the behaviour expected if transverse curvature acts primarily through the thickness effect identified in the MLL analysis.

The same thinning also helps explain why acoustic activity becomes relatively more prominent as the curvature increases, even though it does not control the transition location. A thinner hypersonic boundary layer supports shorter characteristic acoustic scales and therefore shifts Mack-mode activity toward higher frequencies, as confirmed by the SPOD results below. In the present parameter range, however, this increased acoustic prominence does not overcome the reduction in centrifugal amplification. The decisive effect of spanwise convex curvature is therefore the suppression and delay of the G\"ortler/sinuous breakdown sequence, with the acoustic branch remaining secondary in setting the primary transition point. The same thinning that weakens the centrifugal branch also favours higher-frequency acoustic activity. The modal analysis below shows that acoustic modes become more prominent with increasing curvature, but they remain secondary to the G\"ortler/sinuous branch in setting the primary transition location.

\begin{table}
\centering
\caption{Freestream-referenced G\"ortler-number ranges based on full-profile momentum thickness, $G_{\theta}$, evaluated with $h_{\max}=18$ mm.}
\begin{tabular}{lccccc}
\hline
Case & Min $G_{\theta}$ & $x$ at min (m) & Max $G_{\theta}$ & $x$ at max (m) & Range \\
\hline
Flat & 11.98 & 0.100 & 317.63 & 0.358 & 11.98--317.63 \\
LCA  & 11.65 & 0.100 & 277.60 & 0.358 & 11.65--277.60 \\
HCA  & 11.44 & 0.100 & 193.07 & 0.358 & 11.44--193.07 \\
\hline
\end{tabular}
\label{gortlertabel}
\end{table}

\subsection{Spectral analysis of hypersonic boundary layer transition}
\label{subsec:spod_results}

To identify the instability content responsible for the delayed transition, SPOD was applied in two streamwise windows. Region~1 lies upstream of the breakdown zone and is intended to isolate the organized disturbances that develop on the still-coherent boundary layer, whereas Region~2 spans the onset of breakdown and therefore captures the stage at which vortical and acoustic motions coexist most strongly. In all three geometries the leading eigenvalue remains well separated from the remainder of the spectrum, indicating that the dominant dynamics in each window are carried by a comparatively small number of coherent structures rather than by a broadband background alone.

\begin{figure}
     \centering
     \begin{subfigure}[b]{0.42\textwidth}
         \centering
         \includegraphics[width=\textwidth]{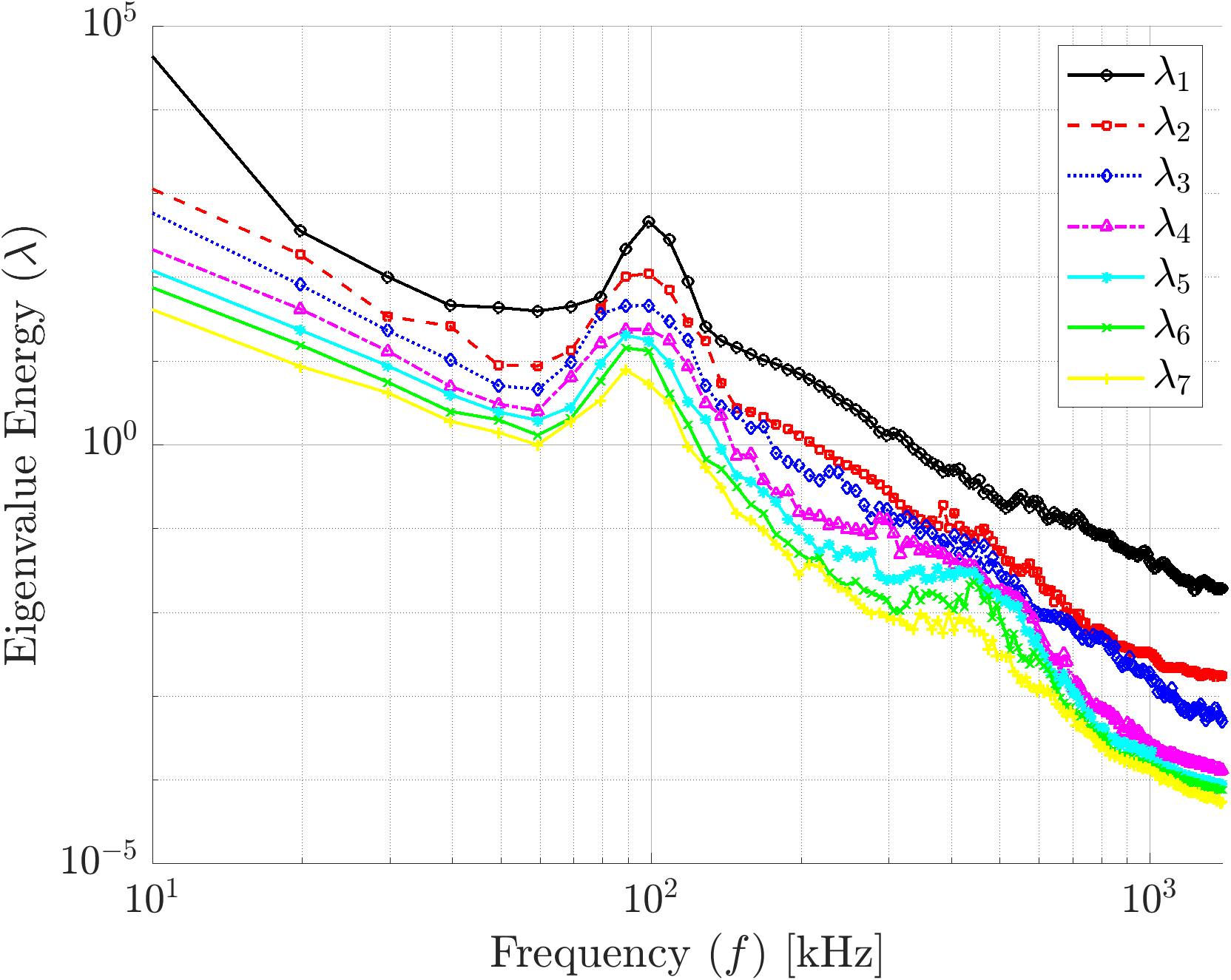}
     \end{subfigure}
     \hfill
     \begin{subfigure}[b]{0.42\textwidth}
         \centering
         \includegraphics[width=\textwidth]{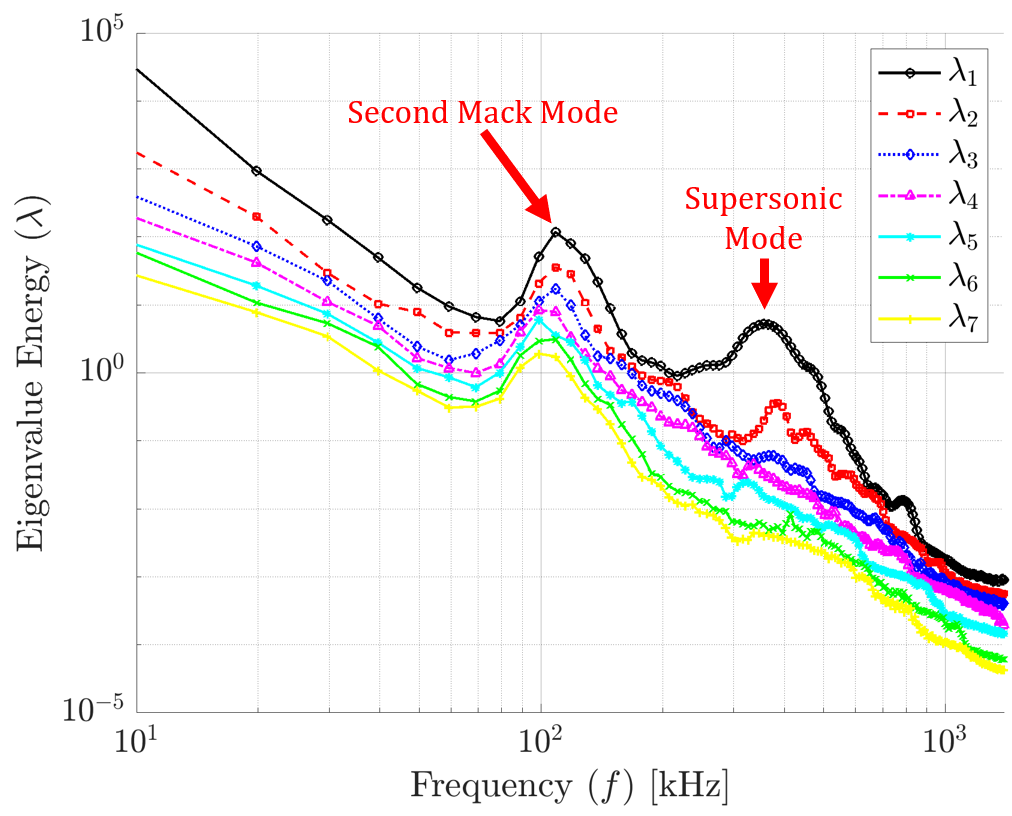}
     \end{subfigure}
     
     \caption{SPOD pressure-eigenvalue energy in Region~1 for the no-curvature case (left) and high-curvature case (right).}
     \label{fig:comparison}
\end{figure}

The spectral results show that both vortical and acoustic branches are present throughout the transition process, but that their relative roles depend on frequency and streamwise position. In Region~1, the pressure spectra contain a prominent high-frequency peak between roughly $88$ and $120~\mathrm{kHz}$ for all three configurations, as illustrated in Figure~\ref{fig:comparison}. This peak is consistent with a second Mack mode and shifts to higher frequency as the spanwise curvature increases. That trend matches the boundary-layer-thinning argument developed above: a thinner hypersonic boundary layer supports shorter trapped-acoustic scales and therefore pushes the dominant acoustic content to higher frequency. At the same time, the spectra retain appreciable energy near zero frequency, indicating that slowly varying or quasi-stationary vortical content remains present even upstream of full breakdown.

\begin{figure}
    \centering
    \begin{subfigure}[b]{0.31\textwidth}
        \centering
        \includegraphics[width=\textwidth]{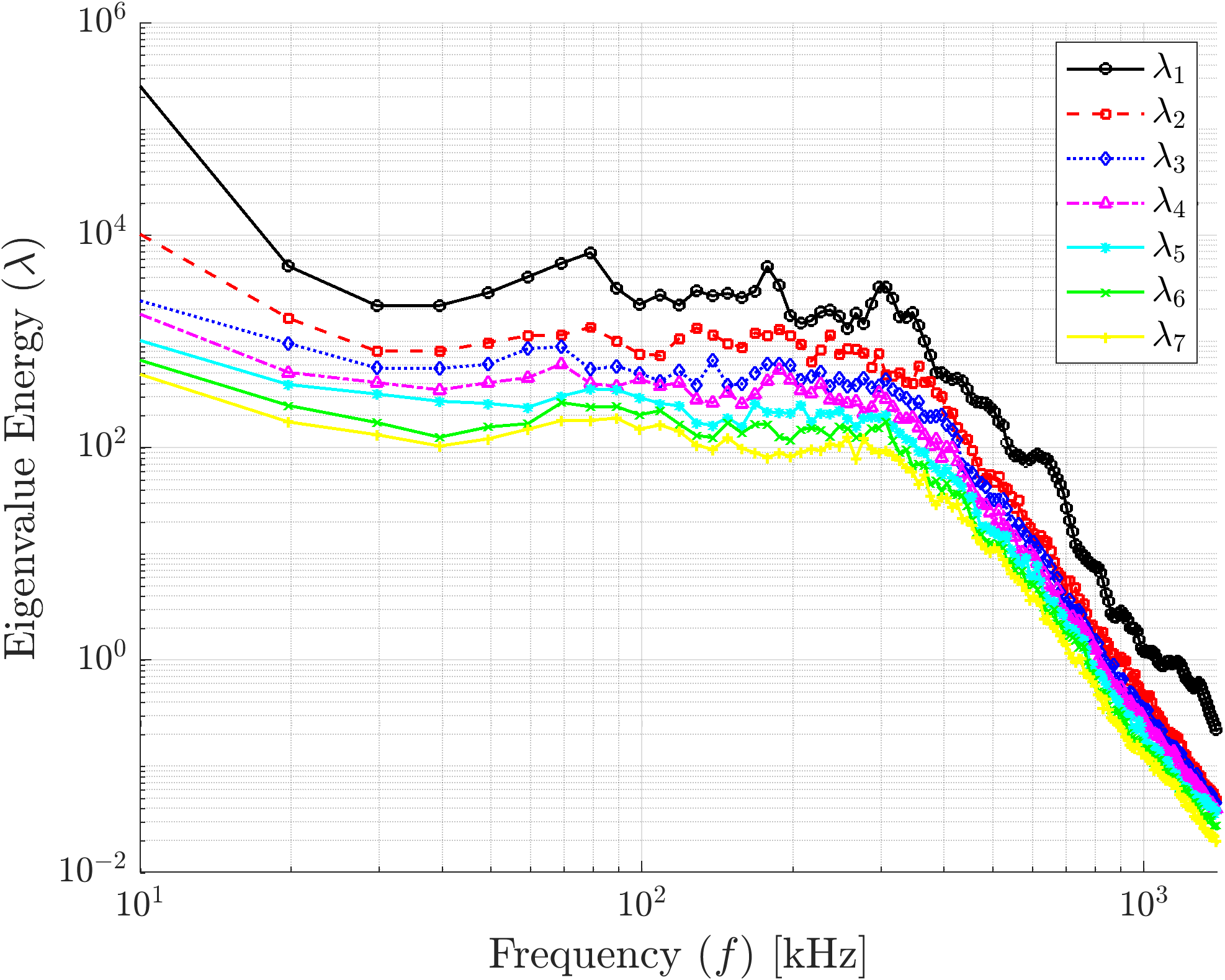}
        \caption{No curvature: $p'$}
    \end{subfigure}
    \hfill
    \begin{subfigure}[b]{0.31\textwidth}
        \centering
        \includegraphics[width=\textwidth]{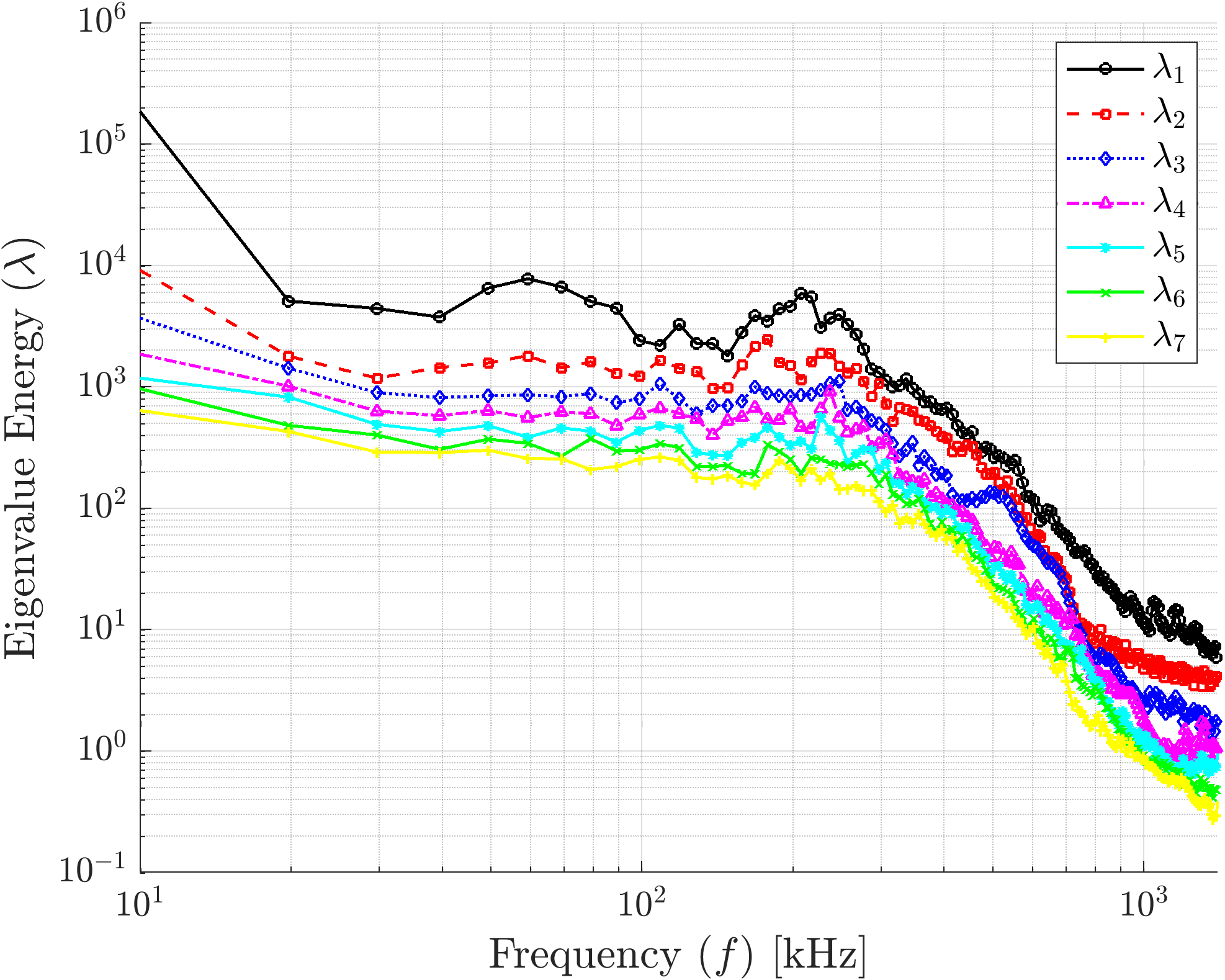}
        \caption{Medium curvature: $p'$}
    \end{subfigure}
    \hfill
    \begin{subfigure}[b]{0.31\textwidth}
        \centering
        \includegraphics[width=\textwidth]{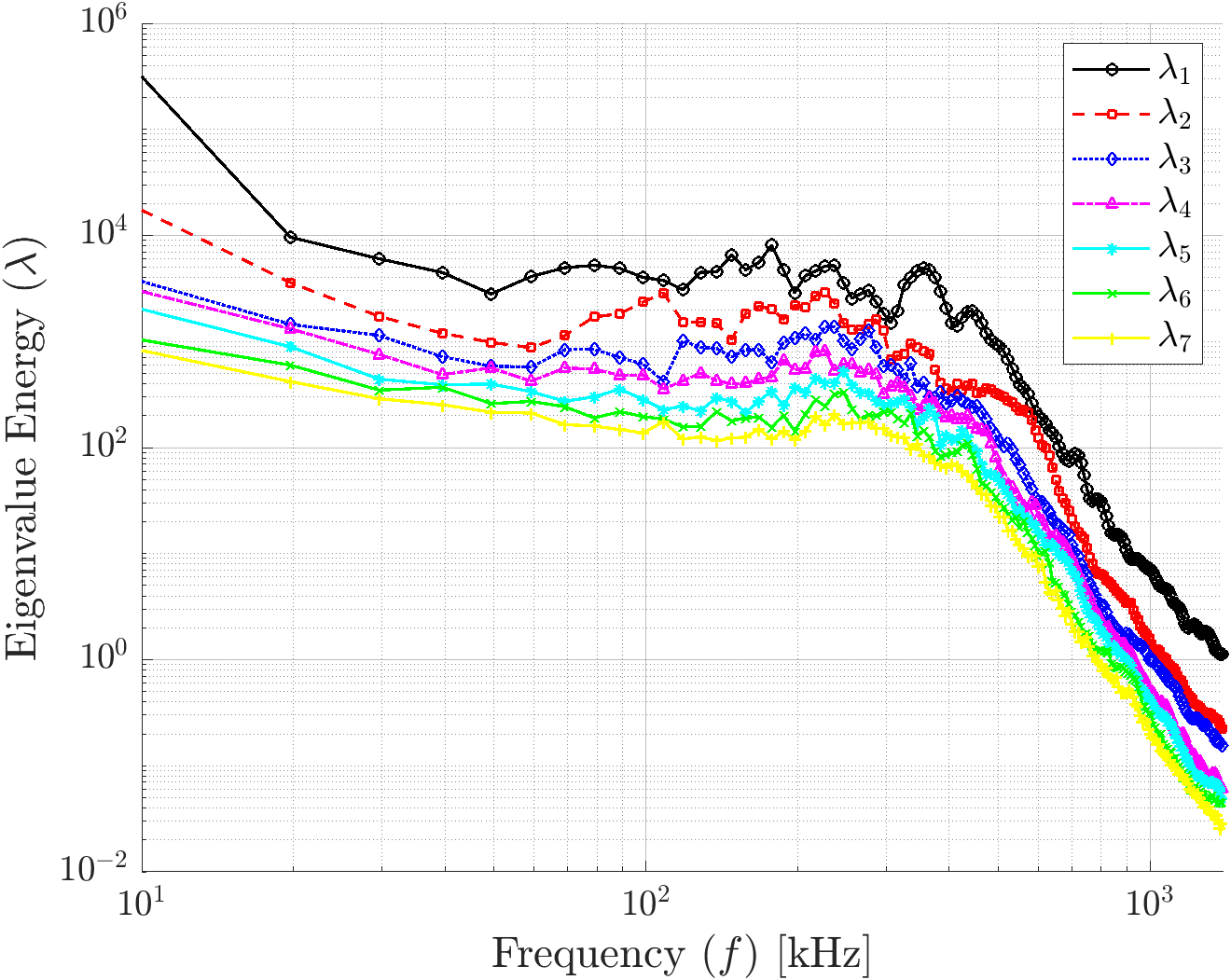}
        \caption{High curvature: $p'$}
    \end{subfigure}

    \caption{SPOD pressure-fluctuation energy spectra in Region~2 across the three geometries.}
    \label{fig:spod_spectra_all}
\end{figure}

The Region~2 spectra in Figure~\ref{fig:spod_spectra_all} clarify how the modal content changes as the flow approaches breakdown. A low-frequency energetic feature appears below about $100~\mathrm{kHz}$ in all three cases. This part of the spectrum is consistent with the sinuous secondary instability of the G\"ortler vortices: sinuous modes displace the vortex cores laterally in an oscillatory manner, whereas varicose modes primarily modulate their thickness and produce streamwise pinching. The sinuous branch is generally the more amplified secondary mode on G\"ortler vortices and is the dominant breakdown mechanism observed here \citep{swearingen1987growth}. Its frequency does not follow the simple monotonic shift exhibited by the Mack-mode peak in Region~1, which suggests that it is controlled mainly by the curvature-modified vortical structure rather than by acoustic scaling alone.

The same Region~2 spectra also contain higher-frequency content extending from roughly $100$ to $300~\mathrm{kHz}$. These peaks are consistent with Mack-type and supersonic acoustic disturbances and, in some cases, exhibit harmonic structure indicative of nonlinear interaction. The important point is that the low-frequency vortical peak and the higher-frequency acoustic peaks coexist in the same breakdown window. Transition therefore cannot be interpreted as a purely centrifugal or purely acoustic event; rather, the centrifugal branch continues to set the principal route to breakdown while the acoustic branch becomes increasingly active as the flow evolves downstream.

\begin{figure}
    \centering
    \begin{subfigure}[b]{0.48\textwidth}
        \centering
        \includegraphics[width=\textwidth]{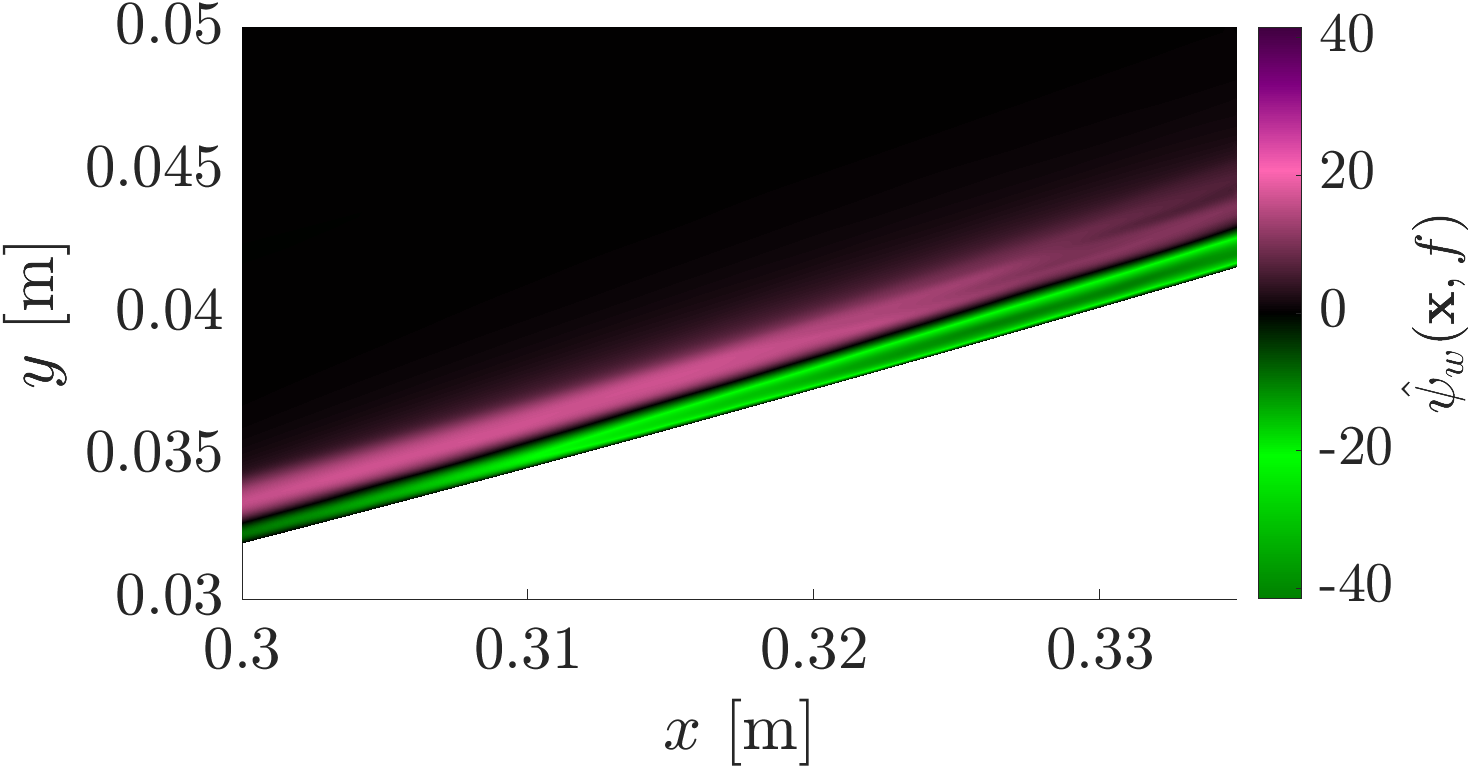}
        \caption{No curvature, $f=0.000$ kHz}
    \end{subfigure}
    \hfill
    \begin{subfigure}[b]{0.48\textwidth}
        \centering
        \includegraphics[width=\textwidth]{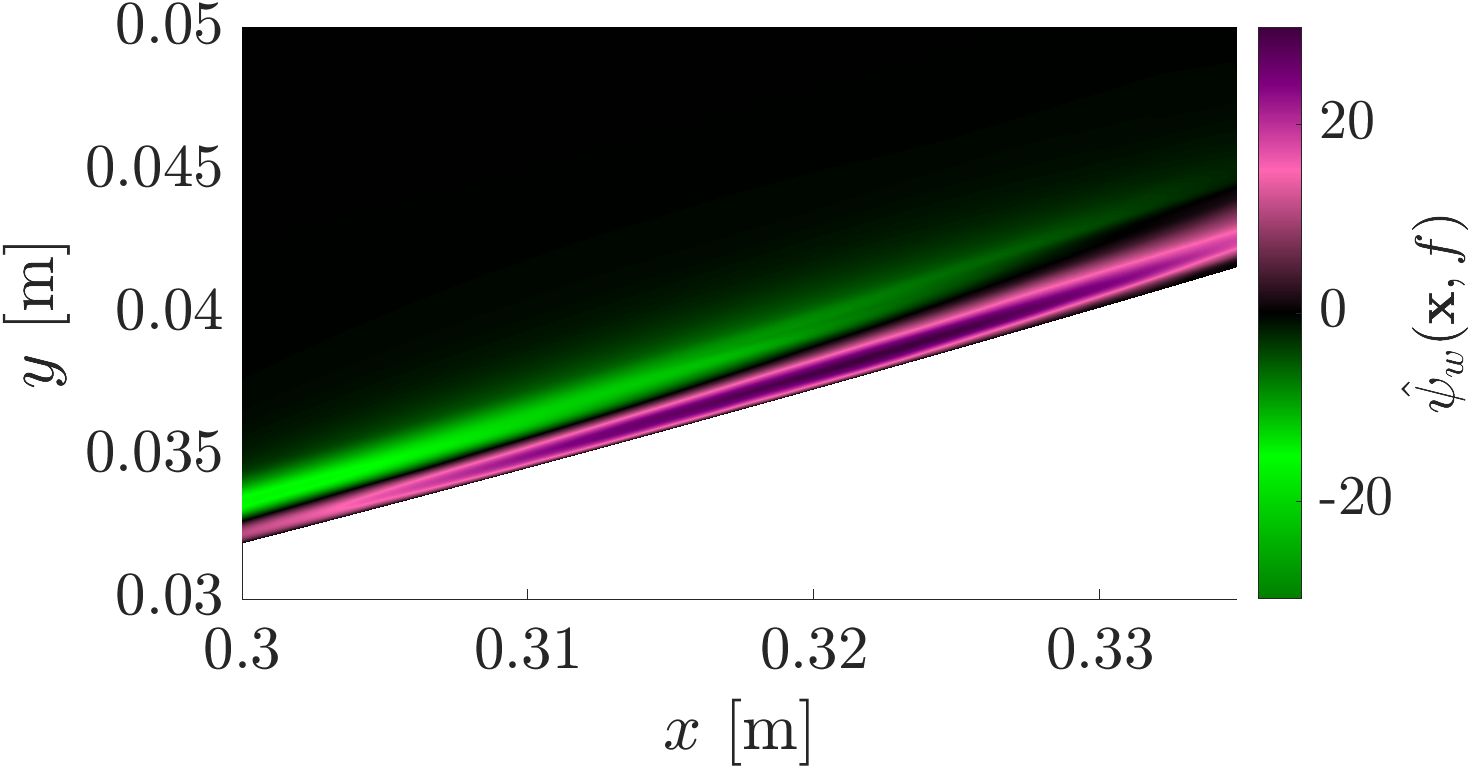}
        \caption{No curvature, $f=9.869$ kHz}
    \end{subfigure}
    \caption{Representative low-frequency $w'$ modes in Region~2, showing quasi-stationary G\"ortler structure.}
    \label{fig:wpod_lowfreq}
\end{figure}

The $w'$ mode shapes confirm that the low-frequency branch is vortical. At the lowest frequencies, the dominant modes capture quasi-stationary or slowly drifting G\"ortler structure, as shown in Figure~\ref{fig:wpod_lowfreq}. At somewhat larger low frequencies the mode shapes become rope-like and oscillatory, which is consistent with sinuous deformation of the vortex cores. These structures are strongest and appear earliest in the flat case, then shift downstream with increasing spanwise curvature. This behaviour mirrors the instantaneous visualizations and the wall signatures: the same vortical mechanism persists in all three geometries, but it is weakened and displaced downstream as the boundary layer becomes thinner.

\begin{figure}
    \centering
    \begin{subfigure}[b]{0.88\textwidth}
        \centering
        \includegraphics[width=\textwidth]{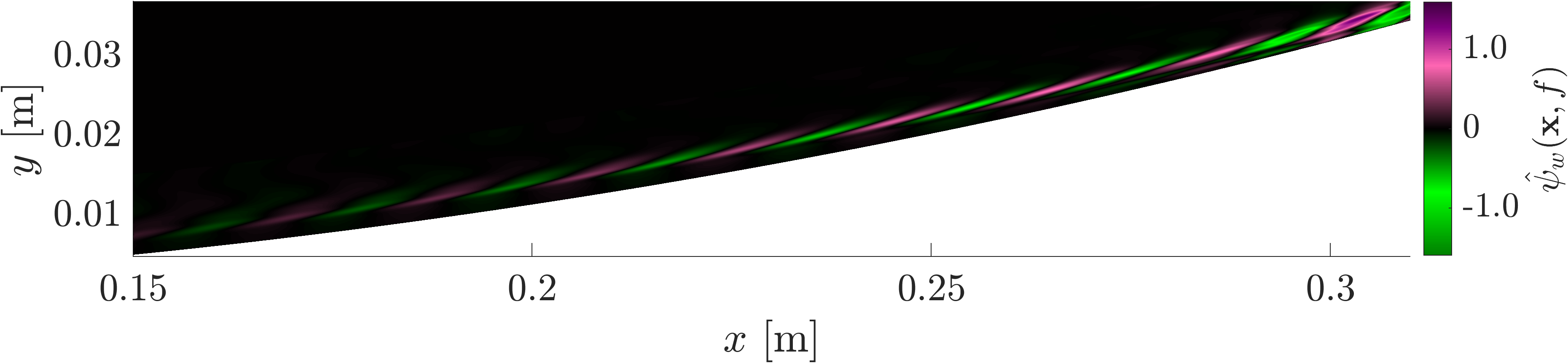}
        \caption{No curvature, $f=49.346$ kHz}
    \end{subfigure}
    \hfill
    \begin{subfigure}[b]{0.88\textwidth}
        \centering
        \includegraphics[width=\textwidth]{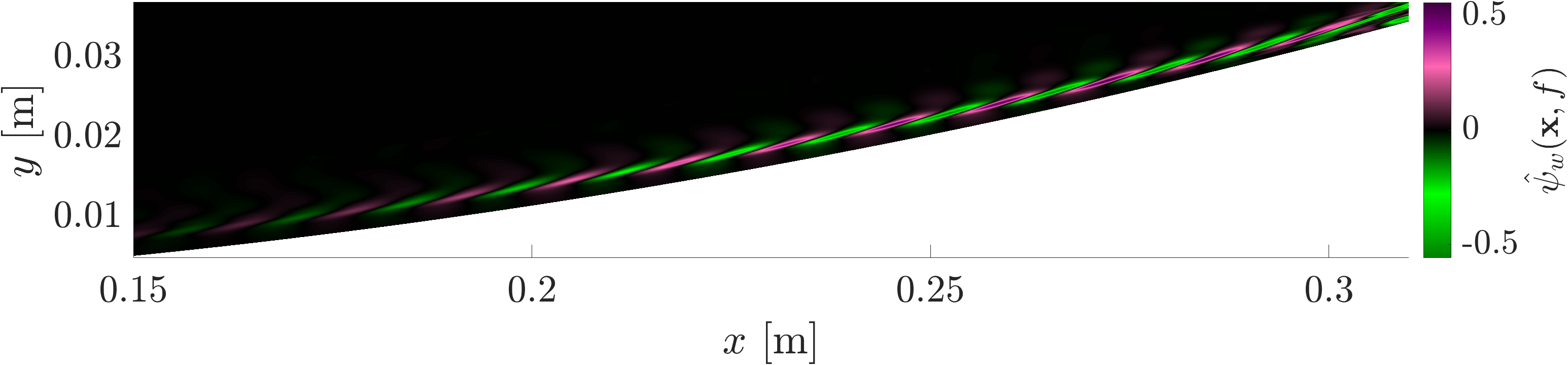}
        \caption{High curvature, $f=69.085$ kHz}
    \end{subfigure}
    \caption{Representative $w'$ modes in Region~1 showing rope-like, oscillatory structures consistent with sinuous secondary instability.}
    \label{fig:wpod_sinuous}
\end{figure}

\begin{figure}
    \centering

    \includegraphics[width=0.88\textwidth]{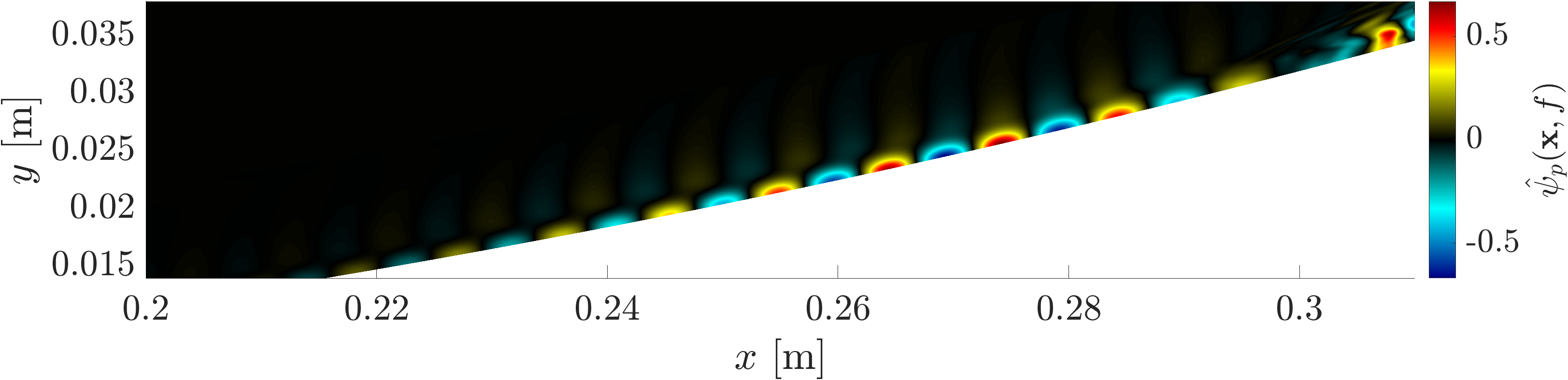}

    \vspace{4pt}

    \includegraphics[width=0.88\textwidth]{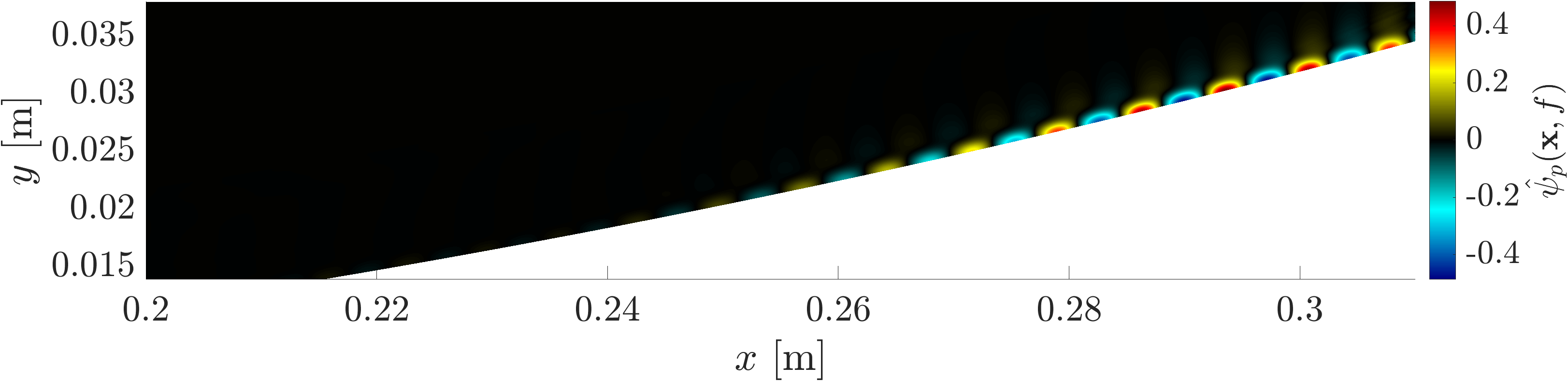}

    \caption{Real part of the pressure mode shape at $z=25$: no spanwise curvature at $f=88.823$ kHz (top) and high spanwise curvature at $f=118.431$ kHz (bottom). Both are sampled in Region~1 and exhibit second Mack-mode structure.}
    \label{fig:combined_modes_quad}
\end{figure}

The pressure modes isolate the acoustic branch more clearly than the velocity modes. Representative pressure eigenfunctions in Figure~\ref{fig:combined_modes_quad} show the structure expected of a second Mack mode in Region~1, and the corresponding frequency increases from the flat configuration to the high-curvature configuration. This shift provides modal confirmation that the acoustic environment is being modified by the same thickness variation that suppresses the centrifugal branch.

\begin{figure}
    \centering
    \begin{subfigure}[b]{0.48\textwidth}
        \centering
        \includegraphics[width=\textwidth]{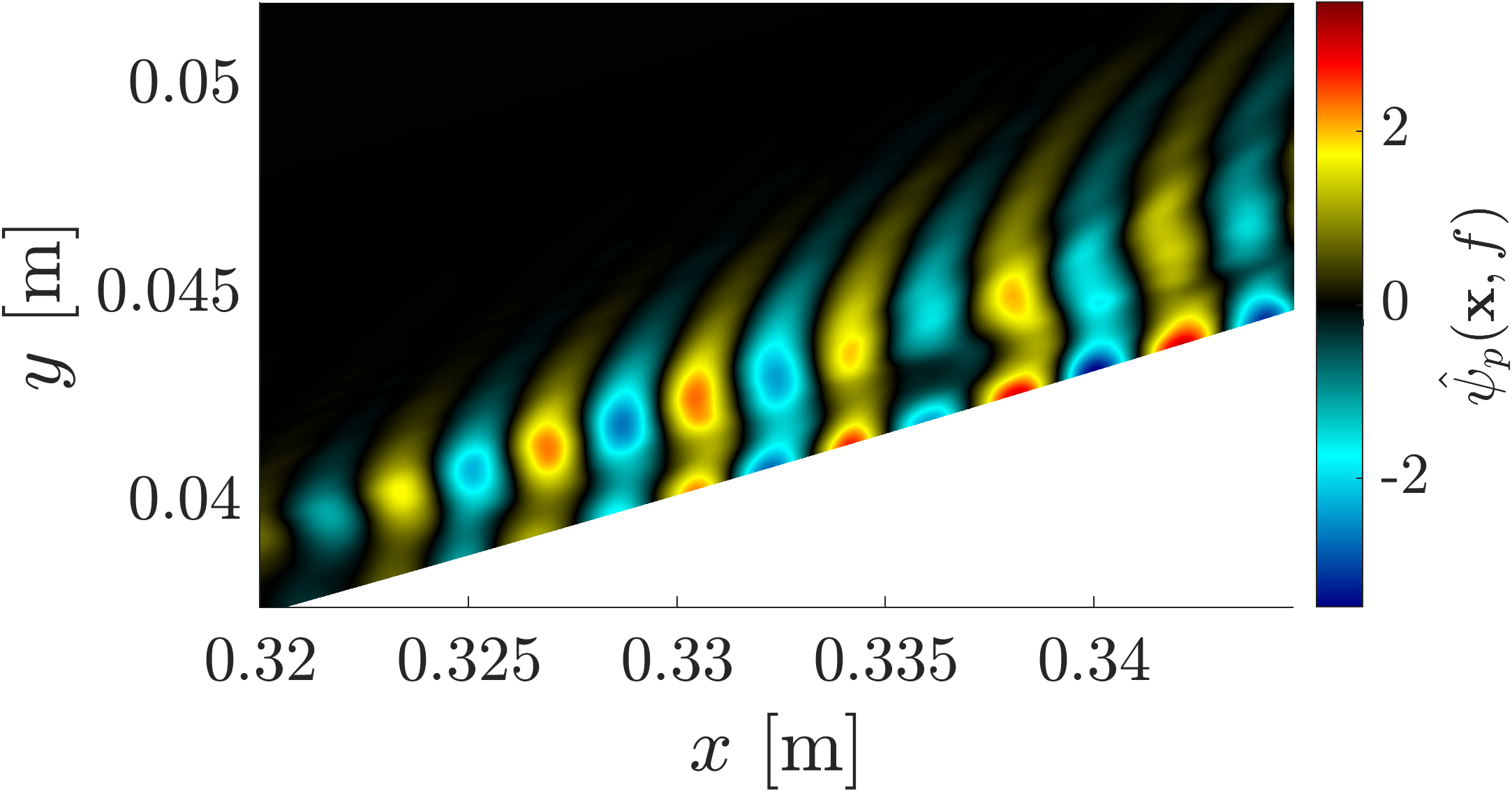}
        \caption{$f=187.516$ kHz}
    \end{subfigure}
    \hfill
    \begin{subfigure}[b]{0.48\textwidth}
        \centering
        \includegraphics[width=\textwidth]{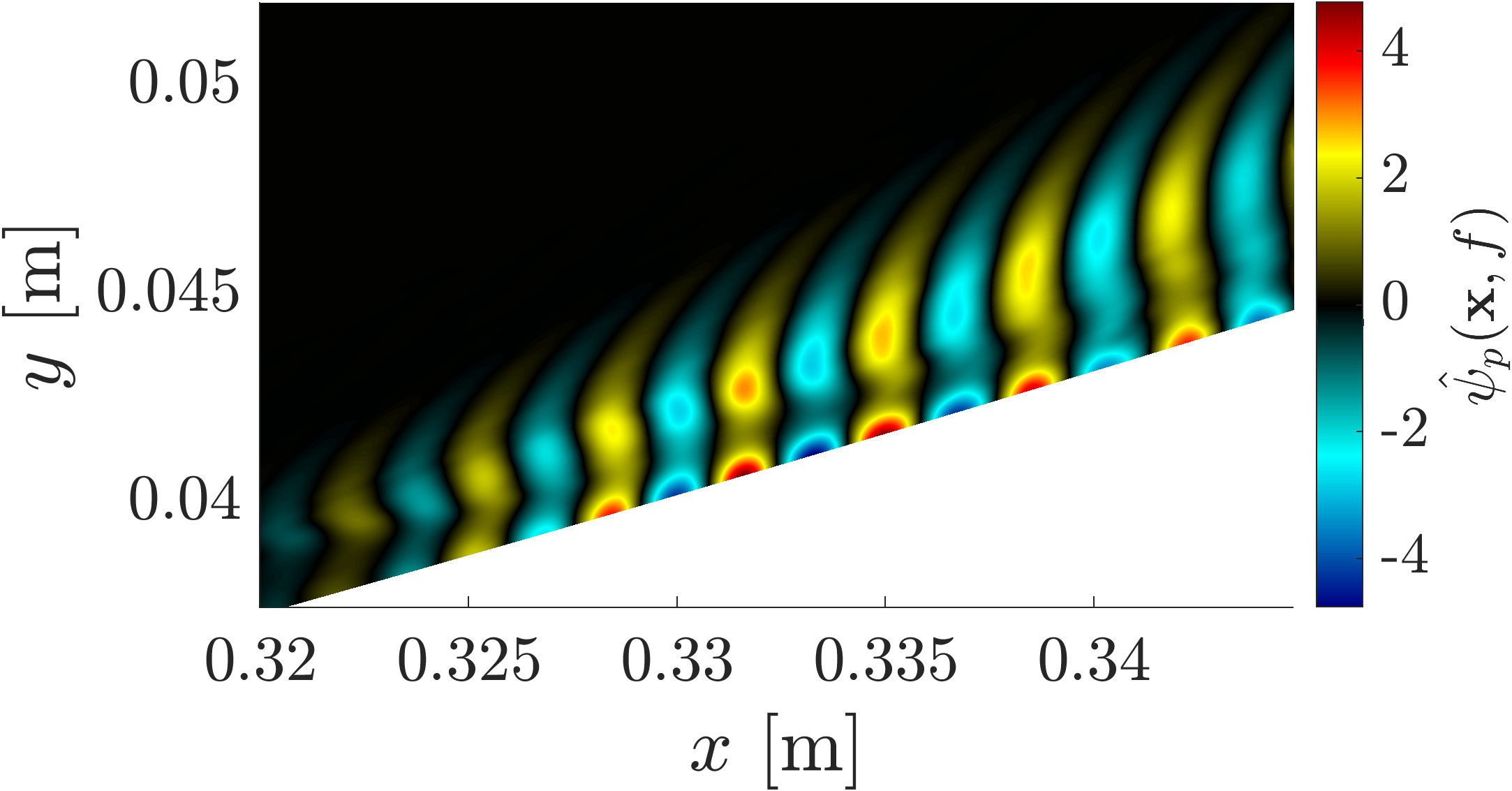}
        \caption{$f=207.254$ kHz}
    \end{subfigure}
    \caption{Representative Region~2 pressure modes for the medium-curvature case, showing increasingly supersonic-mode content near breakdown.}
    \label{fig:LCAmidfreq}
\end{figure}

In Region~2, the dominant higher-frequency pressure modes become more supersonic in character. Representative examples for the medium-curvature case are shown in figure~\ref{fig:LCAmidfreq}. These modes indicate that the acoustic field becomes more strongly expressed near breakdown, especially in the portion of the boundary layer where the vortical structures have already undergone substantial deformation. Even so, their appearance does not coincide with the disappearance of the G\"ortler branch; instead, they emerge alongside it and modify the late stages of the transition process.

\begin{figure}
\centering
    \includegraphics[width=.7\textwidth]{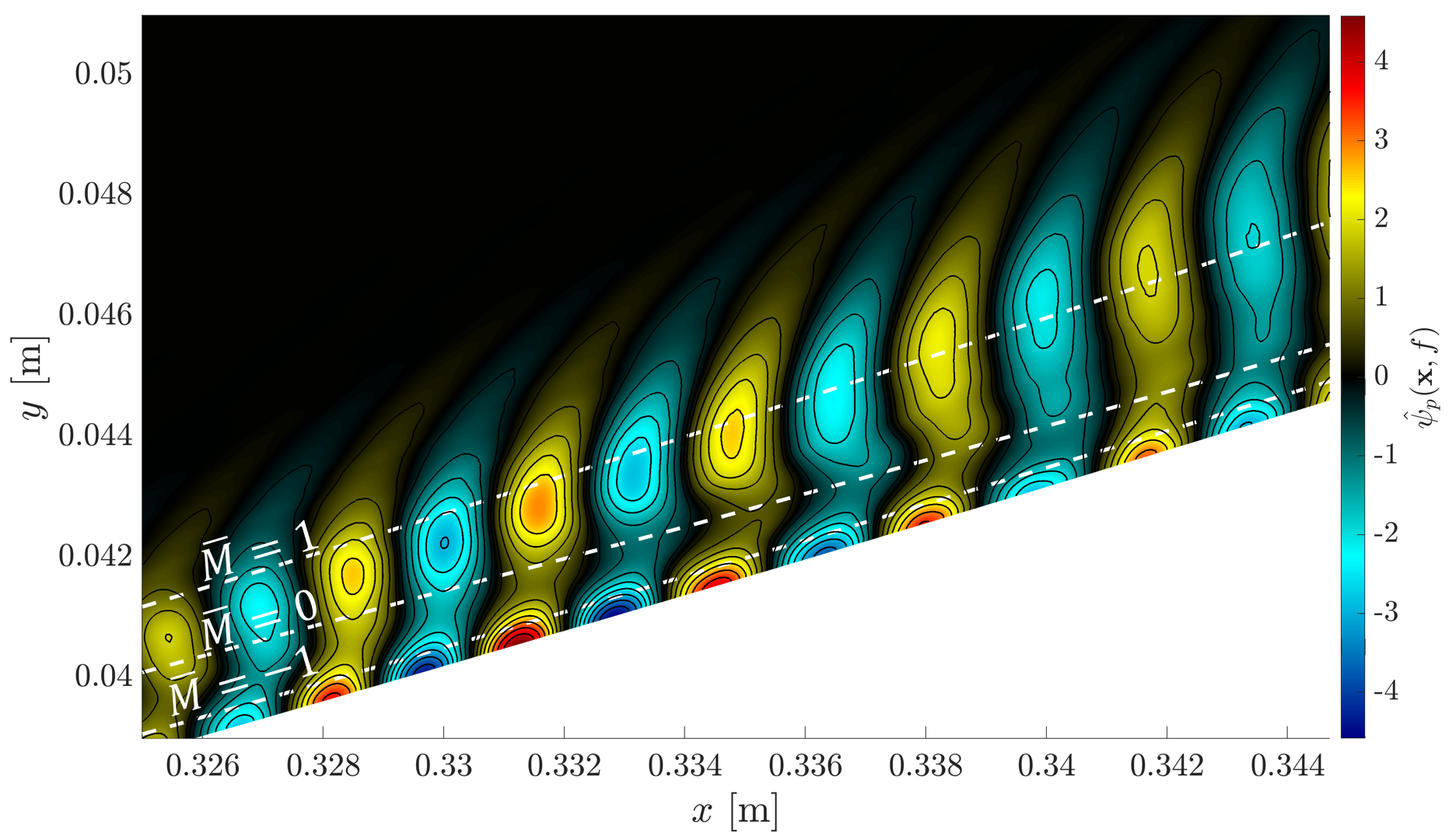}
    \caption{Representative Region~2 pressure mode for the medium-curvature case with relative sonic-line contours, illustrating Mack/supersonic acoustic structure.}
    \label{fig:ultrafigure}
\end{figure}

The relation between the Mack and supersonic acoustic branches is illustrated in figure~\ref{fig:ultrafigure}, which shows the confinement of Mack-type content beneath the appropriate relative sonic boundary together with the extension of the supersonic content into the freestream. This behaviour is consistent with prior work on cold-wall hypersonic boundary layers \citep{doi:10.2514/6.2019-2852} and supports the interpretation that the higher-frequency branch seen in Region~2 is not merely a continuation of the low-frequency vortical dynamics, but a distinct acoustic contribution that becomes more visible as breakdown proceeds.

\begin{figure}
\centering
    \includegraphics[width=.90\textwidth]{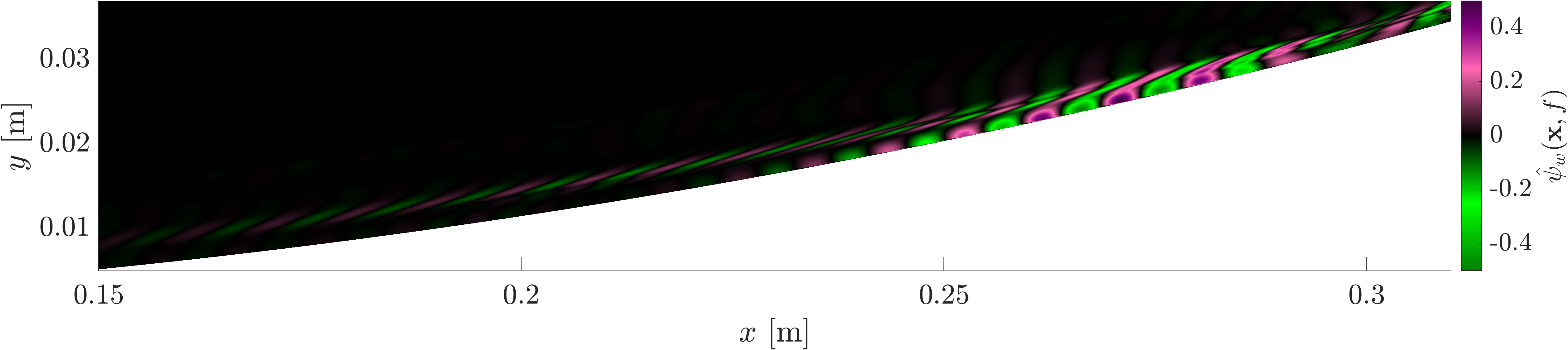}
    \caption{Representative $w'$ mode of the medium-curvature case at $98.693$ kHz showing interaction between G\"ortler structure and higher-frequency acoustic content.}
    \label{fig:interplay}
\end{figure}

Finally, Figure~\ref{fig:interplay} shows that the vortical and acoustic branches are not fully separable near breakdown. Even in a $w'$ mode, higher-frequency oscillations remain superposed on the downstream G\"ortler structure, indicating that acoustic activity is already embedded within a flow field whose large-scale organization is still set by the vortices. The spectral analysis therefore refines, rather than overturns, the picture obtained from the instantaneous and wall-based diagnostics: transition remains primarily G\"ortler- and sinuous-driven, while increasing spanwise curvature shifts the balance toward relatively stronger acoustic participation without replacing the vortical route to breakdown.

Overall, the SPOD results show that the delayed transition produced by spanwise convex curvature is fundamentally a modal reweighting of an otherwise similar instability system. Increasing curvature weakens and displaces the low-frequency vortical branch associated with G\"ortler growth and sinuous secondary instability, while simultaneously shifting the acoustic branch toward higher frequencies and making it relatively more visible near breakdown. The dominant transition pathway nevertheless remains vortical across the entire parameter range considered. Spanwise curvature therefore stabilizes the boundary layer primarily by suppressing centrifugal amplification, not by replacing G\"ortler-driven breakdown with an acoustic transition scenario.

\section{Experimental results}
\label{sec:exp_results}

The experiments provide independent confirmation of the principal computational result: increasing spanwise convex curvature delays transition while preserving a G\"ortler-dominated pathway. Infrared thermography gives the clearest surface evidence of this trend, while Schlieren imaging confirms the associated shock and near-wall disturbance structures.

\subsection{Infrared thermography}
\label{subsec:ir_results}

\begin{figure}
     \centering  
     \begin{subfigure}[t]{0.32\textwidth}
         \centering
         \includegraphics[width=\textwidth]{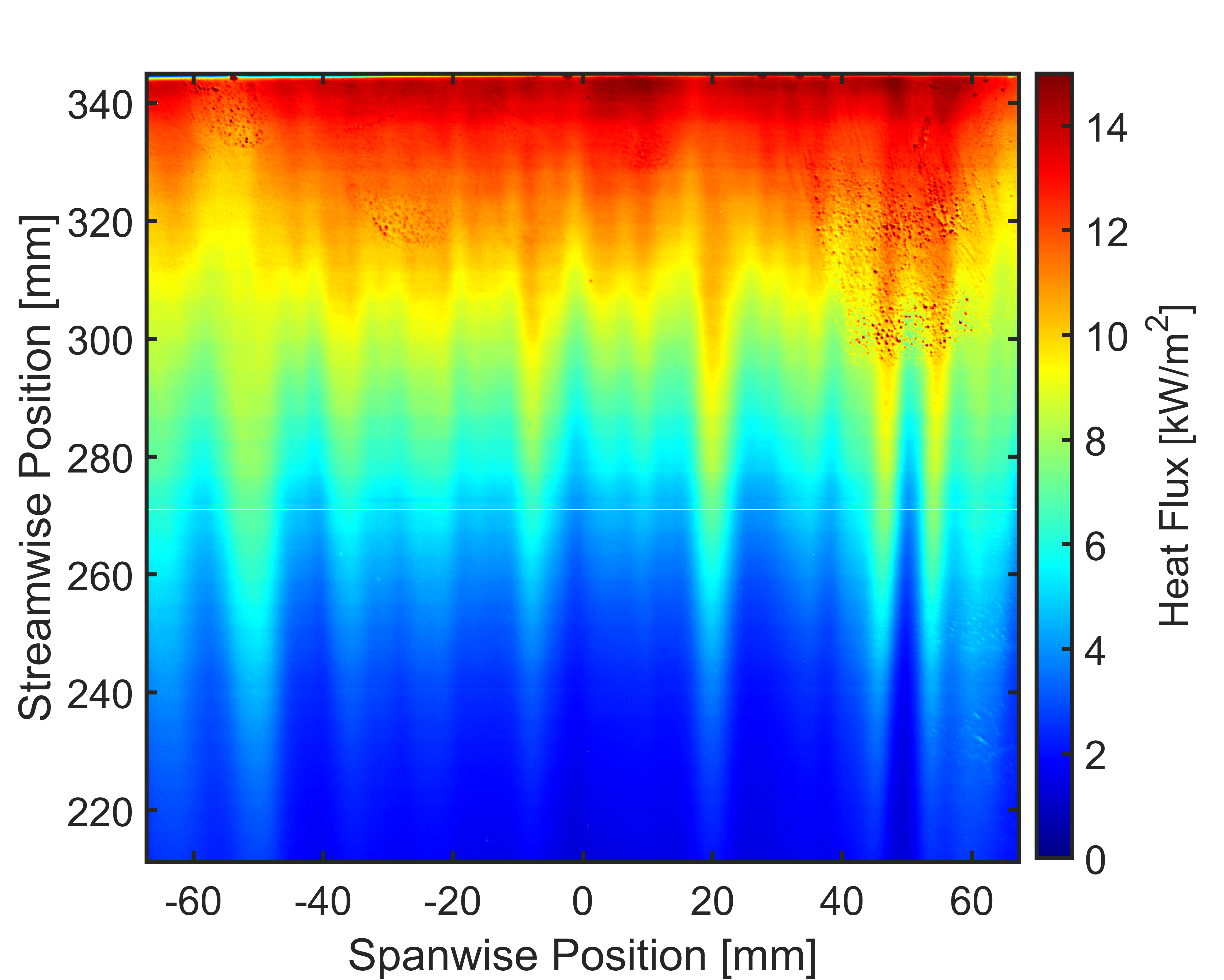}
         \caption{No curvature}
     \end{subfigure}
     \hfill
     \begin{subfigure}[t]{0.32\textwidth}
         \centering
         \includegraphics[width=\textwidth]{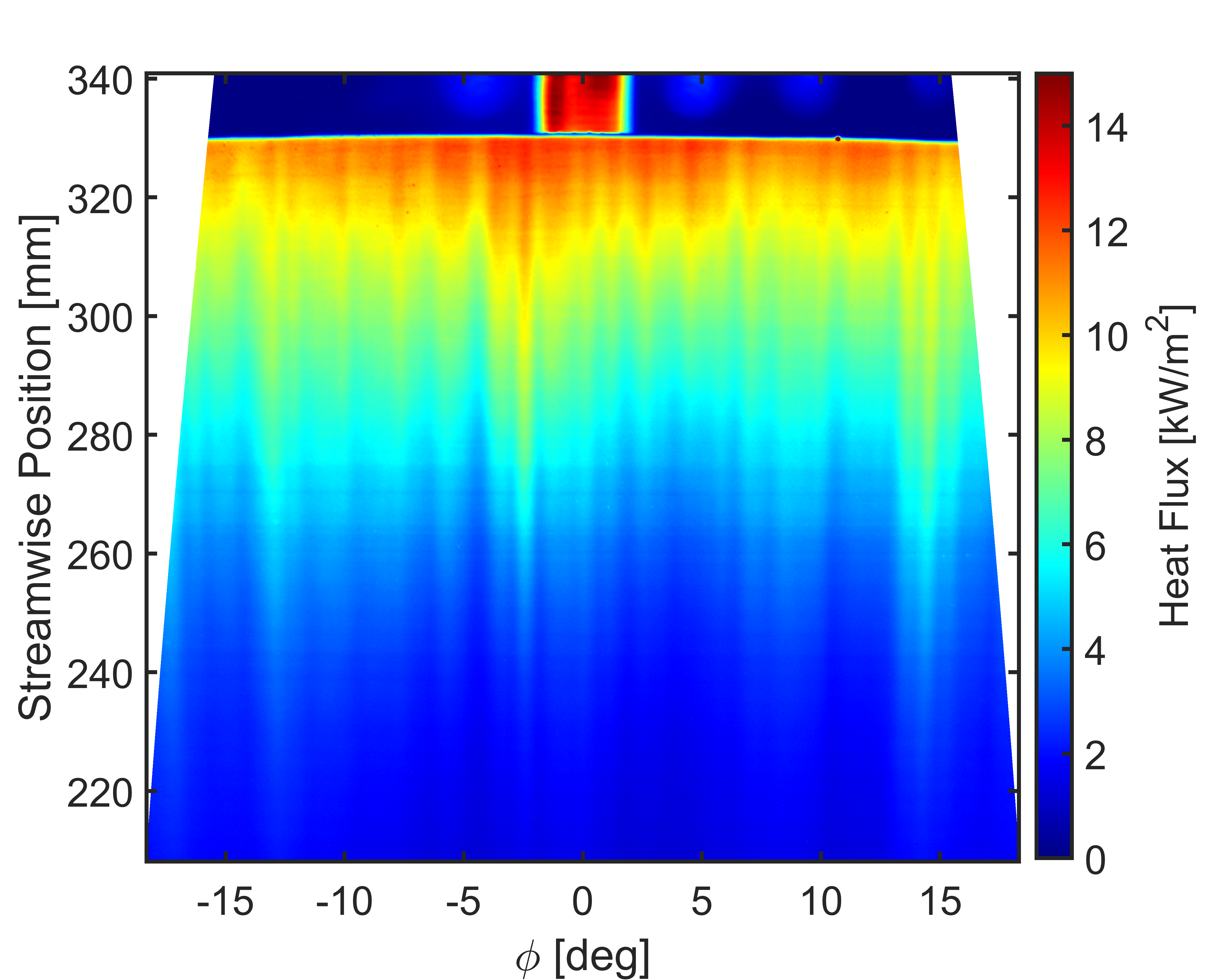}
         \caption{Medium curvature}
     \end{subfigure}
     \hfill
     \begin{subfigure}[t]{0.32\textwidth}
         \centering
         \includegraphics[width=\textwidth]{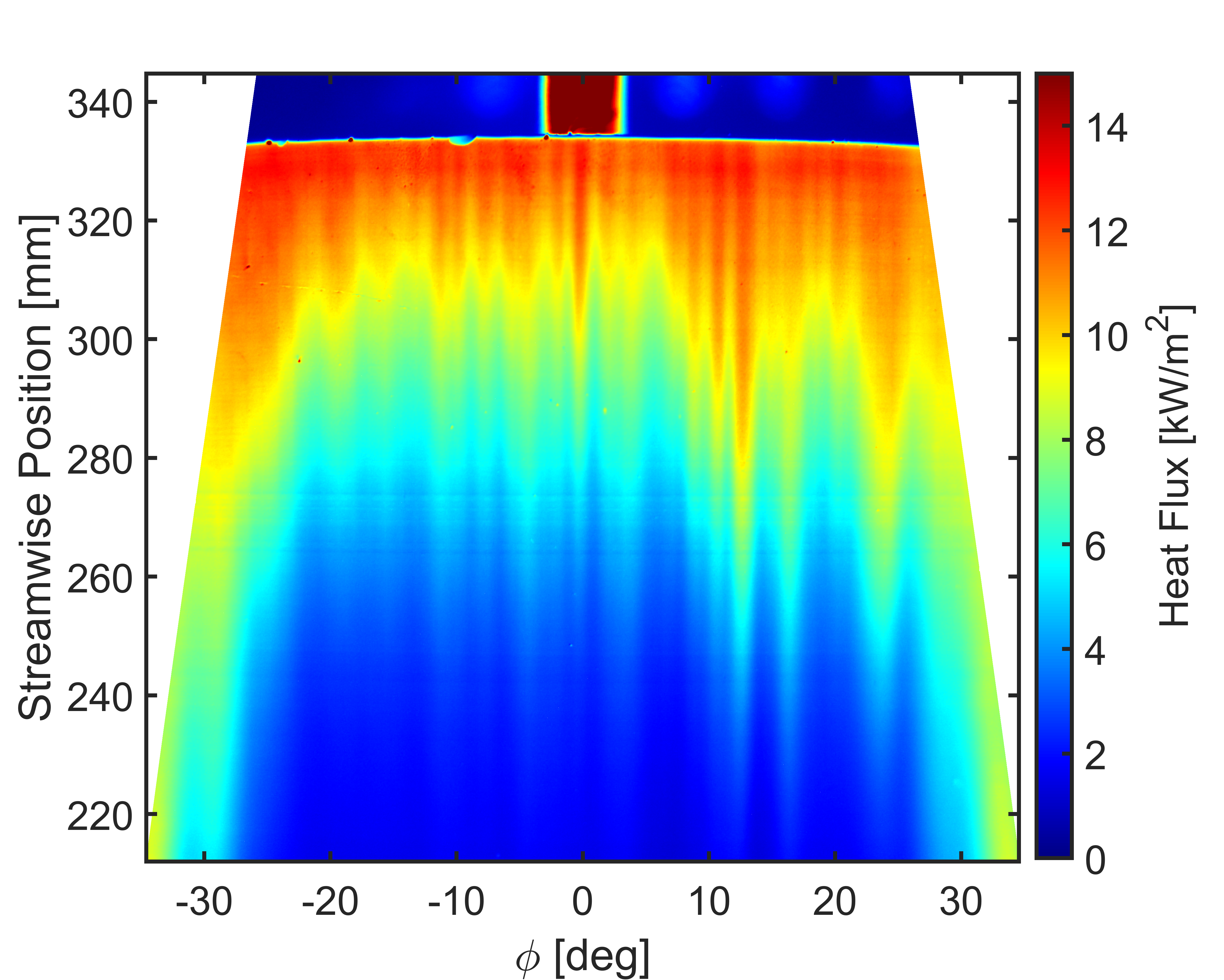}
         \caption{High curvature}
     \end{subfigure}
     
     \caption{Representative downstream infrared thermography for the three geometries. Thermal streaks associated with G\"ortler vortices shift downstream with increasing spanwise curvature.}
     \label{fig:curvature_comparison}
\end{figure}

Infrared thermography shows clear thermal streaks associated with G\"ortler vortices in all three geometries and confirms that increasing spanwise curvature shifts transition downstream. Representative downstream views are shown in Figure~\ref{fig:curvature_comparison}. The flat case exhibits the earliest streak amplification and breakdown, while the medium- and high-curvature cases remain more orderly farther downstream.

\begin{figure}
    \centering
    \includegraphics[width=0.6\textwidth]{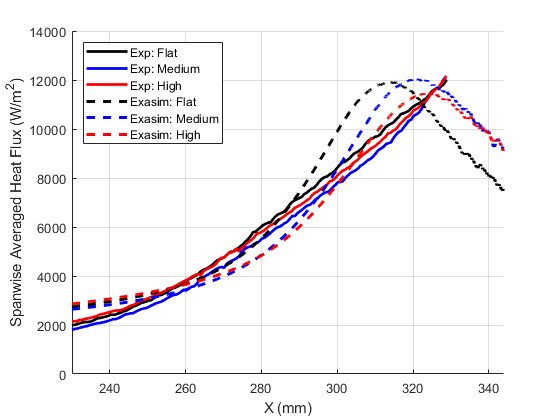}
    \caption{Spanwise-averaged surface heat flux from experiment (solid) and simulation (dashed) for the flat, medium-curvature, and high-curvature cases.}
    \label{fig:heat_flux_comparisonexp}
\end{figure}

A quantitative comparison is given in Figure~\ref{fig:heat_flux_comparisonexp}, which shows the spanwise-averaged surface heat flux from experiment and simulation. In both datasets, the flat case transitions earlier than the medium-curvature case. For the flat case, the experimental transition onset occurs roughly $30\,\mathrm{mm}$ downstream of the simulation, but the similar slope of the rise suggests that both capture the same underlying growth process despite a shifted breakdown threshold. This offset is not unexpected: the disturbance environments differ in kind as well as in level, with the tunnel fluctuations (\S\ref{subsec:ES-Facility}) being predominantly acoustic in content while the synthetic inflow perturbations (\S\ref{sec:synthetic}) are purely vortical, so the two intensities are not directly comparable and quantitative coincidence of the breakdown location is not anticipated. The comparison between experiment and simulation is therefore made at the level of trends. The high-curvature case is more difficult to interpret quantitatively because of edge effects in the experiment.

\subsection{Schlieren imaging}
\label{subsec:schlieren_results}

\begin{figure}
     \centering
     
     \begin{subfigure}[t]{0.32\textwidth}
         \centering
         \includegraphics[width=\textwidth]{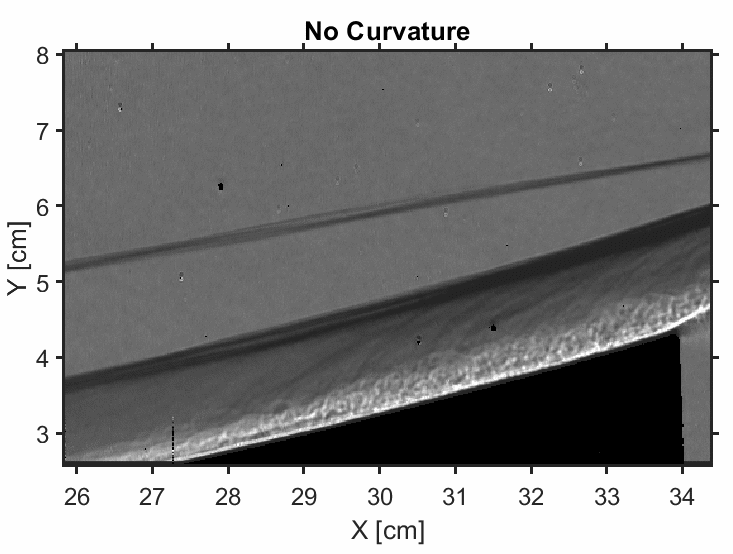}
         \caption{No spanwise curvature}
     \end{subfigure}
     \hfill
     \begin{subfigure}[t]{0.32\textwidth}
         \centering
         \includegraphics[width=\textwidth]{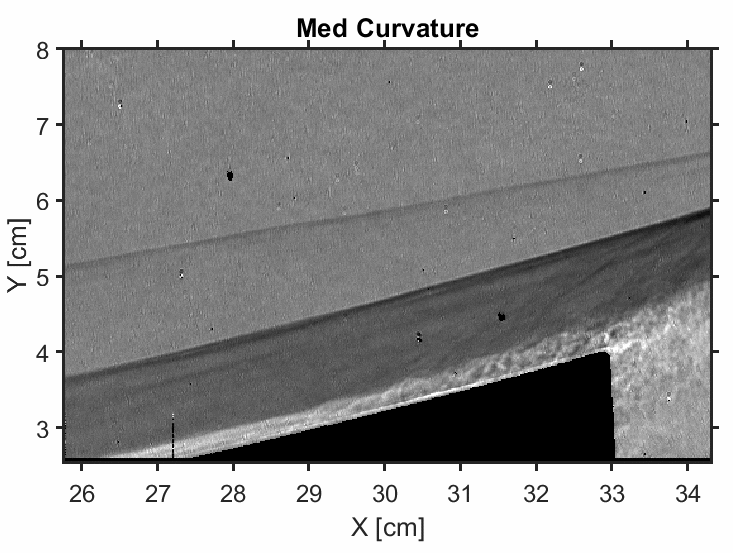}
         \caption{Medium spanwise curvature}
     \end{subfigure}
     \hfill
     \begin{subfigure}[t]{0.32\textwidth}
         \centering
         \includegraphics[width=\textwidth]{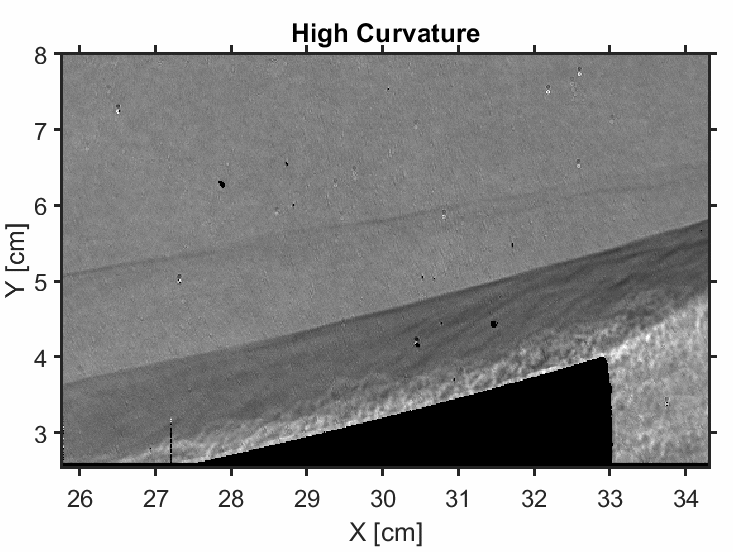}
         \caption{High spanwise curvature}
     \end{subfigure}
     
     \caption{Zoomed instantaneous Schlieren images near the downstream portion of the models. Increasing spanwise curvature delays the onset of strong near-wall disturbance growth.}
     \label{fig:schlieren_zoom_comp}
\end{figure}

Schlieren imaging confirms the same qualitative trend. Figure~\ref{fig:schlieren_zoom_comp} shows zoomed instantaneous images near the downstream portion of the models. The flat case exhibits the earliest near-wall distortion and strongest evidence of disturbance growth, whereas the medium and high-curvature cases remain more coherent farther downstream. The images also capture the leading-edge and compression-shock structure seen in the simulations.
While the no spanwise curvature  case appears to exhibit a stronger compression and leading edge
shock, this is hypothesized to be an artifact of performing Schlieren imaging on a
translationally symmetric surface rather than an axisymmetric one, which can stack multiple
frames together, giving the appearance of a stronger visual.

\subsection{PCB Results}
\label{subsec:pcb_results}

To further investigate the presence of potential second modes, PCBs were mounted onto the midpoint of the zero and high-spanwise curvature models, at 1 and 3 inch locations from the trailing edge. This corresponds to streamwise locations of $x=263.8\,\mathrm{mm}$ and $x=314.6\,\mathrm{mm}$ on the no spanwise curvature model, and $x=253.8\,\mathrm{mm}$ and $x=304.6\,\mathrm{mm}$ on the high spanwise curvature model.

\begin{figure}
  \centering
  
  \begin{subfigure}[b]{0.48\textwidth}
    \centering
    \includegraphics[width=\textwidth]{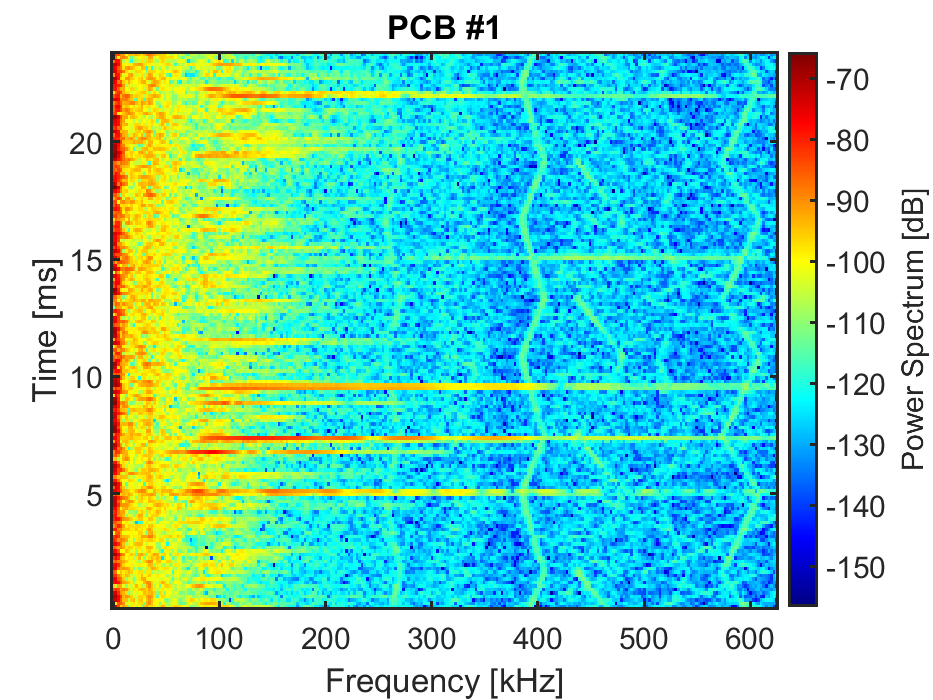}
    \caption{No Curvature: 3 inches from aft}
    \label{fig:sub1}
  \end{subfigure}
  \hfill
  \begin{subfigure}[b]{0.48\textwidth}
    \centering
    \includegraphics[width=\textwidth]{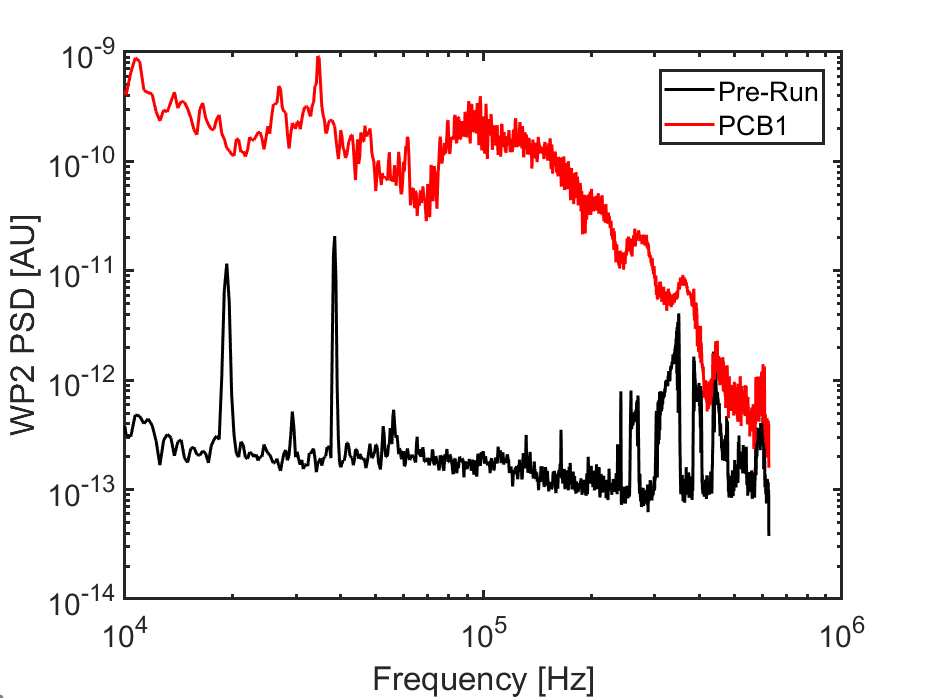}
    \caption{No Curvature: 3 inches from aft}
    \label{fig:sub2}
  \end{subfigure}

  \vspace{1em} 

  \begin{subfigure}[b]{0.48\textwidth}
    \centering
    \includegraphics[width=\textwidth]{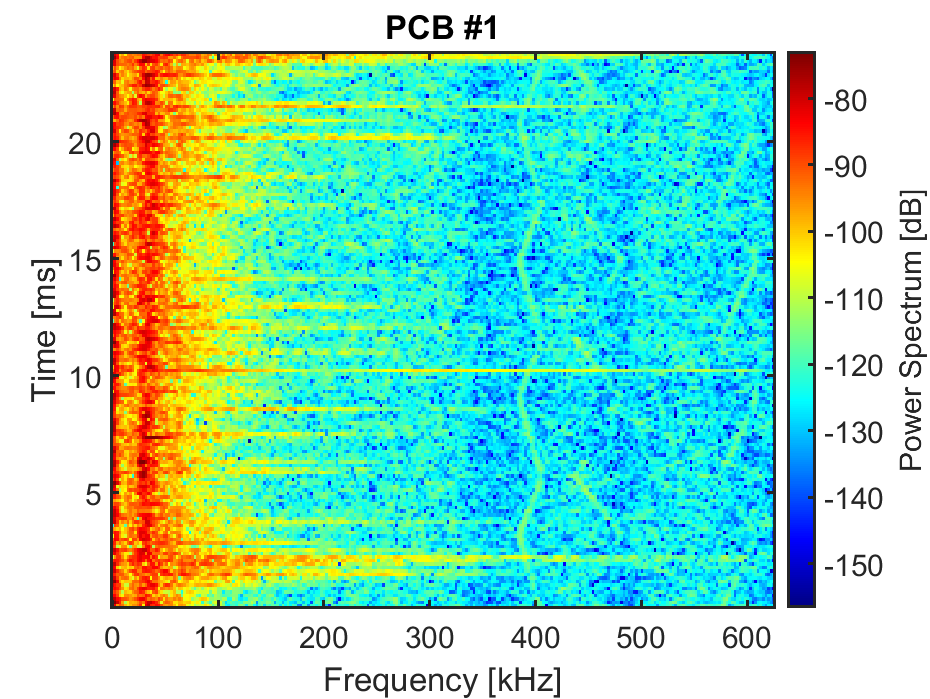}
    \caption{High Curvature: 3 inches from aft}
    \label{fig:sub5}
  \end{subfigure}
  \hfill
  \begin{subfigure}[b]{0.48\textwidth}
    \centering
    \includegraphics[width=\textwidth]{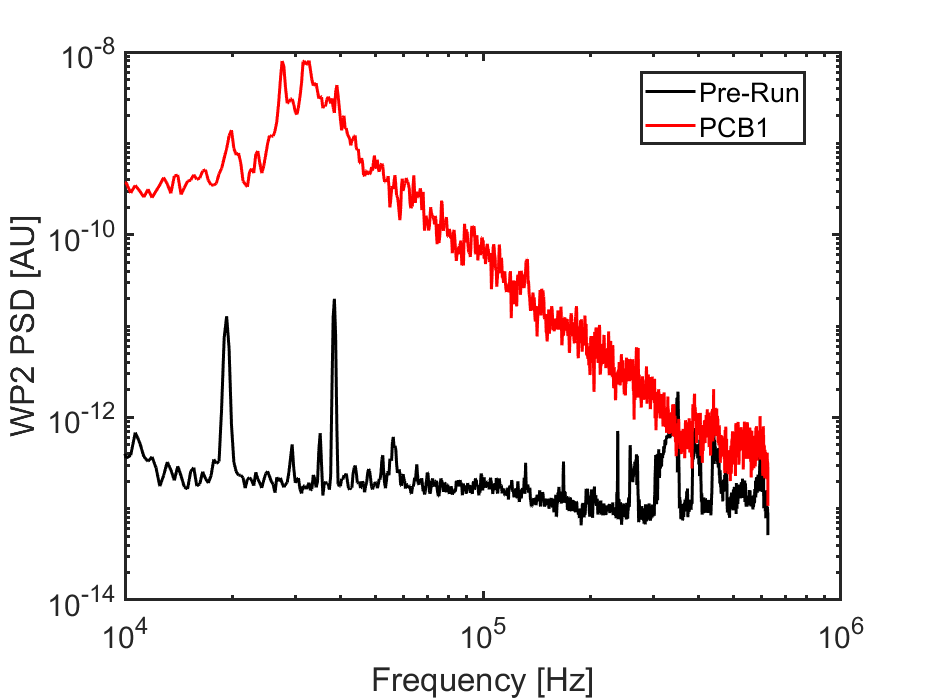}
    \caption{High Curvature: 3 inches from aft}
    \label{fig:sub6}
  \end{subfigure}

  \caption{PCB Spectrogram and Spectrum Data for the no and high curvature models 3 inches from their respective trailing edges.}
  \label{fig:pcb_data3in}
\end{figure}

\begin{figure}
  \centering
  
  \begin{subfigure}[b]{0.48\textwidth}
    \centering
    \includegraphics[width=\textwidth]{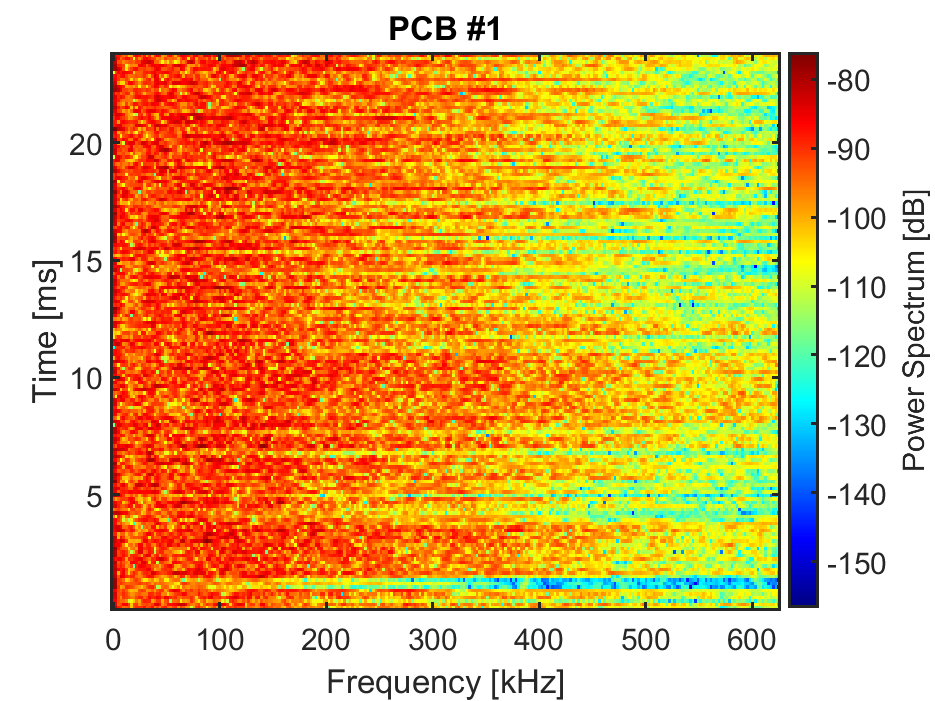}
    \caption{No Curvature: 1 inch from aft}
    \label{fig:sub1}
  \end{subfigure}
  \hfill
  \begin{subfigure}[b]{0.48\textwidth}
    \centering
    \includegraphics[width=\textwidth]{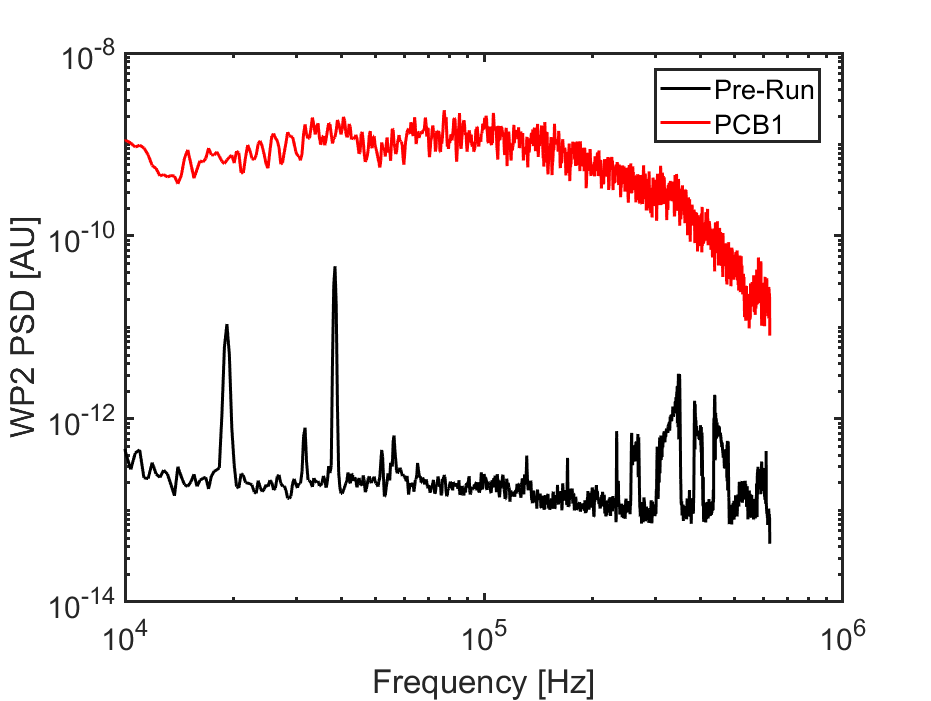}
    \caption{No Curvature: 1 inch from aft}
    \label{fig:sub2}
  \end{subfigure}

  \vspace{1em} 

  \begin{subfigure}[b]{0.48\textwidth}
    \centering
    \includegraphics[width=\textwidth]{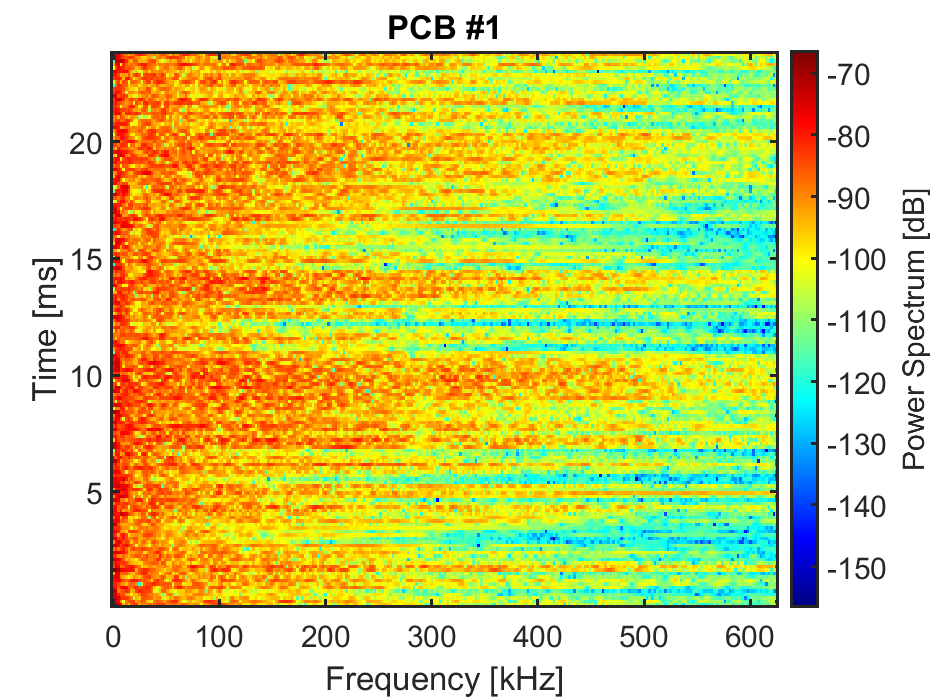}
    \caption{High Curvature: 1 inch from aft}
    \label{fig:sub5}
  \end{subfigure}
  \hfill
  \begin{subfigure}[b]{0.48\textwidth}
    \centering
    \includegraphics[width=\textwidth]{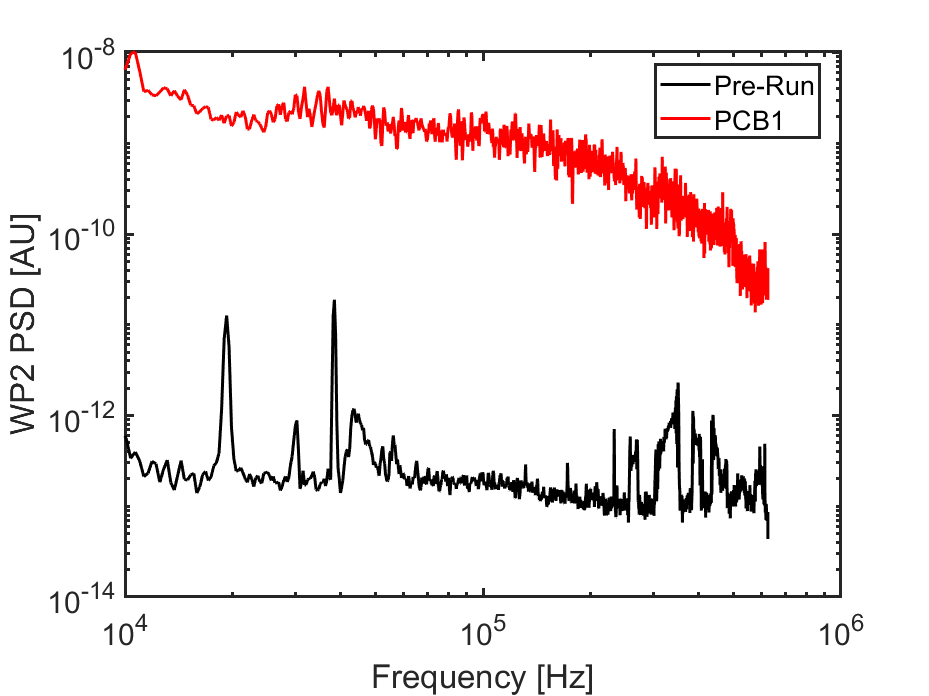}
    \caption{High Curvature: 1 inch from aft}
    \label{fig:sub6}
  \end{subfigure}

  \caption{PCB Spectrogram and Spectrum Data for the no and high curvature models 1 inch from their respective trailing edges.}
  \label{fig:pcb_data1in}
\end{figure}

In Figures \ref{fig:pcb_data3in} and \ref{fig:pcb_data1in}, spectrogram and spectral results from the PCB sampling are presented. In Figure \ref{fig:sub2}, the presence of a weak second mode peak of approximately 85-100 kHz was confirmed in the range defined by \cite{Stetsonetal1991} on the zero spanwise curvature model. This aligns with the frequency range found in the simulations in Figure \ref{fig:combined_modes_quad}. The second mode's presence in both the experiments and the simulations remains quite weak and is not the primary cause of breakdown.

On both models, there was a peak around the 30-40 kHz range which also aligns with the frequency peak detected in the simulations that strongly correlates with the presence of G{\"o}rtler sinuous instability. As the G{\"o}rtler rolls in both experiment and simulation appear to vary simultaneously, it is possible that the PCB during the high curvature case was more conveniently positioned towards a G{\"o}rtler vortex, resulting in stronger magnitudes on the PCB reading. The results of potential edge effects are also not disregarded here.

\section{Conclusions}

The effect of spanwise convex curvature on hypersonic boundary-layer transition over a streamwise-concave surface has been investigated using direct numerical simulation together with complementary Mach~7 experiments. By considering a family of smooth doubly curved geometries in which the streamwise concavity is held fixed while the spanwise curvature is increased, the study isolates the influence of transverse geometric curvature on a G\"ortler-dominated transition process. Across the full set of numerical and experimental diagnostics, increasing spanwise convex curvature delays transition while preserving the same qualitative route to breakdown.

The direct numerical simulations show that the dominant instability pathway remains centrifugal. In all three geometries, the transition process is initiated by the growth of streamwise G\"ortler vortices, followed by secondary deformation and eventual turbulent breakdown. Increasing spanwise curvature weakens and displaces this sequence downstream rather than replacing it with a different mechanism. The instantaneous interior fields reveal that the vortices remain more compact, closer to the wall and coherent over a longer streamwise extent as the curvature increases. The wall heat-flux maps show the same trend through delayed streak amplification, reduced downstream intermittency and a downstream shift of the onset of intense surface heating.

The principal stabilising mechanism is the reduction in effective boundary-layer thickness induced by spanwise convex curvature. The Mangler--Levy--Lees scaling provides the theoretical basis for this effect, and the DNS confirms that the physical boundary layer becomes progressively thinner as the spanwise curvature increases. Because the effective G\"ortler forcing decreases with decreasing boundary-layer thickness, the primary vortices experience weaker centrifugal amplification, their wall-normal penetration is reduced and the onset of sinuous secondary instability is delayed. The corresponding reduction in the freestream-referenced G\"ortler-number range provides a quantitative measure of this suppression.

The spectral analysis further clarifies how the instability system is modified by curvature. SPOD identifies a persistent low-frequency vortical branch associated with quasi-stationary G\"ortler structure and sinuous secondary instability, together with a higher-frequency acoustic branch containing Mack-type and supersonic disturbances. As the spanwise curvature increases, the acoustic content shifts toward higher frequencies and becomes relatively more prominent near breakdown, consistent with the thinner boundary layer. These acoustic disturbances nevertheless remain secondary in setting the primary transition point. The dominant route to transition therefore remains G\"ortler- and sinuous-driven throughout the parameter range considered, even though the late-stage breakdown region exhibits increasing acoustic participation.

The experimental results support the same physical interpretation. Infrared thermography shows that the thermal streaks associated with G\"ortler vortices shift downstream as the spanwise curvature increases, in agreement with the numerical wall-heat-flux fields. Schlieren imaging confirms the associated delay in strong near-wall disturbance growth. Although the available PCB measurements do not provide strong, definitive spectral peaks, the overall experimental trends are consistent with the DNS in showing that spanwise convex curvature stabilizes the boundary layer while preserving a G\"ortler-dominated transition scenario.

Taken together, these results show that spanwise convex curvature constitutes a stabilising geometric influence in hypersonic G\"ortler-driven boundary layers. Its primary role is to modify the mean boundary-layer development, thereby suppressing centrifugal amplification and delaying secondary breakdown, while leaving the fundamental transition pathway qualitatively unchanged. This finding is relevant to practical hypersonic configurations, for which streamwise and spanwise curvature commonly coexist and transition cannot be inferred reliably from streamwise concavity alone. More broadly, the results demonstrate that controlled variation of transverse curvature provides a useful framework for disentangling geometric effects on the coupled vortical and acoustic instability dynamics of three-dimensional hypersonic boundary layers.

\backsection[Acknowledgements]{This research used resources of the Oak Ridge Leadership Computing Facility at the Oak Ridge National Laboratory, which is supported by the Advanced Scientific Computing Research programs in the Office of Science of the U.S. Department of Energy under Contract No. DE-AC05-00OR22725. The authors gratefully acknowledge the computational resources provided by the Tuolumne supercomputer at Lawrence Livermore National Laboratory. These high-performance computing resources were instrumental in enabling the numerical simulations presented in this work.}

\backsection[Funding]{This material is based upon work supported by the U.S. Department of Energy, National Nuclear Security Administration, under Award Number DE-NA0004266.}

\backsection[Declaration of interests]{{} The authors report no conflict of interest.}

\backsection[Data availability statement]{The data that support the findings of this study are openly available in the Exasim GitHub at https://github.com/exapde/Exasim.}

\backsection[Author ORCIDs]{A. Welter, https://orcid.org/0009-0009-6298-6649}




\bibliographystyle{jfm}
\bibliography{jfm}

\end{document}